\documentclass[12pt]{article}
\pdfoutput=1 

\usepackage{amsmath,amsfonts,amssymb}
\usepackage{psfrag}
\usepackage{enumerate}
\usepackage{mathrsfs}
\usepackage{graphicx}
\usepackage{wrapfig}
\usepackage{xcolor}
\usepackage{caption}
\usepackage{amsthm}

\usepackage{jheppub}
\allowdisplaybreaks

\numberwithin{equation}{section}

\makeatletter
\def\@fpheader{\relax}
\makeatother

\newcommand{\be}{\begin{equation}}
\newcommand{\ee}{\end{equation}}
\newcommand{\beq}{\begin{eqnarray}}
\newcommand{\eeq}{\end{eqnarray}}

\def\[{\left [}
\def\]{\right ]}
\def\({\left (}
\def\){\right )}

\def\r2{\sqrt{2}}

\newcommand{\bbibitem}[1]{\bibitem{#1}\marginpar{#1}}

\newcommand{\myeq}[1]{\begin{equation} #1 \end{equation}}
\newcommand{\myal}[1]{\begin{align} #1 \end{align}}

\usepackage{stackengine}
\stackMath

\def\Label#1{\label{#1}%
  \smash{\hbox to0pt{\raise1ex\hbox{\tiny[#1]}\hss}}}
\def\noLabels{\let\Label=\label}
\def\nobbibitem{\let\bbibitem=\bibitem}

\title{On black hole interior reconstruction, singularities and the emergence of time}
\author[a]{Jan de Boer,}
\author[b]{Daniel Louis Jafferis}
\author[c]{and Lampros Lamprou}%

\affiliation[a]{Institute for Theoretical Physics and Delta Institute for Theoretical Physics, University of Amsterdam, PO Box 94485, 1090GL, Amsterdam, The Netherlands}
\affiliation[c]{Center for the Fundamental Laws of Nature, Harvard University, Cambridge, MA 02138, USA}
\affiliation[c]{University of British Columbia, Vancouver, BC V6T 1Z1, Canada}

\abstract{We propose a CFT definition of local observables in both the exterior and interior of bulk black holes, whenever such an interior exists. We achieve this by introducing a small microcanonical black hole as a ``probe'' and using its modular flow to propagate operators from the asymptotic boundary to the interior of other black holes along its worldline, elaborating on the ideas of [2009.04476]. The key conceptual advance is a CFT criterion for selecting states whose modular flow acts as geometric proper time translation in the bulk, which we dub ``local equilibrium'' states. Our interior reconstruction depends on the choice of code subspace but not on the specific black hole microstate and does not suffer from the ``frozen vacuum'' problem of other approaches. By virtue of our construction, the question of firewall typicality reduces to a technical problem we articulate and we identify a CFT correlator that is expected to signal the approach to the black hole singularity via a universal divergence. We end with comments on the utility of our framework to the quest for a quantum description of de Sitter cosmologies.
}

\begin{document}
\noLabels 
\nobbibitem 

\maketitle

\flushbottom
\newpage
\tableofcontents

\section{Introduction}

\subsection{Emergent time}
In a theory of gravity, time is an unphysical, gauge degree of freedom. An exception to this statement is found in spacetimes with boundaries or asymptotic regions where gravitational fluctuations are tamed and an asymptotic clock becomes available. It is hardly a surprise that the quantum theories of AdS gravity we have a non-perturbative definition of, identify the generator of this asymptotic time with the microscopic Hamiltonian. There exists, however, another operationally more relevant concept of time: The proper time, experienced by a particular internal observer of the system, like ourselves, in their rest frame. This semiclassical notion is physical because it is \emph{relational}, following from a comparison of the state of the Universe with the state of the selected observer. In an AdS quantum gravity theory, this notion is, also, \emph{emergent}: Its Hamiltonian is not an operator we get to externally prescribe but should be implicitly determined by the system's state, dynamics and choice of observer, via some appropriate principle. One of the main goals of this paper is to articulate such a principle.

The question of the emergence of time in quantum gravity is not an esoteric philsophical pursuit but, in fact, a practical one, for three important reasons:
\begin{enumerate}
    \item A first-principles CFT construction of an observer's proper time generator $H_{p.t.}$ offers a new, background independent and conceptually richer perspective on the AdS/CFT dictionary: It allows us to \emph{define} local fields deep in the bulk via the propagation of near boundary operators with $H_{p.t.}$, without presupposing knowledge of the background bulk geometry, or the field equations of motion. The latter are, instead, outputs of the construction, i.e. ways to efficiently organize the so-defined bulk operators.
    \item Puzzles associated with the black hole interior, like the nature and resolution of the black hole singularity in the CFT, whether a given black hole microstate has a semi-classical interior or how typical firewall states are, are all questions most naturally articulated in the infalling reference frame. A CFT definition of $H_{p.t.}$ enables the construction of observables in the black hole interior by the same principle as above, helping us illuminate these important issues.
    \item There are gravitational systems of physical interest, e.g. de Sitter cosmologies, for which no spatial asymptotic boundary exists. It is, therefore, conceivable that the only available clocks in their quantum description are emergent ones, meaningful only after separating the system into ``observer'' and ``environment'' subsystems. 
\end{enumerate}
In this paper, we engage directly with problems 1 and 2 above in Sections~\ref{sec:exteriorreconstruction} and~\ref{sec:insidebh}, respectively, and offer some preliminary comments on problem 3 in our final Section~\ref{sec:discussion}. We include a brief outline of our construction and main results in Section \ref{sec:outline}.
\begin{figure}[t!]
    \centering
    \includegraphics[width=6cm]{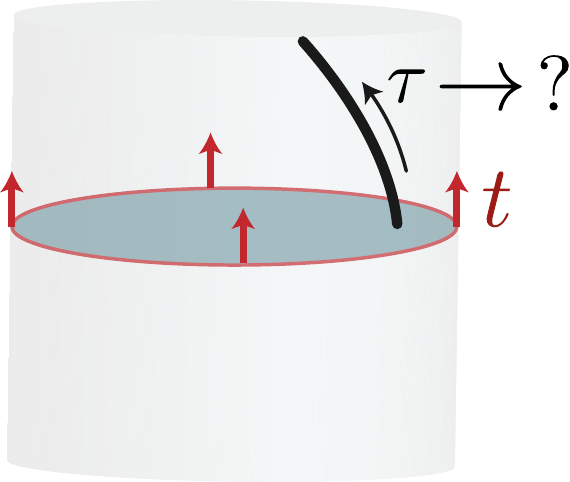}\hspace{1cm}
    \includegraphics[width=5cm]{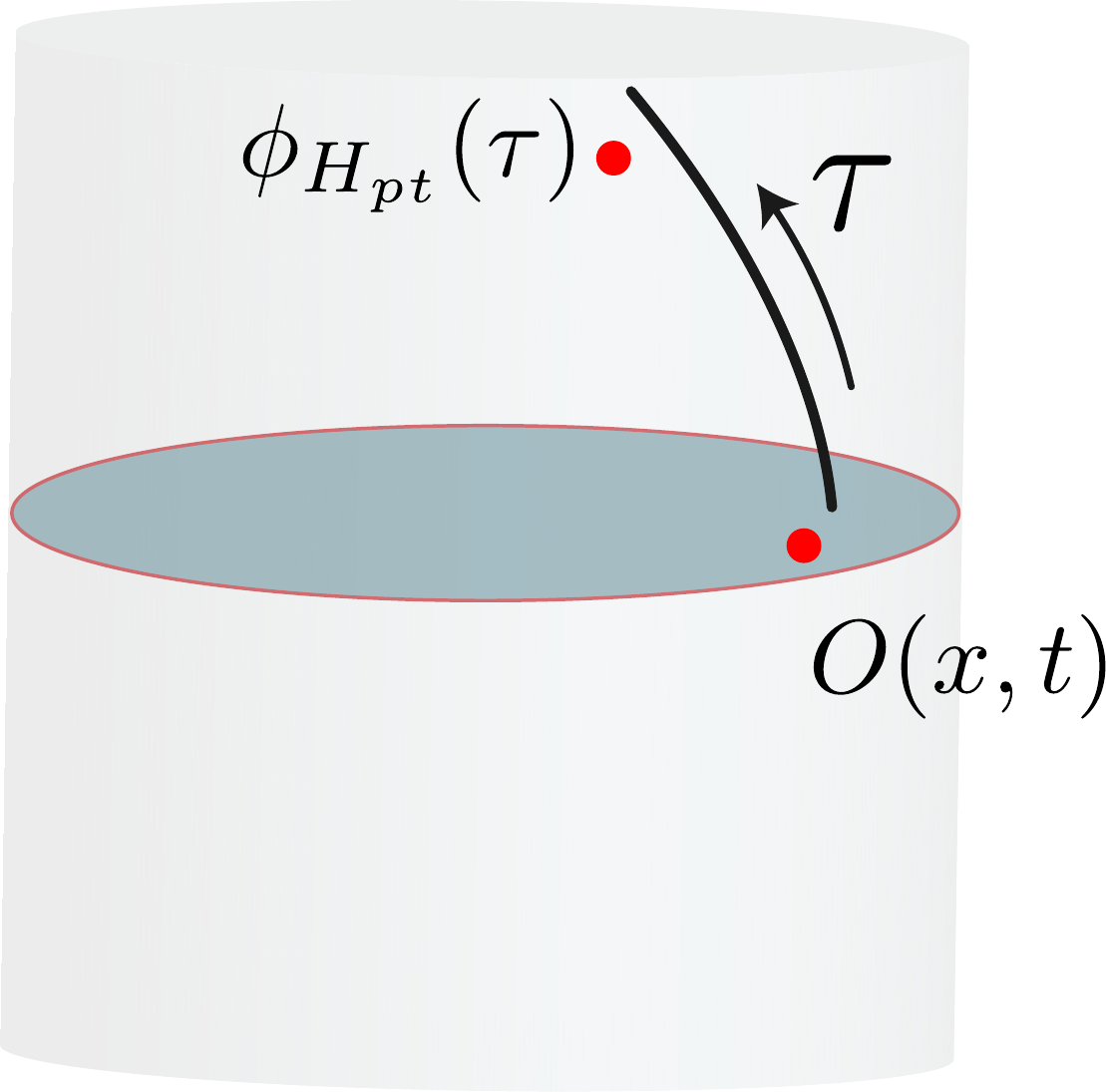}
    \caption{\footnotesize{\textbf{Left:} Two physical notions of time in AdS: Asymptotic time and proper time of a selected observer. The generator of the first is the boundary CFT Hamiltonian while deriving the generator of the second from first CFT principles is the subject of this work. \textbf{Right:} With a CFT definition of the proper time Hamiltonian, $H_{p.t.}$, bulk reconstruction can be expressed in a background independent way as the evolution of boundary observables with $H_{p.t.}$. If the background spacetime contains a black hole, our prescription provides a reconstruction of the interior region that can be causally accessed by infalling observers.} }
    \label{fig:mainpoint}
\end{figure}

\subsection{The last vestiges of the information puzzle in AdS}\label{sec:infoparadox}
One of our primary motivations for this work was to develop a bulk reconstruction technique in AdS/CFT suitable for addressing the following facet of the black hole information problem:

\emph{Given full computational control over a holographic CFT and a state dual to an AdS black hole, what precise CFT operator can unambiguously predict the statistics of bulk field measurement performed by an infalling observer at some behind-the-horizon point? }

It is useful to note that, as of now, there exist no general satisfactory definition of these CFT operators, except in special cases.\footnote{There are of course a number of interesting proposals that inform our approach. Their status and comparison to ours is discussed in Section \ref{sec:discussion}.} This means that even if ${\cal N}=4$ $U(N)$ Super Yang-Mills was successfully simulated on some condensed matter system, we would not be able to tell our experimentalist colleagues what measurement to perform to diagnose the existence or absence of a firewall. It is, therefore, a conceptual problem and, in fact, one of the last vestiges of the information problem in AdS. In view of the recent progress in understanding black hole evaporation, we deem it useful to utilize the remainder of this Section for carefully justifying the last statement.

\begin{figure}
\centering
\includegraphics[width=4cm]{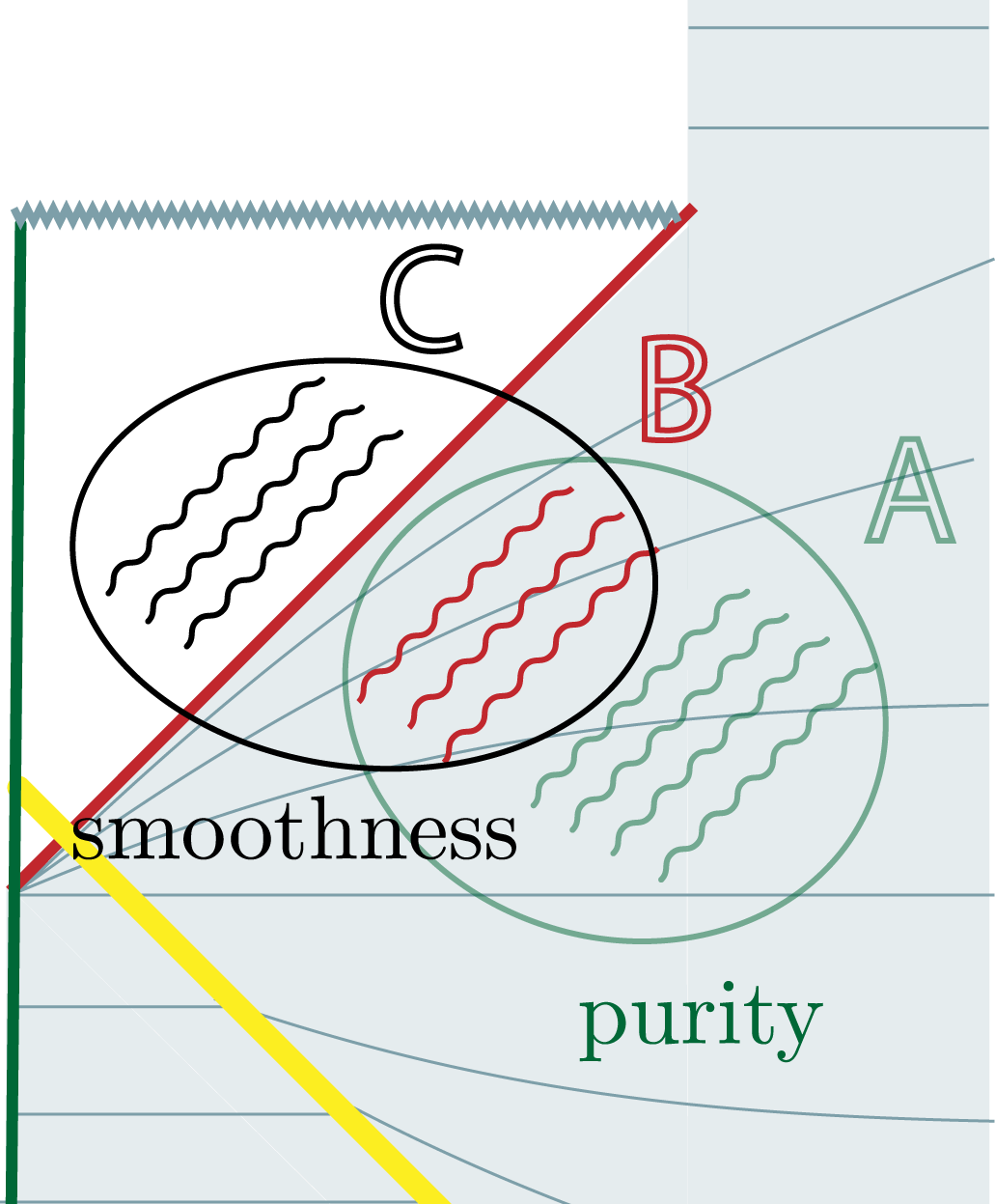}\quad\quad\quad
\includegraphics[width=4cm]{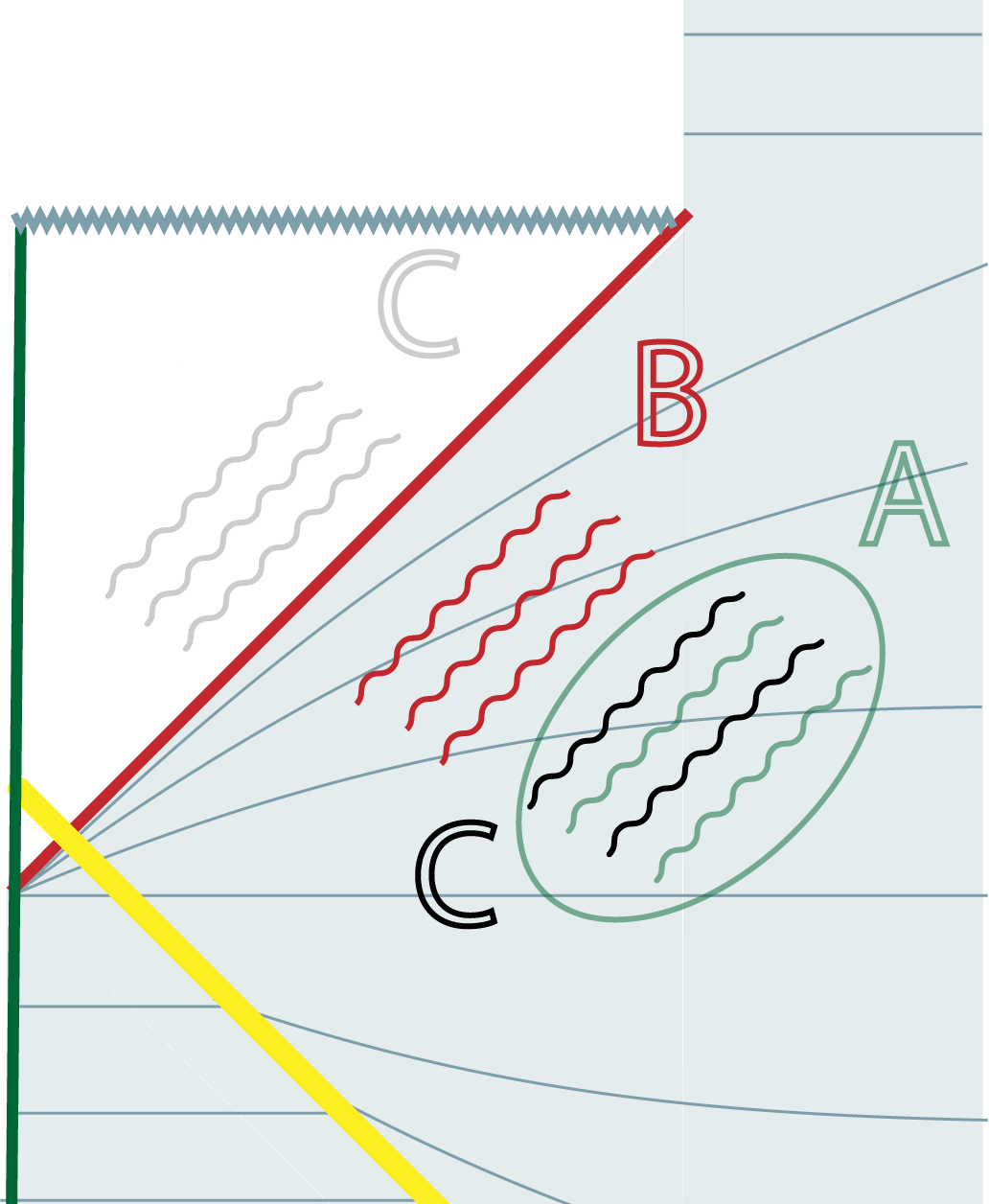}
\caption{\footnotesize{A useful way of articulating the information puzzle is as a tension between the entanglement of subsystems $A$ and $B$ needed for the purity of Hawking radiation and the entanglement of $B$ and $C$ necessary for an uneventful horizon, after the Page time. The logical tension is evaded in the ``entanglement island'' framework where $C$ is identified with a subsystem of $A$ after the Page time.}}
\label{fig:infoproblem}
\end{figure}
The information problem can be viewed from a plethora of angles. For our purposes it is optimal to focus on the tension between the unitarity of black hole evaporation and the smoothness of semiclassical spacetime in the neighborhood of the horizon. The source of the tension is the impossible, by the rules of quantum theory, entanglement pattern required for both unitarity and smoothness to simultaneously hold, after a black hole radiates half of its initial entropy (fig.~\ref{fig:infoproblem}). For AdS black holes, even when they are small and thermodynamically unstable, unitarity of evaporation is a manifest fact in the dual CFT description. This property was recently explicitly confirmed in the bulk for some theories, in computations of the time evolution of the Hawking radiation entropy \cite{Penington:2019npb,Almheiri:2019psf,Penington:2019kki, Almheiri:2019hni} that directly use the Euclidean gravitational path integral. The original information puzzle then gives its place\footnote{A caveat here is that the particular Euclidean saddles that restore unitarity in the entropy computation are only rigorously understood in gravitational theories that have an ensemble interpretation. It is currently unclear to what degree these Euclidean path integral rules depend on the choice of UV theory and the presence or absence of disorder in it.}  to the question of under what conditions ---if at all--- a semiclassical interior is experienced by an infalling observer, as General Relativity suggests, and whether the macroscopic, uneventful black hole interior of the GR solutions is an appropriate description for generic black hole microstates or not. Naively, the statements above imply that this expectation is misguided: time evolution appears to disrupt the entanglement structure at the horizon and a ``wall of fire'' or some other unknown structure awaits the unfortunate astronaut at the, classically uneventful, horizon \cite{Almheiri:2012rt,Almheiri:2013hfa}.

The \emph{logical contradiction} is evaded in the framework of black hole complementarity \cite{Susskind:1993if} due to the \emph{identification} of the black hole interior degrees of freedom with a subset of early radiation modes ---an idea that was sharpened by the identification of entanglement with connections by wormholes \cite{Maldacena:2013xja}, and explicitly demonstrated in the ``entanglement island'' computations \cite{Penington:2019npb,Almheiri:2019psf,Penington:2019kki, Almheiri:2019hni}.\footnote{The fact that the independence of interior and exterior subsystems is the most shaky assumption implicitly used in the firewall argument can already be seen from the semi-classical analysis of \cite{Raju:2021lwh} which is roughly the observation that the Gauss' law requires interior operators to be dressed to the asymptotic region, resulting in non-vanishing commutators between interior and exterior operators controlled by $G_N$. Since the black hole entropy is $O(G_N^{-1})$, it is not obvious there exists a $G_N\to 0$ limit where the subsystems become independent without the black hole entropy ---and hence all evaporation time-scales like the Page time--- becoming infinite and, by extension, obscuring the firewall puzzle. Nevertheless, in order to show that this dependence of interior and exterior modes is such that ensures the radiation entropy follows a Page curve the island argument is required, which is non-perturbative.} This ensures that the entanglement required by unitarity can simultaneously serve as the entanglement necessary for smoothness. Nevertheless, a large amount of entanglement across the horizon, while \emph{necessary} for smoothness, is \emph{not sufficient}; the pattern of entanglement matters as well. Unitaries $U_{in}$ acting in the interior preserve the mutual information between interior and exterior but can dramatically alter the state seen in the infalling frame. The pair of complementary Rindler wedges of a flat space QFT in two distinct but equally entangled states, depicted in figure~\ref{fig:frozenvacuum}, provides an elementary illustration of this point. Any interior reconstruction proposal that is insensitive to the effect of such unitaries would lead to the unacceptable conclusion that all sufficiently entangled states look like the local vacuum state near the horizon ---implying a ``frozen vacuum'' that can never be excited unitarily \cite{Bousso:2013ifa}---which contradicts the ordinary QFT intuition of fig.~\ref{fig:frozenvacuum}. We are thus led to the question: How can we tell whether the interior of given black hole state is excited? Even in view of entanglement islands guaranteeing unitarity of evaporation, the question of firewalls persists!

Diagnosing smoothness requires a CFT probe that is sensitive to the unitary frame of the interior experienced by the infalling observer. In the simple QFT example of fig.~\ref{fig:frozenvacuum}, the prototypical probes of this type are expectation values of \emph{local} operators behind the Rindler horizon, e.g. measuring the local energy-momentum tensor behind the horizon unambiguously informs us about the presence of excitations on the other side. In other words, QFT locality selects a special operator basis for the interior algebra and expectation values of elements of this basis help us attribute physical interpretation to quantum states. Carrying this observation over to our black hole setup, the key to putting the information paradox and its surviving descendants to rest is to identify a natural and physically reasonable \emph{principle} that unambiguously selects those CFT operators\footnote{or radiation operators, depending on the context} which are dual to what a bulk infalling observer interprets as \emph{local fields} behind the black hole horizon, in an arbitrary black hole state. This is what this paper aims to accomplish.

\subsection{Our construction and key results}\label{sec:outline}

We will be able to make progress on this problem by introducing a physical observer that is bound to fall in the black hole of interest and identifying the physical principle that defines the local operator algebra in the vicinity of their worldline. More precisely, we will construct the CFT unitary flow that relates \emph{local bulk operators} at different proper times along the geodesic to each other (fig.~\ref{fig:mainpoint}). Interior operators can then be obtained by applying this proper time evolution to exterior ones. We now summarize how to achieve this, elaborating on \cite{Jafferis:2020ora, Gao:2021tzr}.

\paragraph{An internal clock for QFT subsystems}
A toy version of the problem we want to solve can be articulated in the context of QFT on a rigid Minkowski spacetime. Two natural reference frames in this context are the global frame of an inertial observer and the Rindler frame, describing physics from the point of view of a constantly accelerated observer. Time evolution in the former frame is generated by the global dynamical Hamiltonian of the system $H_g$ which is an external input in the definition of the theory. The Hamiltonian in the latter frame is the Lorentz boost generator $B$ and we do not need to separately define it: It is already implied by (a) the definition of a dynamical system and (b) the selection of the observer, as is made clear by the following abstract definition of $B$. The accelerated observer only has causal access to a subalgebra of the global QFT, i.e. operators localized in their Rindler wedge, for example ${\cal A}_R$. Given this subalgebra and a QFT state $|\psi\rangle$, we can define the modular Hamiltonian $K_\psi$ which generates an inner automorphism of ${\cal A}_R$. When the state $|\psi\rangle$ is chosen to be the vacuum $|0\rangle$ of the global Hamiltonian $H_g$, the modular Hamiltonian $K_0$ coincides with the Hamiltonian of the accelerated observer, a fact that follows from Lorentz invariance. In view of this property, we may then take the vacuum modular Hamiltonian as the fundamental definition of $B$.

\begin{figure}
\centering
\includegraphics[width=4cm]{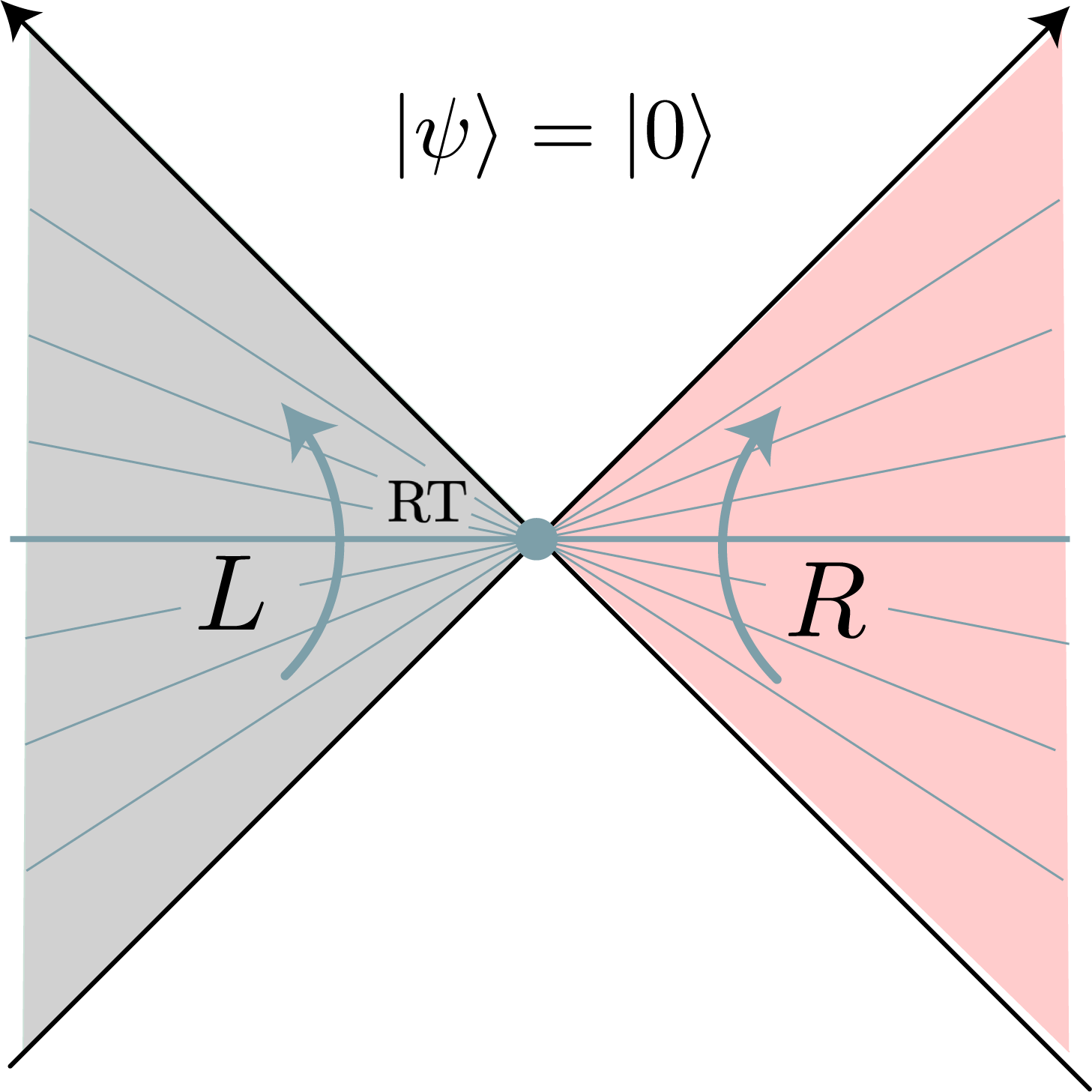}\quad\quad\quad
\includegraphics[width=4cm]{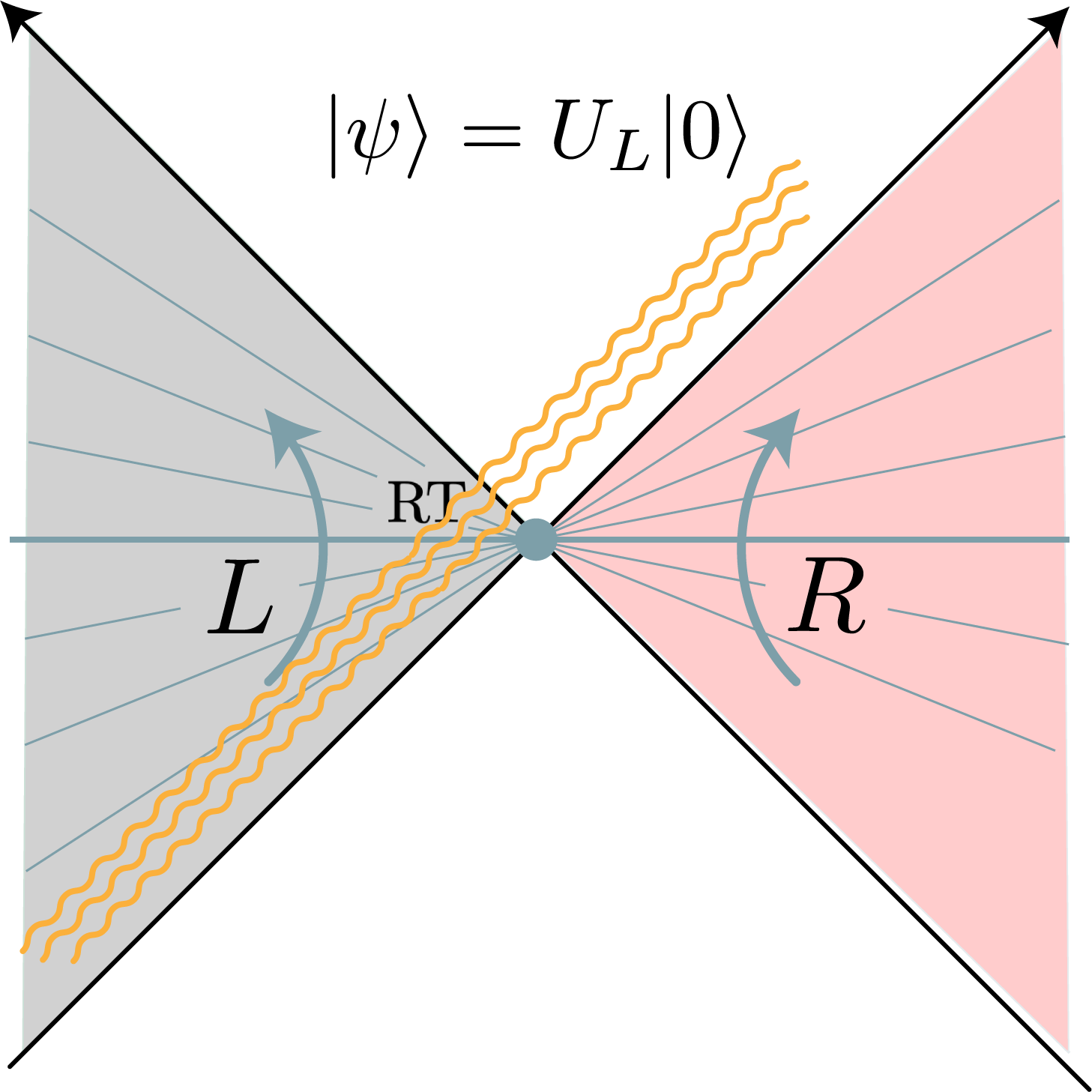}
\caption{\footnotesize{Having the appropriate amount of entanglement between two complementary subregions of spacetime is not enough for ensuring a smooth horizon crossing for an infalling observer from the right wedge: High energy shocks can be generated by unitaries localized in the complementary side (left Rindler wedge in the figure). These preserve the amount of Left-Right entanglement but change its pattern.}}
\label{fig:frozenvacuum}
\end{figure}

\paragraph{An internal clock in the bulk} The central idea of our construction is to leverage the elementary QFT observation above to define an internal clock in the bulk description of a holographic CFT. We will achieve this by introducing a small probe black hole in the bulk which is entangled with an external reference system ---a setup we will occasionally refer to as a ``bulk observer''. We explain how to do this concretely in the CFT in Section \ref{sec:framework}, addressing a number of subtleties. The technical advantage of doing this is that the near-horizon region of our probe black hole looks identical to the pair of complementary Rindler wedges of figure~\ref{fig:frozenvacuum}, where our observer's causal wedge is the rest of the bulk Universe we embedded the black hole in and the complementary wedge is the auxiliary reference system we introduced. Once again, we have a natural operator associated to the subalgebra available to our observer, the modular Hamiltonian $K_\psi$ associated to the state $|\psi\rangle$ we prepared the theory in. In special states, which we dub ``local equilibrium'' states $|\psi_{eq}\rangle$ and may be viewed as the analog of the local vacuum around our observer, the modular Hamiltonian $K_{eq}$ acts geometrically as a near horizon boost, or more precisely Schwarzschild time translation. This modular flow, therefore, propagates operators along the worldline of our small probe black hole while keeping their location relative to its apparent horizon fixed, in the entire code subpace\footnote{The ``code subspace'' is by now the standard nomenclature \cite{Almheiri:2014lwa} for referring to the subspace of the CFT Hilbert space spanned by bulk QFT excitations about a certain semi-classical background, due to its quantum error correcting properties. In this work, these properties do not play a substantial role but the subspace itself does and we will use the same terminology when referring to it.} about the original state. Since the operator $K_{eq}$ can be defined directly in the CFT, this prescription identifies for us the microscopic generator of a bulk observer's proper time. 

This idea, which was first discussed in the earlier work \cite{Jafferis:2020ora}, is reviewed and elaborated on in a number of important ways in Section \ref{sec:framework}. Section \ref{sec:stateprep} details a general Euclidean path integral prescription for introducing the ``observer'' black hole in any semi-classical state of interest, by generalizing the method of \cite{Gao:2021tzr} to CFTs. A number of essential technical subtleties are clarified in Section \ref{sec:microcanonicalbh}, including the use of ``microcanonical'' black holes to probe sub-AdS scales, the corrersponding Gregory-Laflamme instability, the way to localize their bulk wavefunction and, crucially, a double-scaled $G_N\to 0$ limit in which the probe's Schwarzschild radius goes to zero in $L_{AdS}$ units while its scrambling time is kept fixed which, as we explain, is important for probing the interior of bigger black holes. The key argument for the modular flow acting geometrically in \emph{local equilibrium states} is, then, summarized in Section \ref{sec:equilibrium}. 

\paragraph{Identifying local equilibrium states} For the prescription above to be well-defined, we need a CFT criterion that selects the set of local equilibrium states in every code subspace ${\cal H}_{code}$ we want to study. One of the main technical developments in this work is identifying an extremization principle that selects this preferred class of states $|\psi\rangle_{eq} \in {\cal H}_{code}$. The phenomenon we exploit is inherent to our setup. Non-equilibrium states, by definition, contain bulk particle excitations that cross paths with our probe. Because our probe is a black hole these infalling quanta get exponentially blueshifted as they approach it, due to its near horizon geometry. As a result, even the most innocuous infalling excitation incurs a dramatic effect on the modular evolution of any initial operator $\phi$ located near the probe $\phi_{K_\psi}(\tau) = e^{iK_{\psi}\tau} \phi e^{-iK_{\psi}\tau}$.\footnote{The bulk excitations do not strictly speaking need to fall in the probe black hole for the onset of scrambling; skirting the atmosphere and disturbing its local thermal equilibrium suffices. The reason is that such scattering processes are inelastic and the energy deposited in the atmosphere dissipates by getting absorbed by the black hole. It is this infalling energy flux that underlies scrambling.} The key distinction is that when $|\psi\rangle$ is a local equilibrium state, $\phi_{K_\psi}(\tau)$ is just a local bulk operator, transported along a trajectory at a fixed distance from the probe whereas, when it is not, modular flow generates an out-of-time-order product of bulk operators which becomes substantially complex as $\tau \to \log S_{probe}$. Based on this observation we discuss the following two extremization prescriptions for mathematically diagnosing this phenomenon: 
\begin{enumerate}
    \item As long as our probe stays outside of any background black hole horizons, a local equilibrium state $|\psi\rangle_{eq}$ is a state in ${\cal H}_{code}$ whose modular flow $e^{iK_{eq}\tau}$ \emph{maximizes} the correlation function of any bulk operator $\phi_{K_{\psi}}(\tau)$ with boundary operators $O(t)$, in a kinematic limit we explain in detail in Section \ref{sec:nonequilibrium}, \emph{if and only if} the maximal correlators are $O(S_{probe}^0)$. This follows from the observation that for $|\psi\rangle \neq |\psi\rangle_{eq}$ the relevant correlators are out-of-time-order and, in our limit, are parametrically suppressed in $1/S_{probe}$ as compared to equilibrium ones which are in turn time-ordered and remain $O(S_{probe}^0)$. This prescription is technically simple to implement and straightforward to establish but fails when the probe is in the interior of another bigger black hole. 
    \item As a generalization of the previous prescription that continues to work in the interior of background black holes, we conjecture that a local equilibrium state $|\psi\rangle_{eq}$ is one for which the modular flowed operator $\phi_{K_{eq}}(\tau)$ leads to the \emph{minimal increase of state complexity} among all operators $\phi_{K_{\psi}}(\tau)$ for $|\psi\rangle \in {\cal H}_{code}$, as $\tau \to \log S_{probe}$, \emph{if and only if} $\frac{||\phi_{K_{eq}}(\tau_{scr})||}{||\phi||} = O(S_{probe}^0)$, where $||\cdot ||$ a choice of operator norm in ${\cal H}_{code}$. We provide evidence for this conjecture in Section \ref{sec:interiorreconstruction}.
\end{enumerate}
It is important to observe that neither of the criteria guarantees the existence of a local equilibrium state in a given ${\cal H}_{code}$: A solution to the relevant extremization problem always exists but it may not satisfy the requirement that correlators of modular flowed operators be $O(1)$. Such situations are expected to occur when the probe undergoes a high energy collision or encounters a high curvature region, signaling the breakdown of our reconstruction method and potentially of the semi-classical description all together.

Furthermore, note that the second prescription implies the first when restricting to the latter's regime of validity: The $O(1)$ overlap of $\phi_{K_{eq}}(\tau)$ with simple boundary operators implies that we can represent it in the CFT in terms of simple operators. The second prescription above, if correct, resolves in principle the problem of black hole interior reconstruction: The modular Hamiltonian of the infalling observer for a code subspace state that minimizes operator complexity along its flow can be used to define local interior observables in the CFT, offering an unambiguous answer to the information-problem-related question of Section \ref{sec:infoparadox} and allowing us to concretely ask questions about the presence of firewalls, their typicality in the Hilbert space and the behavior near the singularity.

\paragraph{Black hole singularities and firewalls}
In Section \ref{sec:singularity}, we also utilize our method to probe the physics near the black hole singularity from the holographic CFT. We show that as our observer's modular flow transports an operator close to the background black hole's singularity, its two-point function with boundary operators is expected to exhibit a universal logarithmic divergence in a semi-classical ($G_N\to 0$) double scaling limit we discuss. Since the microscopic modular flowed correlator is not expected to be divergent in the finite $N$ theory, the resolution of the divergence above is an important step towards understanding the physics of the singularity. We set up the CFT computation we need to perform for this task in terms of a Euclidean path integral (figure~\ref{fig:twosidedBHreplica}) and identify the CFT data that is required to perform it but postpone its detailed calculation to future work. Importantly, the modular time at which this divergence is encountered serves as a measure of the geometric size of the black hole interior region, due to its relation to the proper time the infalling probe takes to hit the singularity. Therefore, the typical value of this modular time and its variance among single-sided black hole states in a given energy window offers a well-defined diagnostic of the typicality of firewalls.

\begin{figure}
\centering
\includegraphics[width=6cm]{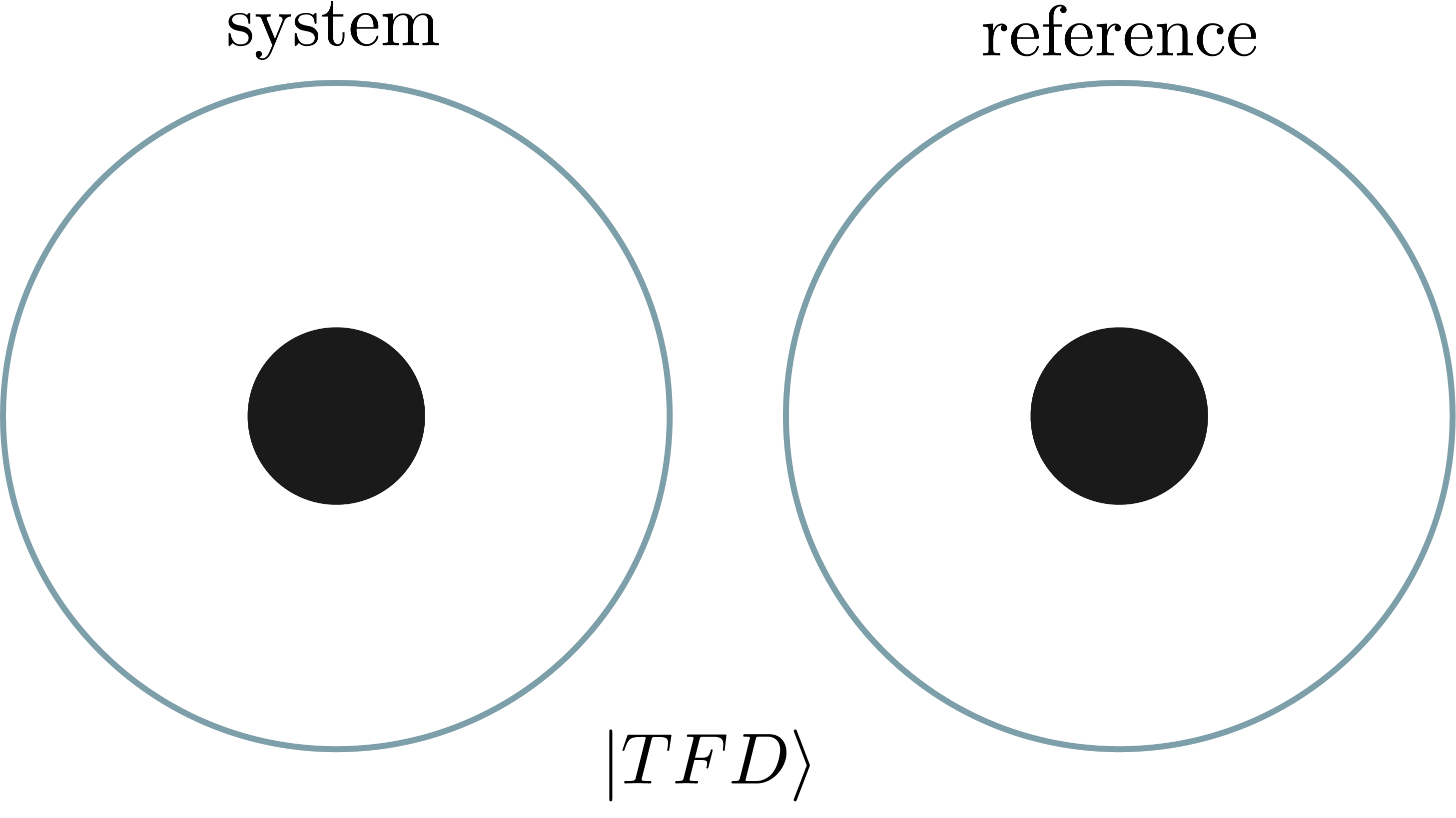}\quad\quad\quad
\includegraphics[width=6cm]{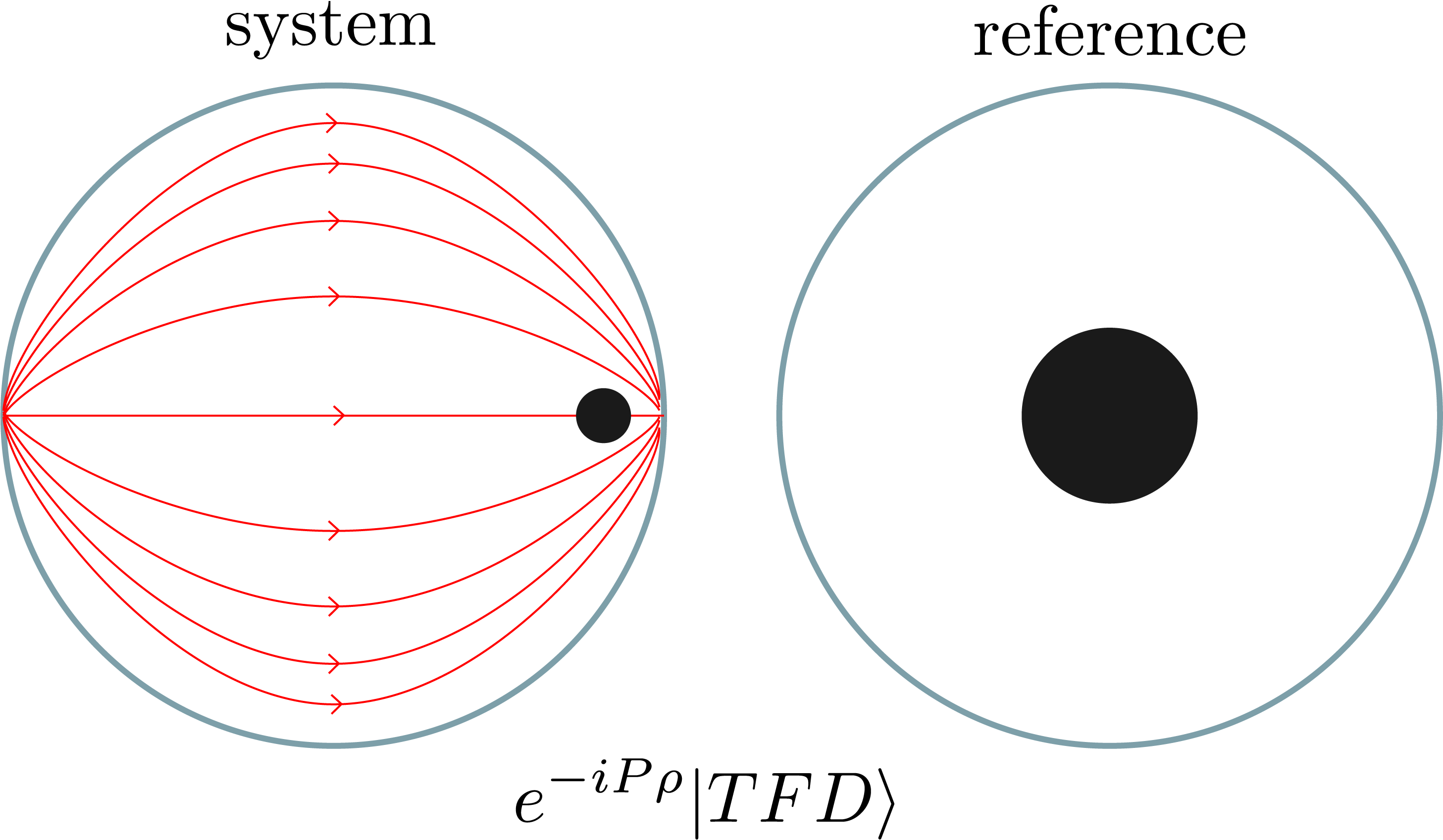} \\
\includegraphics[width=6cm]{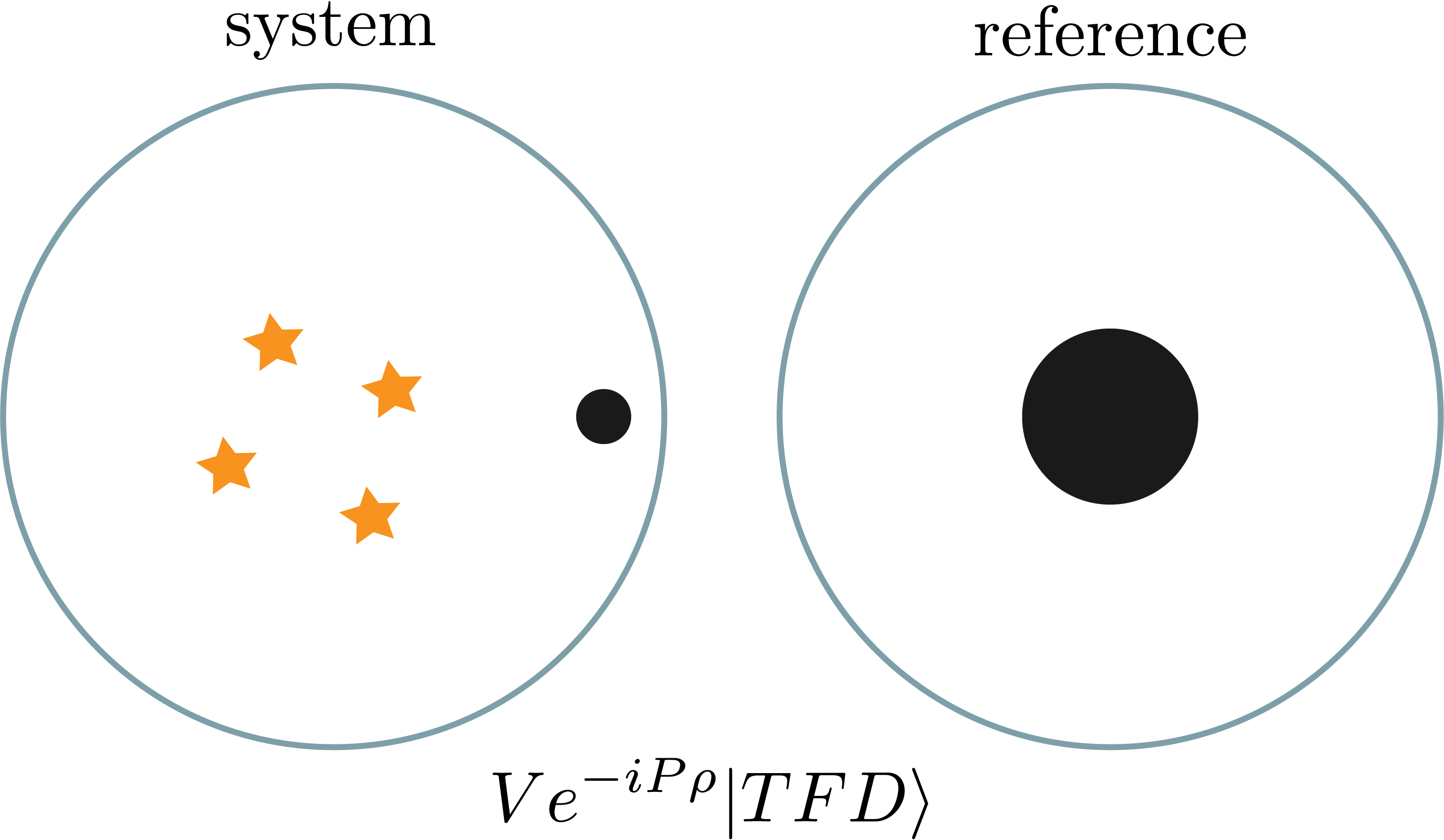}\quad\quad\quad
\includegraphics[width=6cm]{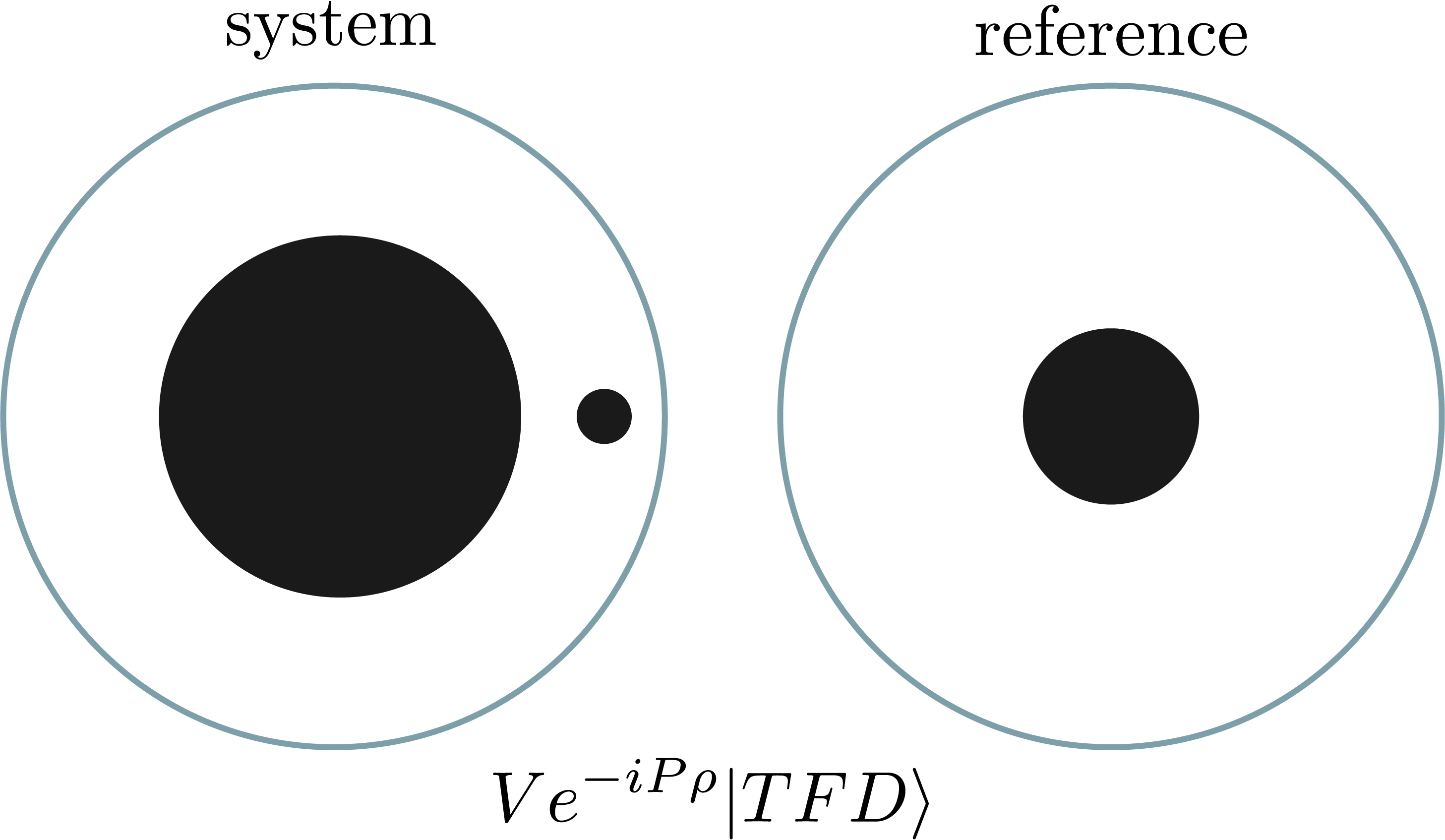} 
\caption{\footnotesize{Preparation of a CFT state containing a bulk probe black hole, following the steps detailed in the main text. \textbf{Top left:} Our holographic CFT and the auxiliary ``reference'' system are initiated in the thermofield double state,  describing a pair of bulk black holes connected by a wormhole (not explicitly shown in the figure). The system black hole is our ``observer''. \textbf{Top right:} We act on the system with a conformal transformation that shifts the ``observer'' towards the boundary. \textbf{Bottom panel:} We then act on the previous state with a unitary at a spacelike separation from the black hole that prepares the bulk configuration of interest, e.g some configuration of stars (\textbf{left}) or a background black hole (\textbf{right}).}}
\label{fig:stateprep}
\end{figure}

\section{The probe black hole framework}\label{sec:framework}
\subsection{Preparing the probe state}\label{sec:stateprep}
We begin this section by briefly summarizing the basic framework developed in \cite{Jafferis:2020ora}, referring to this earlier work for details. A bulk observer is generally a localized, semi-classical configuration of the bulk degrees of freedom with a certain mass. The latter is reflected in the long range gravitational field it sources. The most universal such configuration is a black hole of equal mass initialized in the same kinematical state. We may, in fact, regard the black hole as a physical version of the familiar pointlike approximation to the probe which also respects the laws of gravity that prohibit mass from being concentrated in spatial regions smaller than its Schwarzschild radius. We will choose such a black hole to be our ``observer'' or probe and identify the rules that enable us ``see'' the hologram from its point of view.

Our first task is to prepare the CFT state that describes such a black hole probe propagating in a general asymptotically AdS Universe. A pedagogical step by step procedure is depicted in (fig.~\ref{fig:stateprep}):

\begin{enumerate}
\item Consider the CFT dual to the AdS Universe we wish to explore, which we call the \emph{system}, as well as a copy of it we refer to as the \emph{reference}. Initiate the pair of CFTs in the \emph{thermofield double state} 
\begin{equation}
|\beta\rangle_{sys+ref} = \frac{1}{{\cal Z}^{\frac{1}{2}}} \sum_n \,e^{-\frac{\beta}{2}E_n} |E_n\rangle_{sys} |\bar{E}_n\rangle_{ref} \label{tfd}
\end{equation}
Holographically, this describes a static AdS black hole in the dual of $CFT_{sys}$ connected via a short Einstein-Rosen bridge to a similar static black hole in the reference, with the global geometry that of the eternal AdS-Schwarzschild. The \emph{system} black hole will serve as our \emph{observer or probe}. In the initial state we have chosen, this probe sits in an empty AdS Universe and is thermally entangled with the reference which we use as a convenient way to ``tag'' it from the outside. In the ultimate formulation of our state preparation, we also project (\ref{tfd}) onto a narrower energy window, switching to a ``microcanonical'' thermofield state that allows us to build semi-classical black holes that are smaller than $L_{AdS}$ in size. We postpone a detailed discussion of this to Section \ref{sec:microcanonicalbh}. 

\item Act on the system CFT with an asymptotic AdS translation $e^{-iP_{sys}\rho}$ for some large $\rho$, where $P_{sys}$ is the appropriate element of the boundary conformal algebra, in order to move the probe far out towards the AdS boundary. Then act with a unitary $V_{sys}$ on the resulting state to produce the bulk configuration of interest
\begin{equation}
|\psi\rangle = V_{sys}\, e^{-iP_{sys}\rho} |\beta\rangle_{sys+ref} \label{genstate}
\end{equation}
The unitary $V$ could be generating any non-trivial arrangement of stars and galaxies, or perhaps a much larger black hole our probe will ultimately be absorbed by. The latter is of course the case of greatest interest, since  the black hole interior is where our other bulk reconstruction tools fail.

It is worth noting that creating an ambient single-sided black hole that our probe will fall in (fig.~\ref{fig:stateprep}) is fairly simple from the CFT point of view. A unitary excitation $V$ selected randomly from the ensemble of all unitaries with fixed asymptotic charges creates a state which is with very  high probability a bulk black hole with those charges, assuming its asymptotic energy is sufficiently high. This follows simply from the thermodynamic dominance of black holes over other configurations of the same energy, which implies that the amplitudes of non-black hole branches of the wavefunction will be non-perturbatively small $\sim e^{-S_{BH}/2}$.
\end{enumerate}

\begin{figure}
\centering
\includegraphics[width=11cm]{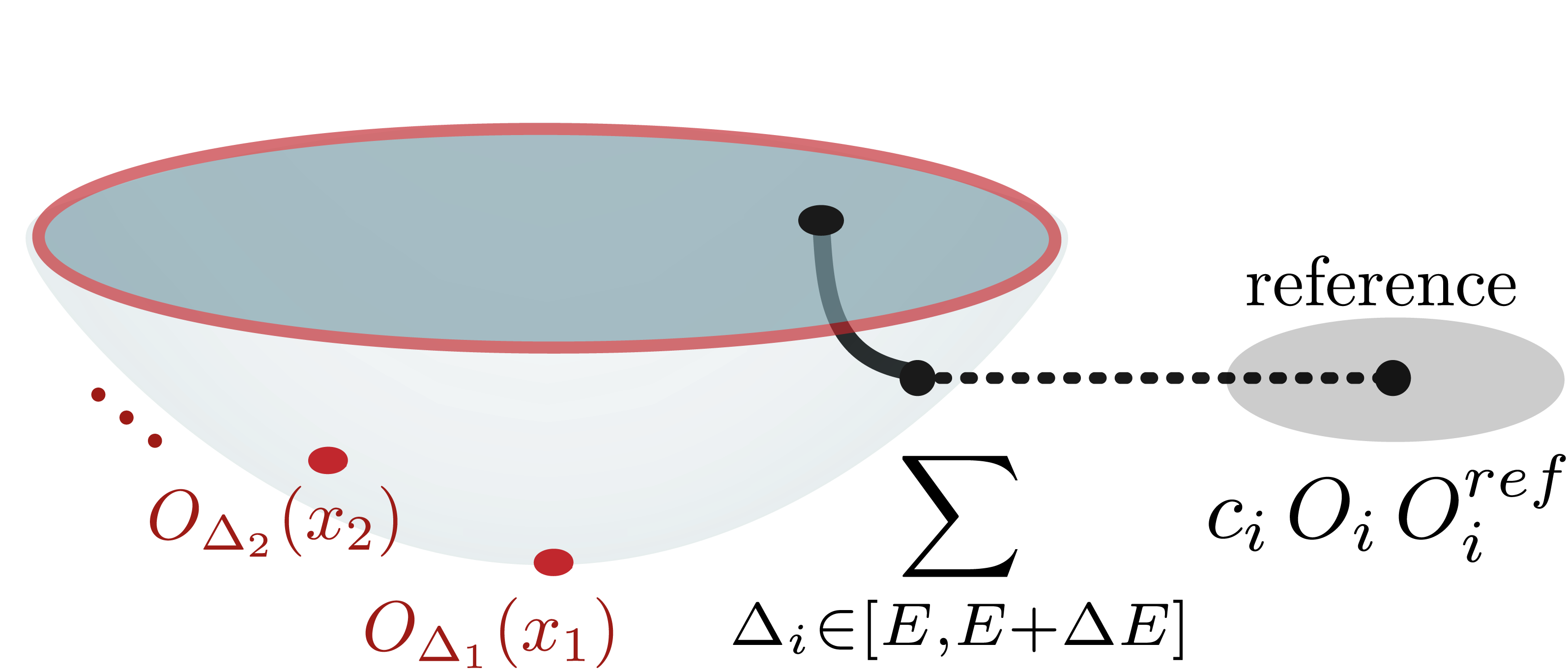}
\caption{\footnotesize{Preparation of a semi-classical CFT state that contains a bulk probe black hole, using a Euclidean path integral. The ``observer'' black hole is inserted by the sum over heavy bi-local operators that entangle the system and the reference CFTs. From the bulk point of view, this operator creates a wormhole connecting the observer black hole in the system with an identical copy in the reference. The bulk location of the observer is determined by the location of the Euclidean path integral insertion. The red operator insertions prepare the background state of interest, e.g. a background black hole.}}
\label{fig:euclideanprep}
\end{figure}

\paragraph{A Euclidean path integral preparation} It is useful both for computational purposes and for generalization to express the construction in terms of a Euclidean CFT path integral (fig.~\ref{fig:euclideanprep}). The key for doing so is to notice that the probe wormhole of the previous construction can be produced by inserting a sum over operators $O_i^{sys}O_i^{ref}$ in the Euclidean path integral that prepares the tensor product of the system and reference vacuum states, by virtue of the operator-state correspondence. The general prescription then is the following: Starting with the CFT that holographically describes our system, its Euclidean path integral with operator insertions and classical sources prepares a general semi-classical state of the bulk gravity theory. For example, placing a large ambient black hole at the AdS center is simply achieved by inserting some $O_{\Delta}(0)$ with $\Delta\sim O(N^2)$ at the origin of the Euclidean plane. We then introduce our probe into the system by (a) assuming a similar Euclidean path integral preparation of a state in the reference and (b) inserting a bilocal Euclidean coupling between the system and the reference
\begin{equation}
    \Sigma_{sys,ref} = \sum\limits_{\Delta_i\in [E_0,E_0+\Delta E]} c_i \,O^{sys}_{\Delta_i}(x) O^{ref}_{\Delta_i}(0) \label{wormholeop}
\end{equation}
The coefficients $c_i\in \mathbb{R}$ are there for normalizing the corresponding states and the location and width of the energy window $E_0,\Delta E$ are chosen from an appropriate parametric regime we discuss in detail in Section \ref{sec:microcanonicalbh}. As a consequence of the thermodynamic dominance of black holes, the insertion of $\Sigma_{sys,ref}$ introduces a black hole into the system which is entangled with the reference and has a size controlled by $E_0$.\footnote{This is a microcanonical black hole whose properties we discuss in detail in Section \ref{sec:microcanonicalbh}.} The latter is our probe. This path integral construction, while at face value different from the one descibed above, prepares an identical semi-classical bulk state. This preparation of the probe was used in \cite{Gao:2021tzr} to successfully explore the interior of the eternal AdS$_2$ black hole directy from the SYK system.

\begin{figure}
\centering
\includegraphics[width=7cm]{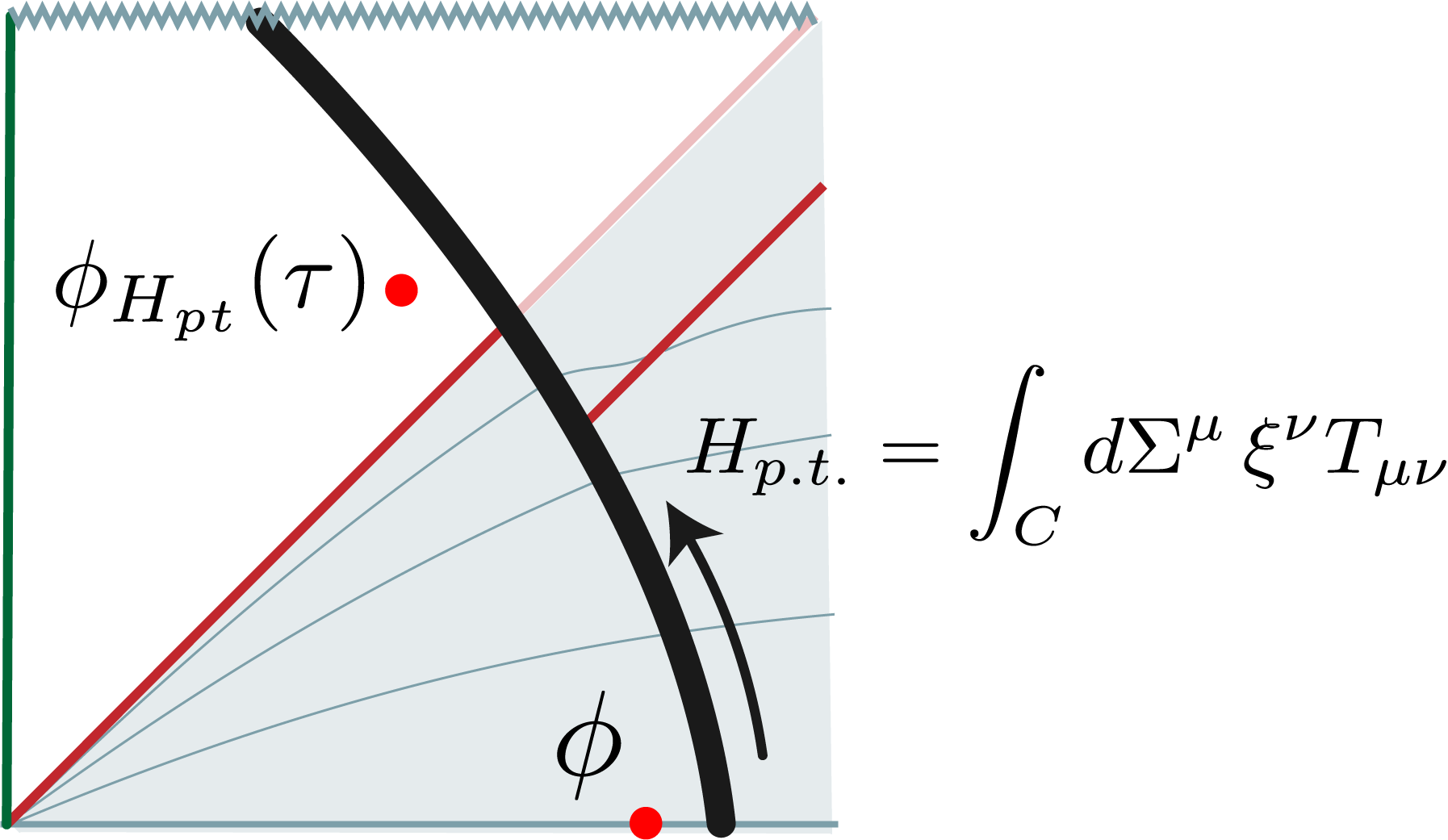}
\includegraphics[width=7.5cm]{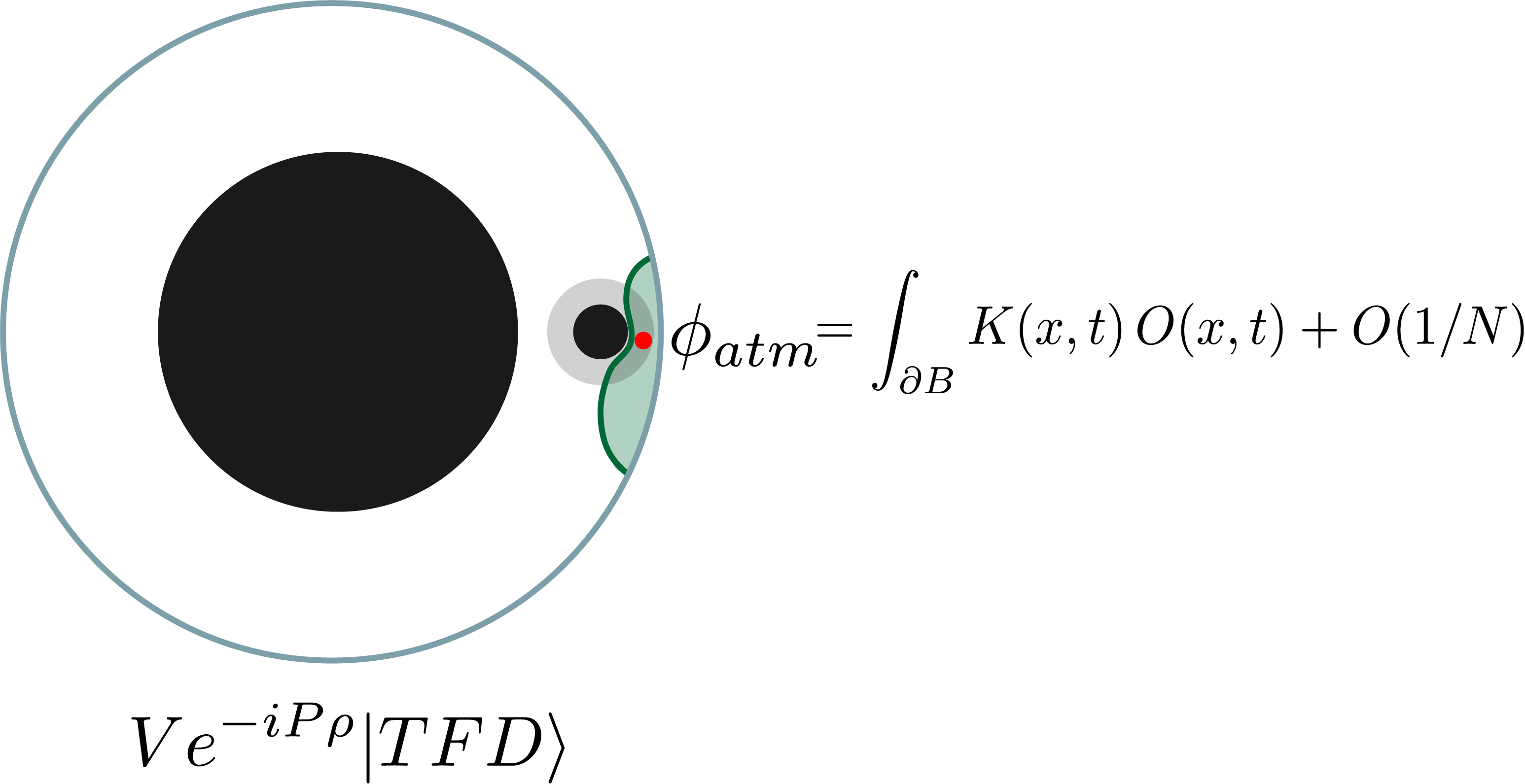}
\caption{\footnotesize{\textbf{Left:} Lorentzian spacetime generated by the initial conditions of fig.~\ref{fig:tfdobserver} or \ref{fig:euclideanprep}, in the approximation where the probe black hole is much smaller than the background one. From the outside, this describes a black hole merger. In the frame of the probe black hole, however, nothing special happens when we cross the background black hole horizon. The probe's Schwarzschild time evolution can then propagate bulk fields from the exterior of the background black hole to its interior. \textbf{Right:} Knowledge of the CFT representation of a bulk operator in the atmosphere of the probe at some initial moment in time is an input in our construction. This can be done by usual HKLL or, more generally, entanglement wedge reconstruction, since this is a bulk operator near the AdS boundary. This operator is then transported along the infalling black hole's Schwarzschild time using modular flow in a ``local equilibrium'' state (Section \ref{sec:nonequilibrium}).}}
\label{fig:history}
\end{figure}

\paragraph{The Lorentzian spacetime history} The resulting bulk configuration provides the initial condition for the Einstein's equations whose solution determines the rest of the spacetime history dual to the state $|\psi\rangle$ and its CFT time evolution. While the precise geometry may be fairly complicated, for a probe with Schwarzschild radius $R_{probe}$ much smaller than the local curvature radius of the ambient spacetime (e.g. the Schwarzschild radius of the big ambient black hole $R_{ambient}$ in fig.~\ref{fig:stateprep}), we can approximate it by a point-like particle propagating along a time-like geodesic in the background geometry, as shown in fig.~\ref{fig:history}.

When the probe black hole is inserted in a much bigger black hole background spacetime, the Lorentzian evolution looks, in the asymptotic AdS frame, like a black hole merger and the spacetime equilibrates to an AdS black hole with a slightly larger boundary energy, after a period controlled by the scrambling time of the ambient black hole $t_{scr}=\beta \log S$. On the other hand, in the probe black hole's frame, e.g. from the point of view of a planet orbiting the probe at a reasonably close distance, nothing remarkable happens at the moment it crosses the ambient event horizon, when $R_{probe}\ll R_{ambient}$. The small black hole continues to propagate in the interior of the larger one until it gets close to the singularity. 

\paragraph{The probe black hole's time} At any point along the evolution of fig.~\ref{fig:history}, in the infalling black hole's rest frame, the geometry around the probe is to a good approximation AdS-Schwarzschild,\footnote{Here we assume that the charges and angular momentum of the probe are zero for simplicity} up to potential small corrections from the backreaction of infalling matter whose energy is assumed for now to be small compared to the probe's rest mass. This local geometry has an approximate time-like killing vector $\xi^\mu$ generating Schwarzschild time translations near the observer's horizon. This defines a \emph{canonical clock} in the observer's frame which is in correspondence with the proper time along the infalling geodesic in the point-particle approximation. We will denote its quantum mechanical generator by $H_{p.t.}$. Semiclassically, this is simply $ H_{p.t.}= \int_C d\Sigma^\mu \xi^\nu T_{\mu\nu}$ where $T_{\mu\nu}$ is the bulk stress tensor and $C$ a bulk Cauchy slice. There is, of course, generally no unique extension of the vector field everywhere in spacetime, so this time evolution is only unambiguously defined within our probe's atmosphere. 

Our central objective is to obtain this internal Hamiltonian from first CFT principles. With access to $H_{p.t.}$, bulk reconstruction proceeds as follows: We start by assuming we have access to a local bulk operator near our probe at the initial slice, when it is conveniently located near the asymptotic AdS boundary. Such a local exterior field $\phi$ can be expressed as a CFT operator via any of the well-understood reconstruction techniques. We simply need to find a boundary subregion whose entanglement wedge contains $\phi$ and the existence of a CFT dual is guaranteed by \cite{Dong:2016eik}. Knowledge of this initial operator in the CFT is an \emph{input} in our construction. We then utilize the local Schwarzschild Hamiltonian $H_{p.t.}$ to propagate it along the infalling observer's trajectory anywhere in the bulk, including the ambient black hole interior (fig.~\ref{fig:history}). The entire approach, thus, hinges on identifying the correct CFT operator $H_{p.t.}$.

\subsection{Microcanonical description of small probes and a double-scaling limit} \label{sec:microcanonicalbh}
Before diving into the main part of our discussion, we need to address a very important subtlety: The size of our probe. The previous Section summarized a sequence of steps for preparing a CFT state containing a probe black hole entangled with a reference, living in some otherwise general asymptotically AdS space. The probe was introduced by entangling the system and reference CFTs in a thermofield double-like state. The dual of such states, however, is an AdS black hole only for temperatures above the Hawking-Page transition $\beta_{probe}\geq L_{AdS}$. Such black holes are known to have cosmological size, with the smallest accessible Schwarzschild radius set by $L_{AdS}$. Below the Hawking-Page temperature, the dominant branch of the bulk thermal state describes instead a thermal gas of particles in AdS. 

The ``probes'' discussed in the previous Section are, therefore, somewhat undeserving of their name. The situation is even worse if we intend to utilize such probe black holes to explore the interior of another ambient AdS black hole. This is due to the fact that the proper time $\tau_{sing}$ between crossing the horizon and hitting the singularity of a large AdS black hole along an infalling timelike geodesic is independent of the black hole radius and set by the cosmological constant $\tau_{sing} \sim L_{AdS}$. This is a serious problem for us, since this is equal to the thermal time of the smallest probe prepared via the thermofield double state. In contrast, as we explain in detail in Section \ref{sec:exteriorreconstruction}, for our prescription to work we need access to at least a scrambling time amount of proper time evolution within an AdS length, $\tau_{scr}^{probe} =\beta_{probe} \log S_{probe} \ll L_{AdS}$.

Fortunately, there is a simple fix. It is well-known \cite{Horowitz:1999uv, Marolf:2018ldl} that even though only AdS black holes with $R\geq L_{AdS}$ can ever be described in the \emph{canonical} thermal ensemble, switching to the \emph{microcanonical} ensemble allows us to describe black holes of \emph{parametrically} smaller sizes. A simple estimate \cite{Horowitz:1999uv, Jafferis:2020ora} reveals that black holes \emph{entropically} dominate\footnote{Despite having larger free energy} over a thermal gas of the same AdS energy for Schwarzschild radii as small as:
\myeq{ \frac{R_{micro}}{L_{AdS} } \approx  \left( \frac{L_{AdS}}{L_{pl}}\right)^{\frac{1-d}{2d-1}}\underset{N\to \infty}{\longrightarrow}0\label{microcanonicalR}}
It is clear that both the radius and the scrambling time of such microcanonical black holes, measured in $L_{AdS}$ units can be taken to zero at the $ \frac{L_{AdS}}{L_{pl}} \to \infty$ limit, for all spacetime dimensions $d+1>2$. Microcanonical probes can, therefore, fit comfortably inside large AdS black holes and allow for the reconstruction technique developed in this work to access their interiors. We can then proceed with the construction of the previous Section by simply changing the initial state of our ``observer'' to the microcanonical thermofield double:
\myeq{|E_0,\sigma \rangle = \frac{1}{{\cal Z}^{1/2}} \sum_{n} f(E_n|E_0,\sigma) e^{-\frac{\beta}{2} E_n } |E_n\rangle_{sys} |\bar{E}_n\rangle_{ref} \label{microtfd}}
where $f(E_n|E_0,\sigma)$ is some enveloping function that effectively restricts the sum over $E_{n}$ to a microcanonical window of width $\sigma$ centered around energy $E_0$. A natural choice is a Gaussian $f(E|E_0,\sigma) = \exp[ -\frac{(E-E_0)^2}{2\sigma^2}]$ but any sufficiently smooth and fast-decaying function works. The resulting reduced state on each side describes a black hole in equilibrium with its Hawking radiation.

\subsubsection*{Clarifying a few subtleties}
There are three subtleties regarding microcanonical black holes. One was discussed in detail in \cite{Marolf:2018ldl} and it concerns the width of the microcanonical window. According to the analysis of \cite{Marolf:2018ldl}, in order for $|E_0,\sigma \rangle$ to describe a small semi-classical wormhole, it is necessary for $\sigma$ to be $1\ll \sigma \ll G_N^{-1/2}$. The upper bound is set by the width of the canonical ensemble, whereas the basic argument for the lower bound is that, by virtue of the uncertainty principle, making the width too small results in large quantum fluctuations in the Schwarzschild time difference between the near horizon regions of the system and the reference. This de-correlates the two sides, leading to a quantum mechanical wormhole. A more relevant for our purposes ---yet related--- argument for this lower bound can be made by recalling that we intend to use the microcanonical black hole as a \emph{clock}, with reference to which we will define bulk operators. For a quantum system to be a reliable clock with resolution $\delta \tau \ll 1$, its energy wavefunction needs to span a range $\delta E\sim \delta \tau^{-1}\gg 1$.

The second remark is about the localization properties of the probe. Since (\ref{microtfd}) describes black holes with $R_{micro}\ll L_{AdS}$, it is effectively a wavefunction of a black hole in approximately flat space. As in ordinary particle quantum mechanics, such wavefunctions tend to spatially spread diffusively, over a time scale set by the probe's mass. The equilibrium state $|E_0,\sigma\rangle$, therefore, does not describe a localized black hole wavepacket but rather one that is spread over an $L_{AdS}-$sized bulk region. An extra step is then required to localize the initial black hole configuration. It is easiest to do this in step 1 of the preparation procedure of Section \ref{sec:stateprep} when our black hole observer lives in an empty AdS Universe. We can then localize the wavefunction by coherently constraining the AdS momentum of the state. If $P$ is the conformal charge dual to radial momentum, a wavepacket localized within a proper size $\delta\ell$ region from the center of AdS reads:
\myeq{|E_0,\sigma, \delta\ell \rangle= e^{-\delta \ell^2 P^2} |E_0,\sigma\rangle \label{localmicrotfd}}

Putting everything together, by replacing $|\beta\rangle_{sys+ref}$ with our new state $|E_0,\sigma, \delta\ell \rangle$ in step 1 of Section \ref{sec:stateprep} and leaving the rest unchanged we arrive at the state
\myal{|\psi\rangle&={\cal Z}^{-1/2} \sum_{n} e^{-\frac{(E_n-E_0)^2}{\sigma^2}-\frac{\beta}{2} E_n }  \left( V_{sys} e^{-\delta \ell^2 P^2 -iP \rho}   |E_n\rangle_{sys}\right) |\bar{E}_n\rangle_{ref} \label{genstate2}\\
\text{with: } &1\ll \sigma\ll G_N^{-\frac{1}{2}} \, , \quad \delta \ell \ll 1 \, , \quad \rho \gg 1}
with our new probe black hole being parametrically smaller than $L_{AdS}$, a fact that will eventually allow us to reliably peek behind large black hole horizons.

The third subtlety is about the Gregory-Laflamme instability of smaller than $L_{AdS}$-sized microcanonical black holes. This issue was studied numerically for black holes in $AdS_5\times S^5$ in \cite{Dias:2016eto}. As the asymptotic energy of a microcanonical black hole is lowered below the energy of the Hawking-Page critical point, the dominant bulk configuration remains a 5-dimensional AdS black hole for a small range of energies, but they become ``lumpy'' along the compact manifold. Eventually, and at a critical energy $E_c\sim O(N^2)$, the dominant bulk saddle switches to a 10-dimensional black hole localized on the $S^5$. This is referred to as the Gregory-Laflamme instability in the literature \cite{PhysRevLett.70.2837}. It is these 10-dimensional probes that are described by our microcanonical thermofield double state (\ref{microtfd}) for the largest part of the parameter space that corresponds to the black hole phase below the Hawking-Page transition. As a consequence, besides the localization in the bulk AdS manifold in (\ref{localmicrotfd}) we also need to localized them in the $S^5$ by constraining the $R$-charge of the wavefunction in an analogous way. 

From the point of view of the Euclidean preparation of the CFT state containing the probe (figure~\ref{fig:euclideanprep}), the above constraints on the width of the energy, AdS momentum and $R$-charge windows, introduced to ensure a properly localized probe black hole, characterize the ensemble of heavy CFT operators $O_{\Delta_i}$ that are to be used in the construction of the ``wormhole insertion operator'' $\Sigma_{sys,ref}$ in (\ref{wormholeop}).

\subsubsection*{A double scaling limit}
In our subsequent discussion, we will be considering correlation functions of bulk operators in the state (\ref{genstate2}), in both the large $N$ and the small probe black hole limits. Given that the parametric dependence of the Schwarzschild radius $R/L_{AdS}$ on $N$ is not fixed, we must explain how to take the limit. The limit relevant for our analysis is a double scaling limit where $N,L_{AdS}/R \to \infty$ but we keep the probe's scrambling time $t_{scr}/L_{AdS}$ fixed. Since $\beta_{probe}\to 0$, the relevant time unit is our probe's scrambling time. This limit is taken explicitly by introducing an auxiliary parameter $\alpha$ and choosing:
\begin{align}
   R&= e^{\frac{2\alpha}{d-1}} \ell_{pl}\qquad \text{with: }\quad \alpha, N \to \infty\,\, , \qquad \frac{e^\alpha \alpha^{\frac{d-1}{2}}}{N}: \text{fixed} \label{doublescaling}
\end{align}
This is a family of microcanonical black holes with radii that are shrinking in AdS units but growing in Planck units:
\begin{align}
\frac{R}{L_{AdS}}&\sim \frac{1}{\alpha} \to 0\\
\frac{R}{\ell_{pl}} &\sim e^\frac{2\alpha}{d-1} \to \infty
\end{align}
so that $t_{scr}/L_{AdS}$ remains finite. Their asymptotic energy goes to $\infty$ in the double scaling limit as 
\begin{equation}
E\sim \frac{N^2}{\alpha^{d-1} }\sim e^{2\alpha}\to \infty.
\end{equation}
In short, we will be considering small black holes which, nevertheless, remain macroscopic for all $N$, with an energy which is a small (and slowly decreasing) fraction of $N^2$. Note that this family of probes are, for every $N$, \emph{larger} than the smallest dominant microcanonical black hole (\ref{microcanonicalR}) for which $t_{scr}/L_{AdS} \to 0$. They can, therefore, indeed be described by the state (\ref{genstate2}).

\subsection{Local equilibrium states and proper time Hamiltonian}
\label{sec:equilibrium}

Having prepared a CFT state describing a small probe black hole, initially localized somewhere near the asymptotic AdS boundary of some general bulk spacetime, e.g. a large black hole at the center of AdS, we are now ready to turn to the main question of interest: What is the CFT dual of the local Schwarzschild time translation generator $H_{p.t.}$ near our probe's horizon, that will allow us to move along its bulk worldline? Stated more carefully, since the bulk Schwarzschild clock is unambiguously defined only in the neighborhood of the probe's worldline, we are after a CFT operator, $H_{p.t.}^{cft}$, which when acting on (the CFT duals of) bulk fields in the probe's atmosphere, $\phi_{atm}$, inside low energy correlation functions, it is equivalent to the action of a translation in the local bulk Schwarzschild time, at leading order in $1/N$:
\myeq{ \langle \phi(x_1) \cdots [H_{p.t.}^{cft},\phi_{atm}(x_i) ] \cdots \phi(x_k) \rangle = i\partial_{t_{sch,i}} \langle \phi(x_1) \phi_{atm}(x_i) \cdots \phi(x_k) \rangle +O(N^{-1})} 
The correspondence we seek is, therefore, weaker than an operator equality and we will denote it for convenience as:
\myeq{H_{p.t.}^{cft} \hat{=} H^{bulk}_{p.t.} \label{corfunequality}}

We can get significant mileage towards answering this question by making a further assumption about the state $|\psi\rangle$ our system is prepared in: We will assume, for now, that the atmosphere of our probe black hole in $|\psi\rangle$ is in local thermal equilibrium. This is indeed the case for the state $|E_0,\sigma\rangle$ (\ref{microtfd}) dual to a small black hole in an empty universe. It is, however, a very restrictive assumption for a general state (\ref{genstate2}); hence, Section \ref{sec:nonequilibrium} is devoted to the explanation of how to remove it. The notion of local equilibrium around our probe, however, will remain a central conceptual element of our construction. 

By local equilibrium states we mean states $|\psi_{eq}\rangle$ in which 2-point functions of bulk operators localized in the atmosphere of our probe black hole satisfy the KMS condition:
\begin{align} \langle \psi_{eq}| \phi^\dagger_{atm,1}( t_{sch,1}+i\beta_{probe})\, \phi_{atm,2}( t_{sch,2}) |\psi_{eq}\rangle &=  \langle \psi_{eq}| \phi_{atm,2}( t_{sch,2}) \,\phi^\dagger_{atm,1}( t_{sch,1}) |\psi_{eq}\rangle + O(N^{-1})\label{localKMS}
\end{align}
where $\beta_{probe}$ is the inverse temperature of the probe. This condition is of course satisfied by thermofield double black holes and their microcanonical cousins, but it is much more generally applicable. Due to the dissipative nature of black hole horizons, any state will satisfy this condition locally, if enough Schwarzschild time $\sim \frac{\beta_{probe}}{2\pi} \log S_{probe}$ is allowed to pass after the absorption of the last infalling particle. Thus, physically, condition (\ref{localKMS}) selects those states $|\psi_{eq}\rangle$ in which the probe black hole propagates undisturbed from infalling matter. We may think of these states as the analog of the \emph{local vacuum} of an idealized, point-like observer traveling along the same geodesic.

Local equilibrium states are special because condition (\ref{localKMS}) allows us to immediately identify the CFT operator that generates $t_{sch}$ translations about these states. This operator is the \emph{modular Hamiltonian} of $CFT_{sys}$ in the state $|\psi_{eq}\rangle$, obtained by tracing out the reference CFT our probe is entangled with, via the formula 
\myeq{K_{\psi_{eq}}^{sys}=-\frac{1}{2\pi}\log \text{Tr}_{ref}[ |\psi_{eq}\rangle\langle \psi_{eq}| ].}
It is a well known fact in algebraic QFT that the modular Hamiltonian can more formally be defined as\footnote{More precisely, here we refer to the \emph{full} modular Hamiltonian, i.e. the difference between the modular Hamiltonians of the system and the reference $K_{\psi_{eq}} = K_{\psi_{eq}}^{sys} -K_{\psi_{eq}}^{ref}$. This distinction is inconsequential in our discussion and we will thus omit it for simplicity.} \cite{RevModPhys.90.045003}
\myeq{ \langle\psi_{eq} | O_1^\dagger \exp[-2\pi K_{\psi_{eq}}] O_2 |\psi_{eq}\rangle= \langle \psi_{eq}|O_2 \, O_1^\dagger |\psi_{eq}\rangle\quad \forall O\in {\cal A}_{CFT_{sys}} \label{modularKMS}  }
namely, it generates a KMS transformation about the given state $|\psi_{eq}\rangle$. In contrast to (\ref{localKMS}) which is an approximate property of atmosphere correlators in special states, equation (\ref{modularKMS}) is an \emph{exact} statement about $K_\psi$ valid \emph{for all} operators in $CFT_{sys}$ about any state $\psi$ and it determines the action of the relevant modular flow in the entire CFT Hilbert space.\footnote{when the state $|\psi\rangle$ is cyclic.} In other words, $K_\psi$ is always defined as the (generally non-local) ``Hamiltonian'' for which our observer in the state $\psi$ is in thermal equilibrium. 

Conditions (\ref{modularKMS}) and (\ref{localKMS}) together imply that, in states obeying (\ref{localKMS}), the modular Hamiltonian $K_{\psi_{eq}}$ acts on operators $\phi_{atm}$ in the atmosphere of the probe black hole like the near horizon Schwarzschild Hamiltonian 
\myeq{2\pi K_{\psi_{eq}} \hat{=} \beta_{probe} H_{p.t.}^{bulk} \label{insidethehologram}}
The identification (\ref{insidethehologram}) was one of the main ideas of this paper's prequel \cite{Jafferis:2020ora}. It teaches us that, as long as the neighborhood of our observer remains in its ``local vacuum state'', time measured by the observer in their rest frame can be identified with their modular time in the microscopic CFT description via the conversion
\myeq{ t_{sch} = \frac{\beta_{probe}}{2\pi} \tau_{mod} \label{conversion}}
In view of the double scaling limit we introduced in the previous section, where $\beta_{probe}\to 0$ but its scrambling time $t_{scr}=\frac{\beta_{probe}}{2\pi}\log S_{probe}$ remains fixed in $L_{AdS}$ units, the large $N$ limit of (\ref{conversion}) is taken by defining the rescaled modular time $s_{mod}$ via $\tau_{mod} = s_{mod} \log S_{probe}$ and the conversion to bulk Schwarzschild time becomes:
\begin{equation}
    t_{sch} = s_{mod} \,t_{scr} \Label{conversion2}
\end{equation}
Bulk time is, thus, a geometric manifestation of a purely quantum mechanical notion of time in the CFT, stemming from the entanglement between our observer's microstates, described here by the reference, and the rest of the Universe.

The goal of Section~\ref{sec:general} is to explain how to generalize this correspondence in states that do not necessarily satisfy the local equilibrium condition (\ref{localKMS}). This provides us with a useful CFT tool for seeing inside the hologram which can penetrate bulk horizons and appears to resolve a number of conceptual puzzles regarding black hole interior reconstruction (Section \ref{sec:insidebh}).

\section{Emergent time from modular flow}\label{sec:exteriorreconstruction}

\subsection{Two problems and a path forward}
\label{sec:general} 
A naive generalization of the correspondence (\ref{insidethehologram}) of the previous Section would be the statement that for every state $|\psi\rangle$ of the system-reference pair, the observer's modular flow in the CFT coincides with the bulk time evolution in their rest frame (\ref{insidethehologram}). However, this fails in a general state for two key reasons.

\begin{figure}
\centering
\includegraphics[width=15cm]{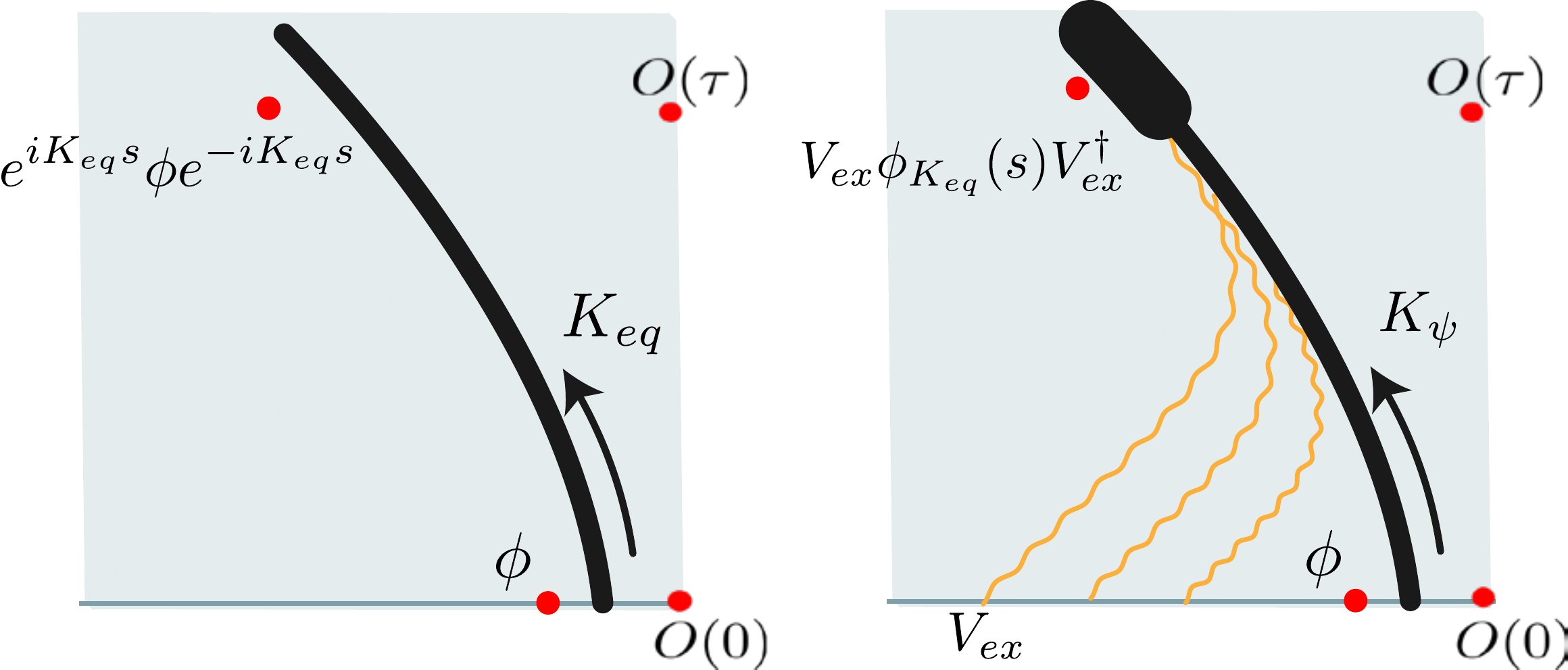}
\caption{\footnotesize{Equilibrium (left) vs non-equilibrium (right) modular flow. In the former case, the modular flow is geometric in the bulk, generating Schwarzschild time translations near the probe's horizon. In the latter, modular flow becomes non-local after crossing the lightcone of the infalling matter. The exponential blueshift of the infalling matter leads to strong gravitational interactions with the time translated field $\phi_{H_{p.t.}}(\tau)$ which in the CFT translates to the scrambling of the dual operator. This scrambling phenomenon is the key to diagnosing whether the probe is in local equilibrium or not.}}
\label{eqvsnoneq}
\end{figure}

\paragraph{Non-locality} The first is that the presence of excitations that hit our probe's atmosphere leads to a non-local modular flow $e^{-iK_{\psi}\tau}$ even in the atmosphere region. Understanding this implication is fairly straightforward: Suppose we start with a local equilibrium state $|\psi_{eq}\rangle$ and act on it with a $CFT_{sys}$ unitary $|\psi_V\rangle = V_{ex}|\psi_{eq}\rangle$ (fig.~\ref{eqvsnoneq}) which excites bulk QFT degrees of freedom at a spacelike separation from the probe's atmosphere at the initial moment $[V_{ex},\phi_{atm}]\hat{=}0$,\footnote{This is also not an operator identity since radial commutativity in the bulk can only hold within small CFT subspaces, i.e. the so-called code subspace ${\cal H}_{code}\subset {\cal H}_{CFT}$} and introduces some $O(1)$ amount of energy that will be absorbed by our probe. The new modular flow of the initial atmosphere operator then reads:\footnote{We are dropping the subscript $atm$ on $\phi_{atm}$ to reduce clutter.}
\myal{ \phi_{K_{\psi_V}} (\tau) &= e^{iK_{\psi_V}\tau} \,\phi e^{-iK_{\psi_V}\tau} = V_{ex}\,e^{iK_{\psi_{eq}}\tau} \,\phi e^{-iK_{\psi_{eq}}\tau}\,V_{ex}^\dagger \nonumber \\
&\hat{=} V_{ex}\,\phi_{H_{p.t.}}(\frac{\beta \tau}{2\pi})\,V_{ex}^\dagger \label{noneqmodflow}}
This coincides with the local geometric flow generated by the Schwarzschild Hamiltonian $H_{p.t.}$, as long as the Schwarzschild time-translated field $\phi_{H_{p.t.}}(t_{sch})$, $t_{sch}=\frac{\beta \tau}{2\pi}$ remains spacelike separated from $V_{ex}$, but it becomes a complicated non-local operator upon crossing its bulk lightcone. The geometric interpretation of modular flow is, therefore, lost in a general state.

\paragraph{Non-linearity} The second problem is that each state  $\psi$ defines a distinct modular operator $K_\psi$. An identification of $K_\psi$ with $H_{p.t.}$ would then imply that the bulk proper time Hamiltonian is a non-linear operator on the CFT Hilbert space, taking us beyond the standard rules of quantum theory. Even if we accept a degree of non-linearity in the holographic dictionary, we may further argue that such a sensitive dependence of $H_{p.t.}$ on the quantum state is incorrect for a simpler bulk reason: Proper time evolution is a property of a given background geometry, hence, it should be generated by the same operator at least on all quantum states that share the same semiclassical spacetime. In holography, this suggests that the non-linearity of the prescription should be limited to, at most, a code subspace dependence and not a state-dependence. 

\subsection*{Towards the resolution} 
There is a simple way to iron these wrinkles while staying true to the essence of our proposal: $H_{p.t.}$ should be identified not with modular Hamiltonian $K_\psi$ of the state $\psi$ we want to explore but, instead, with that of some \emph{fixed} state $\psi_{eq}$ within the code subspace ${\cal H}_{code}^\psi$ that contains the state of interest $\psi$. But which state $\psi_{eq}$ should we use for this role? The result of our observer's measurements in $\psi$, as well as the notion of locality in the vicinity of our probe will depend crucially on this choice. The key question is whether a canonical choice for $\psi_{eq}$ exists. 
Note that the local equilibrium state will be unique only in its restriction to the operators in the atmosphere region. If one applies a unitary that is spacelike to the observables being discussed, this will not affect the modular flow of the atmosphere operators. 

Our approach to this question derives inspiration from the algebraic definition of the Lorentz boost generator $B$ in QFT on Minkowski space. The procedure for the latter is as follows: First, we use the global Hamiltonian $H_g$ to identify the state in the Hilbert space that minimizes the expectation value of the energy, namely the vacuum $|0\rangle$. Then we decompose the system into a pair of complementary Rindler wedges, $W_L$ and $W_R$ and construct the \emph{modular Hamiltonian}, $K_0$, associated to this decomposition. The boost is then defined as $B=K_0$. Notice that the first step in this process, i.e. identifying the QFT vacuum, is indispensable. While any state in the Hilbert space defines a modular Hamiltonian for this Rindler decomposition, its flow will not be geometric, but rather some complicated non-local unitary transformation satisfying (\ref{modularKMS}). It is only for the special state that minimizes $\langle H_g\rangle$ that the modular flow becomes a local geometric transformation.

Our problem is directly analogous to the elementary QFT example above. The bulk Schwarzschild Hamiltonian $H_{p.t.} =\int_C T_{\mu\nu}\xi^\mu n^\nu$ we want to define is the analog to the boost generator in our Rindler example. The special states satisfying the local equilibrium criterion (\ref{localKMS}), whose modular Hamiltonians locally generate Schwarzschild time are the analog of the QFT vacuum above ---at least in the observer's vicinity. All other states ought to be thought of as excitations of the latter. What is missing in completing the correspondence is the analog of the minimal energy condition $\min_\psi \langle H_g\rangle_\psi$ we used in our QFT example to \emph{identify} the vacuum state. Is there a canonical choice for the local equilibrium state in every code subspace or is this choice a fundamental ambiguity in the holographic map? The rest of this paper is devoted to answering this question. And the answer is yes.

\subsection{Proper time flow in backgrounds without ambient black holes}\label{sec:nonequilibrium}

Suppose $K_\psi$ is the modular Hamiltonian of the system in $|\psi\rangle$, a state of CFT$_{sys}\times$CFT$_{ref}$ that describes a small probe black hole in a general asymptotically AdS Universe that contains no other black holes ---and which, as always, is entangled with the referece. Furthermore, $\phi_{atm}$ is an initial bulk operator in the probe's atmosphere whose CFT representation we considere known and $H$ is the system CFT Hamiltonian. With this input, we are interested in finding the local equilibrium state $|\psi_{eq}\rangle$ in the code subspace ${\cal H}_{code}^\psi$ built around $|\psi\rangle$. The modular Hamiltonian $K_{eq}$ will then be the proper time evolution generator we want. We will arrive at such a prescription by utilizing the fast scrambling property of our probe black hole.

Let us first split the boundary operator algebra into operators that are spacelike separated from $\phi_{atm}$
\begin{equation}
    {\cal S}=\{O\in {\cal A}_{CFT}: \, [O,\phi_{atm}]\hat{=}0\}
\end{equation}
and those that are timelike:
\begin{equation}
    {\cal T}=\{O\in {\cal A}_{CFT}: \, [O,\phi_{atm}]\hat{\neq}0\}
\end{equation}
Practically, ${\cal S}$ consists of all local operators in a boundary time band $t\in [-T_1(x),T_2(x)]$ and ${\cal T}$ of those in $(-\infty, T_1(x)]\cup [T_2(x),\infty)$. If an equilibrium state $|\psi_{eq}\rangle\in {\cal H}_{code}^\psi$ exists, it can be expressed as $|\psi_{eq}\rangle = U_{eq}|\psi\rangle$ for some choice of unitary $U_{eq}:{\cal H}_{code}^\psi \to {\cal H}_{code}^\psi$ that satisfies $[U_{eq}, \phi_{atm}]\hat{=}0$. This is the unitary that annihilates all particles present in $|\psi\rangle$ that can fall in our probe black hole and disturb local equilibrium in its future. $|\psi_{eq}\rangle$ is, therefore, a particular case of the class of states $|\psi_U\rangle = U|\psi\rangle$ where $U$ is of the form
\begin{align}
    U&=\exp\left[ i\int J^{(1)}(x) O_i(x) +i\int J^{(2)}(x,y) O_i(x)O_j(y) +\dots \right]\text{  for } O_i(x)\in S \label{Vdef}
\end{align}
with arbitrary smooth sources.\footnote{$O_i$s are assumed to be Hermitian} But what exactly makes the unitary $U_{eq}$ mapping $|\psi\rangle$ to $|\psi_{eq}\rangle$ special among all possible $U$'s above? The answer lies in the behavior of the modular flow $\exp[ -iK_U \tau]$, as we now explain.

\paragraph{Behavior of modular flowed correlators as equilibrium diagnostic} Consider, first the two-point function of $\phi_{atm}$ with the corresponding boundary single-trace $O(t)$ for $t>T_2(x)$. The operator $\phi_{atm}$ here is an arbitrary bulk supergravity field and the argument below holds universally for all of them. This correlation function is a ---typically exponentially--- decreasing function of boundary time, as a consequence of quasi-normal decay in the probe black hole's background. We can define $T_{th}$ to be the boundary timescale for which
\begin{equation}
    \langle \psi| \phi_{atm}\,\, O(T_{th}) |\psi\rangle = O(S_{probe}^{-1}) \label{seq0F}
\end{equation}
For a black hole of $L_{AdS}$ size in an empty universe, the time-scale $T_{th}$ is simply the boundary scrambling time. For the sub-AdS scale black holes we use as probes in general spacetimes, $T_{th}$ is still related to the scrambling time but in a more indirect way. If (\ref{seq0F}) was a two-point function within the $r\sim O(R_{probe})$ atmosphere region, the time-scale for the entropic suppression above would indeed be $t_{scr}=\beta_{probe}\log S_{probe}$, simply based on quasi-normal decay. $T_{th}$, however, is a boundary time-scale. The geometric way to understand it is as the time at which an outgoing light-ray sent from $r\sim R_{probe}$, at time $t_{scr}$ after the $\phi_{atm}$ insertion, reaches the asymptotic boundary, $r_{\partial B}$. For a canonical ensemble black hole $R_{probe}\sim O(L_{AdS})$ and $r_{\partial B} \sim O(L_{AdS})$, thus the distinction between the scrambling time and $T_{th}$ is parametrically immaterial. For small microcanonical black holes, on the other hand, the two timescales can parametrically differ.

Now we ask, what will happen to (\ref{seq0F}) if we use the modular Hamiltonian $K_U$ of the state $|\psi_U\rangle$ above to evolve $\phi_{atm}$? The answer crucially depends on whether $|\psi_U\rangle$ is $|\psi_{eq}\rangle$ or not! By the argument of the previous Section, $K_{eq}$ generates a local geometric flow in the bulk, translating $\phi_{atm}$ in Schwarzschild time. The effect of modular flow will, therefore, counter the decorrelation effect caused by the boundary Hamiltonian evolution and the two-point function will start increasing as a function of modular time  until it reaches an $O(S_{probe}^0)$ value around $\tau=O(\log S_{probe})$:
\begin{equation}
    F_{eq}(\tau \to \tau_{scr}, t\to T_{th}) = \Big|\langle\psi|\phi_{K_{eq}}(\tau) \,O(t) |\psi\rangle \Big|_{\tau\to \tau_{scr},\,t\to T_{th}} =O(S_{probe}^0) \label{eqflow}
\end{equation}
where we denoted the evolution of $\phi_{atm}$ with $K_{eq}$ as $\phi_{K_{eq}}(\tau)$ and defined $\tau_{scr}=\log S_{probe}$.

In contrast, for any other choice of $K_U$, the modular flowed correlator never reaches an $O(S_{probe}^0)$ value due to the scrambling phenomenon. This follows from the observation that, since $|\psi\rangle = U_{eq}^\dagger|\psi_{eq}\rangle$ and $K_U = U K_{\psi}U^\dagger = UU_{eq}^\dagger \,K_{eq}\, U_{eq} U^\dagger$ from our definitions above, the $K_U$-flowed two-point function becomes  
\begin{align}
    F_U(s,t) &= \left|\langle \psi | \phi_{K_U}(\tau) \,O(t)|\psi\rangle\right| \nonumber\\
    &=\left|\langle \psi_{eq}| U_{eq} UU_{eq}^\dagger\phi_{K_{eq}}(\tau) U_{eq} U^\dagger O(t) U_{eq}^\dagger|\psi_{eq}\rangle\right| \label{bulktoboundary}
\end{align}
The correlation function (\ref{bulktoboundary}) is an out-of-time-order correlator (OTOC) \emph{unless} $U=U_{eq}$! As is well-understood, OTOCs decay exponentially in Schwarzschild time and reach parametrically small values when the time separation becomes of order the scrambling time \cite{Shenker:2013pqa, Maldacena:2015waa}. The behavior of correlators of this kind is reviewed in Appendix (\ref{app:modularflow}). In summary, the modular flowed two-point function behaves as:
\begin{equation}
    F_U(\tau \to \tau_{scr}, t\to T_{th}) = \begin{cases} O(S_{probe}^0) &\text{if  } U=U_{eq} \\
    O(S_{probe}^{-1}) &\text{if  } U\neq U_{eq} \label{FofU}
    \end{cases}
\end{equation}
We have, therefore, arrived at a robust criterion for selecting the desired equilibrium state in the code subspace ${\cal H}_{code}^\psi$ of interest: 
\paragraph{Definition of local equilibrium states:} \emph{Let $|\psi\rangle$ be a state in ${\cal H}_{sys} \otimes {\cal H}_{ref}$ containing a probe black hole entangled with the reference, as in Section \ref{sec:stateprep} (fig.~\ref{fig:euclideanprep}), ${\cal H}_{code}^\psi$ a code subspace containing $|\psi\rangle$ and $\phi_{atm}$ the CFT dual of an arbitrary local bulk operator near the probe at a given moment in time. If the background does not contain other black holes, then the local equilibrium state in $ {\cal H}_{code}^\psi$ is the state $|\psi_{eq}\rangle = U |\psi\rangle$ for the unitary $U$ of the form (\ref{Vdef}) that maximizes the magnitude of the modular flowed correlator (\ref{FofU}), for any choice of $\phi_{atm}$, i.e.\footnote{The function $\text{argmax}_U(\cdot )$ returns the operator $U$ for which the argument of the $\text{argmax}$ function is maximized.}
\begin{align}
  |\psi_{eq}\rangle = U_{max}|\psi\rangle \quad&\text{   for   }\quad U_{max}= \text{argmax}_{U} F_U(\tau\to \tau_{scr},t\to T_{th}) \nonumber\\
  &\text{among all }U \text{ of the form (\ref{Vdef})} 
\end{align}
if and only if $F_{U_{max}} \sim O(N^0)$.}

The requirement of the maximum to be $O(N^0)$ is crucial. This is because there exist code subspaces which contain no local equilibrium states. These describe backgrounds in which our probe collides with another massive object with large center of mass energy, passes  through a high curvature region, e.g. a singularity, or experiences some other ``dramatic'' event. In these scenarios, evolution along the probe's trajectory becomes meaningless after the dramatic encounter and the search for equilibrium state must fail. The maximization requirement alone cannot achieve this since there will always be a state which maximizes $F_U$. The remedy is to further demand $F_{U_{max}} \sim O(N^0)$. Cases with no local equilibrium states will either involve large amounts of matter accreted by our black hole that cannot be removed by the simple unitaries $U$ of the form (\ref{Vdef}) or they will be situations in which the probe hits a singularity, both situations that result in $F_{U_{max}} \sim O(N^{-1})$. Interestingly, the latter case appears to also leave a distinct universal signature in modular flowed correlators which we explain in Section \ref{sec:singularity}, allowing us to distinguish it from other non-equilibrium situations.

An alternative way to understand the local equilibrium condition in the
absence of causal horizons is that the correctly reconstructed bulk
operators must have an expression in terms of simple boundary operators
via the bulk Heisenberg evolution. Although we do not take the form of
the evolution as an input, since it depends on the details of the bulk
theory and bulk semi-classical spacetime which is what we are attempting
to reconstruct, it implies that $\phi_{atm} \hat{=} A$, an element of
the simple space of operators. However, the modular flowed atmosphere
operators will fail to obey this property after the probe scrambling
time if the state is not in local equilibrium, due to the growth of the
bulk operators.

Finally, note that the maximization of (3.9) uniquely fixes the atmosphere modular flow associated to the selected local equilibrium state, but it does not uniquely determine the state itself, since transformations away from the region being probed do not affect these quantities. This is to be expected for a notion of {\it local} equilibrium. 

\subsection{Reflections on our proposal as a bulk reconstruction technique} \label{sec:philosophy}

We have arrived at a CFT prescription for defining the time evolution of an internal bulk observer. This allows us to revisit the problem of holographic reconstruction and propose a new, background independent method: Local observables deep in the bulk can be constructed within a code subspace ${\cal H}_{code}$ and at leading order in $N$ by propagating boundary operators with the modular flow of a probe (constructed in Section \ref{sec:framework}) in a local equilibrium state $\psi_{eq} \in {\cal H}_{code}$ (constructed in Section \ref{sec:exteriorreconstruction}).

\paragraph{The method's philosophy} Local bulk fields are mathematical representations of the possible responses different kinds of spatially localized detectors can have. In every-day physics we can often suppress this operational starting point because the local degrees of freedom of a system are usually manifest so they are simply taken as theoretical input. This is, however, merely an \emph{empirical} property of our world. In AdS/CFT, a preferred set of local operators exists only in the CFT. The bulk fields are collective CFT excitations and the existence and meaning of a ``local'' frame for the relevant operator algebra is unclear, at least without any further input. In our approach we are \emph{defining} local bulk fields as those CFT operators that are local relative to a particular detector ---our probe black hole. The choice of probe, therefore, determines the notion of bulk locality in the code subspace! The state $\psi_{eq}$ used to define the proper time generator $K_{eq}$ in a given code subspace is part of the definition of our detector: It is its ground state, the state in which it is calibrated to yield no response. As a result, it is this choice of observer, including the reference state $\psi_{eq}$, that assigns a bulk interpretation to the CFT operator algebra and the corresponding wavefunctions, and selects the dynamical evolution laws in that frame. 

In our approach, the equilibrium criterion that selects $|\psi_{eq}\rangle$ can be intuitively expressed as follows. Due to the chaotic nature of the holographic CFT, both the boundary Hamiltonian evolution and the observer's modular flow scramble whatever operator they act on for a sufficient amount of time. For observers in local equilibrium $H$ and $K_{eq}$ scramble ``in the same way'', i.e. by largely preserving the correlations between the degrees of freedom they are applied to ---since they both correspond to geometric flows in the dual gravity picture. In stark contrast, the presence of any amount of infalling energy in the initial state will cause the trajectory in operator space that modular flow generates to exponentially deviate from the one defined by the boundary Hamiltonian. Non-equilibrium states, in other words, have modular flows that scramble operarators not only relative to the original local operators but also relatively to the boundary time-evolved degrees of freedom, leading to the destruction of the correlations with time-evolved boundary fields.

\begin{figure}
    \centering
    \includegraphics[scale=0.8]{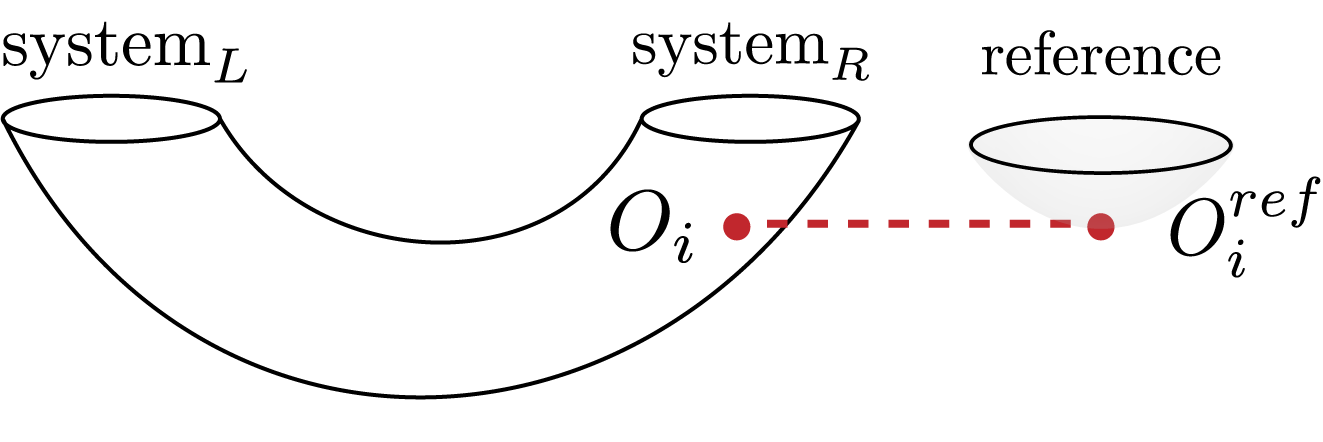} \hspace{1cm}
    \includegraphics[scale=0.5]{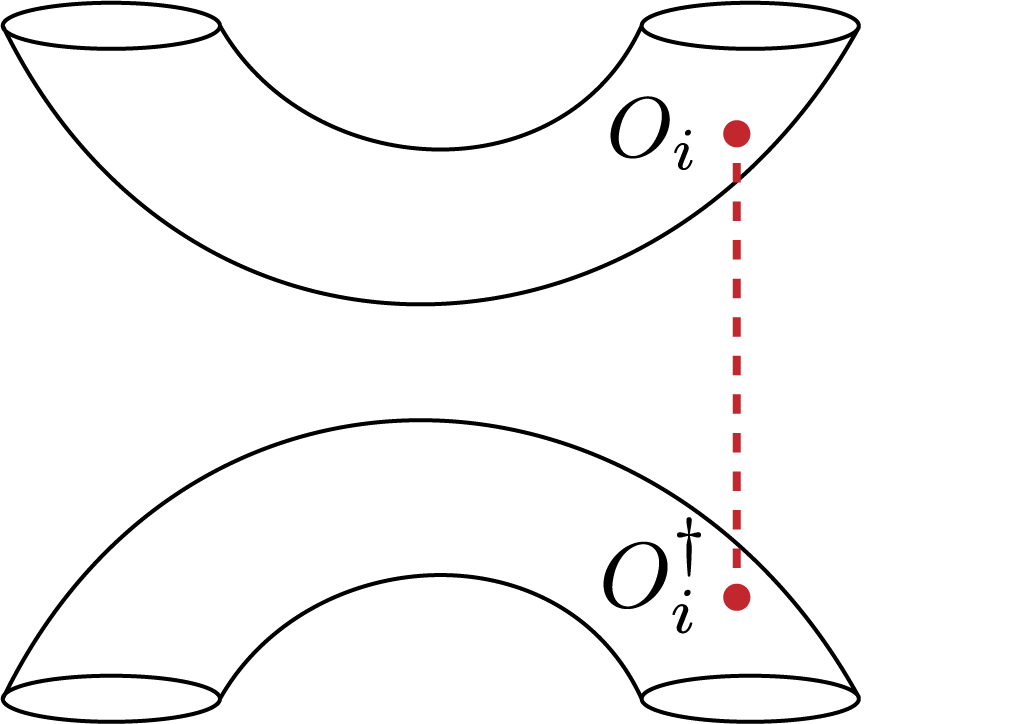}
    \caption{\footnotesize{\textbf{Left:} Eucludean preparation of a state with a small probe black hole ``observer'' in the right exterior of an eternal AdS wormhole with two asymptotic regions. The dashed red line denotes the sum over all operators with conformal dimensions $\Delta_i \in [E, E+\Delta E]$ in the microcanonical black hole window discussed in Section~\ref{sec:microcanonicalbh}. \textbf{Right:} Reduced density matrix of the \emph{system} CFT after tracing out the \emph{reference}. The logarithm of this operator is the modular Hamiltonian generating proper time translations along the trajectory of the probe black hole and into the two-sided interior.}}
    \label{fig:tfdobserver}
\end{figure}

\paragraph{On the subtle role of the CFT dynamics} The relation between the boundary CFT clock and our probe's internal proper time which we use to propagate operators deep in the bulk is an interesting feature of our construction. At first glance, the two notions of time are distinct. The former is part of the dynamical definition of the CFT and provides the spacetime's asymptotic clock while the latter is the clock in the probe's reference frame determined by the quantum correlations between the probe and the rest of the world in a given state. This could lead to a puzzle: A proper time Hamiltonian determined entirely by a state does not appear to take into consideration what boundary conditions we have chosen for the spacetime, e.g. whether we have included in the Hamiltonian sources that affect the background our observer propagates in. The proper time evolved field would appear to always be the same, regardless of the Hamiltonian we used to evolve the boundary system. The puzzle is resolved in the proposal presented in this Section by the fact that the boundary Hamiltonian appears explicitly in the prescription for choosing the local equilibrium state $|\psi_{eq}\rangle$. In this subtle but important way, the boundary dynamics affect the choice of modular Hamiltonian that we will interpret as the probe's proper time generator in the relevant bulk spacetime.

\section{Inside holographic black holes} \label{sec:insidebh}
We now aim our favorite holographic reconstruction tool at the thorniest aspect of AdS/CFT: The CFT definition of observables in the black hole interior. The main objective of this Section is to provide a CFT criterion for selecting ``local equilibrium'' states for probes that fall inside AdS black holes, by generalizing the prescription of Section \ref{sec:nonequilibrium} in the appropriate way. We do this in Section \ref{sec:interiorreconstruction}. This will yield an unambiguous and physically motivated prescription for predicting the experience of an infalling observer.

Before diving into this topic, however, there are two interesting points we can discuss without relying on knowledge of the local equilibrium states. We have already arrived at a technically rather non-trivial statement. If we prepare a black hole together with an infalling observer in a state $|\psi\rangle$ described by the path integral of fig.~\ref{fig:tfdobserver}, then operators of the form
\begin{equation}
    \phi_{K_\psi}(\tau)= e^{iK_\psi \tau} \phi e^{-iK_\psi \tau} \label{interiorops}
\end{equation}
where $K_\psi$ is the modular flow of the system-reference bi-partition and $\phi$ some exterior local bulk operator near the boundary, generate the algebra of the black hole interior degrees of freedom after a finite modular time $\tau$ ---albeit $ \phi_{K_\psi}(\tau)$ are not necessarily local fields themselves. Moreover, if $|\psi\rangle$ has a semiclassical interior, there exists a $|\psi_{eq}\rangle$ in the code subspace ${\cal H}_{code}^\psi$ built around $|\psi\rangle$ for which $K_{eq}$ produces the local interior observables via (\ref{interiorops}) and the rules for obtaining this state will be discussed momentarily. As we now discuss, this general statement already allows us to identify two CFT computations that can contain information about two central aspects of the black hole interior: the typicality of firewalls and the approach of the black hole singularity. We discuss this point in Section \ref{sec:singularity}.

It is worth pointing out, that even the general statement above is not an obvious fact at the technical level. It was, however, explicitly demonstrated recently in \cite{Gao:2021tzr}, for the simplest example of a black hole state available: The eternal AdS$_2$ black hole in Jackiw-Teitelboim gravity. Indeed, for this state, we do not need to worry about particle excitations disturbing the local equilibrium of our infalling observer; its modular flow is, then, expected to be geometric in the bulk, by the general argument of Section \ref{sec:equilibrium}. Remarkably, a detailed SYK computation of the relevant modular flowed correlators produced results consistent with the interpretation of $K_{eq}$ as translation along the infalling geodesic. The analogous higher dimensional computation would be important progress but we do not attempt it here.

\subsection{Probing the black hole singularity}\label{sec:singularity}

Let us begin our exploration with the eternal AdS$_{d+1}$ black hole, produced holographically by the thermofield double of a pair of CFTs. A black hole ``observer'' is introduced on this background via the state preparation of Section \ref{sec:stateprep} which, after tracing out the reference, produces the CFT density matrix of fig.~\ref{fig:tfdobserver}. As in the 2D example of \cite{Gao:2021tzr}, there is no need to look for an equilibrium state any further: By construction no excitations have been inserted in the bulk system other than our observer. Unlike the 2D example, however, the infalling observer will hit a curvature singularity at a finite proper time. What signature does such a dramatic event leave in our modular flow? Is our method useful for understanding the emergence of the singularity in the semiclassical limit and, by extension, its resolution in the finite $N$ theory?

This question can be broken down to two parts. The first is how close to the singularity our modular flow can reliably take us. Let the radius of our probe black hole be $R_p$. As long as the curvature radius of the ambient geometry is much larger than $R_p$ the neighborhood of the probe is approximately Schwarzschild and its modular flow can be reliably represented geometrically in this region. It follows that we lose control over our modular flow inside the ambient black hole interior when the probe reaches $r_{min}\sim R_p$, where $r$ is the time-like AdS-Schwarzschild coordinate in the background black hole's interior. As explained in Section \ref{sec:microcanonicalbh}, this is always a macroscopic distance as compared to the Planck length but can be made parametrically smaller than $L_{AdS}$ by choosing our probe to be a microcanonical black hole and taking the double scaling limit (\ref{doublescaling}):
\begin{equation}
    \text{double scaling: }  \quad\frac{R_p}{L_{AdS}}\sim \frac{1}{\alpha} \, \, , \quad  \frac{R_p}{L_{pl}} \sim e^{\frac{2\alpha}{d-1}} \quad \text{with } \alpha \to \infty.  \label{doublescaling2}
\end{equation}
As we will see momentarily, when taking this limit, reaching a distance $r_{min}\sim R_p$ from the singularity is close enough to get a non-trivial singularity signature in the CFT.

The second part is understanding what correlation functions of a bulk field approaching a black hole singularity are expected to behave like, according to the bulk QFT analysis. For the BTZ black hole, the computation can be performed explicitly as in Appendix \ref{app:singularity}, or by employing the method of images as was done in \cite{Hamilton:2006fh, Hamilton:2007wj}. The result for the bulk-to-boundary propagator for a massless scalar field in the $\frac{r}{R_0}\to 0$ limit, where $R_{0}\sim O(L_{AdS})$ the Schwarzschild radius of the black hole background, reads:
\begin{equation}
    \langle TFD|\, \phi(r\ll R_0, t, \theta) O(t',\theta')\,|TFD\rangle \to \log\frac{R_{0}}{r}\, \sinh^{-2}(\frac{t-t'}{R_{0}}) \label{singularity}
\end{equation}
This result has two obvious but noteworthy features: (a) It decays exponentially in time, as expected of perturbations of a black hole's state, reaching an $O(N^{-1})$ value for $|t-t'|\sim R_0\log S$, and (b) It blows up logarithmically as $r\to 0$. Both feautres will be important below. Interestingly, while the computation can only be performed exactly for the BTZ, our analysis in Appendix \ref{app:singularity} provides compelling evidence for the universality of the result in higher dimensional black holes. As we demonstrate, the scalar propagator in the 5D AdS-Schwarzschild geometry exhibits the same logarithmic divergence as $r\to 0$. The computation of its precise time dependence cannot be done without numerics ---though all equations to be solved are derived in the Appendix--- but an exponential decay in time is the natural expectation for $|t-t'|\gg R_0$, given the quasinormal dissipation of black hole excitations.\footnote{A more geometric argument for an exponential decay in time is the linear growth of the length of spacelike geodesics connecting interior points to the boundary at different times.}

\paragraph{A CFT signature of the singularity} We now recall that, according to our prescription, the modular flowed correlator
\begin{equation}
    F(\tau,t,\theta_0,\theta)= \langle e^{iK \tau} \phi_0(r_0,0,\theta_0) e^{-iK\tau} \,O(t,\theta)\rangle \label{modflowedint}
\end{equation}
is expected to match the bulk-to-boundary propagator computed in the bulk QFT approximation about the black hole background
\begin{equation}
    F(\tau,t,\theta_0,\theta) = \langle  \phi_0(r_\gamma(\tau),t_\gamma(\tau),\theta_0) \,O(t,\theta)\rangle  +O(1/N) \label{correspondence}
\end{equation}
Here $(r_\gamma(\tau),t_\gamma(\tau))$ is a bulk trajectory at a fixed location relative to the infalling probe's apparent horizon, expressed in the background's Schwarzschild coordinates, obeying $r_\gamma(0)=r_0,\,t_\gamma(0)=0$. It is parametrized by modular time $\tau$, which is related to the probe's Schwarzschild time by (\ref{conversion2}). By virtue of our previous observations about the behaviour of the right hand side of (\ref{correspondence}), we conclude that by taking the double scaling limit (\ref{doublescaling}) of the correlator (\ref{modflowedint}) we can probe the black hole singularity, and in particular the logarithmic divergence it introduces.

\begin{figure}
    \centering
    \includegraphics[scale=0.5]{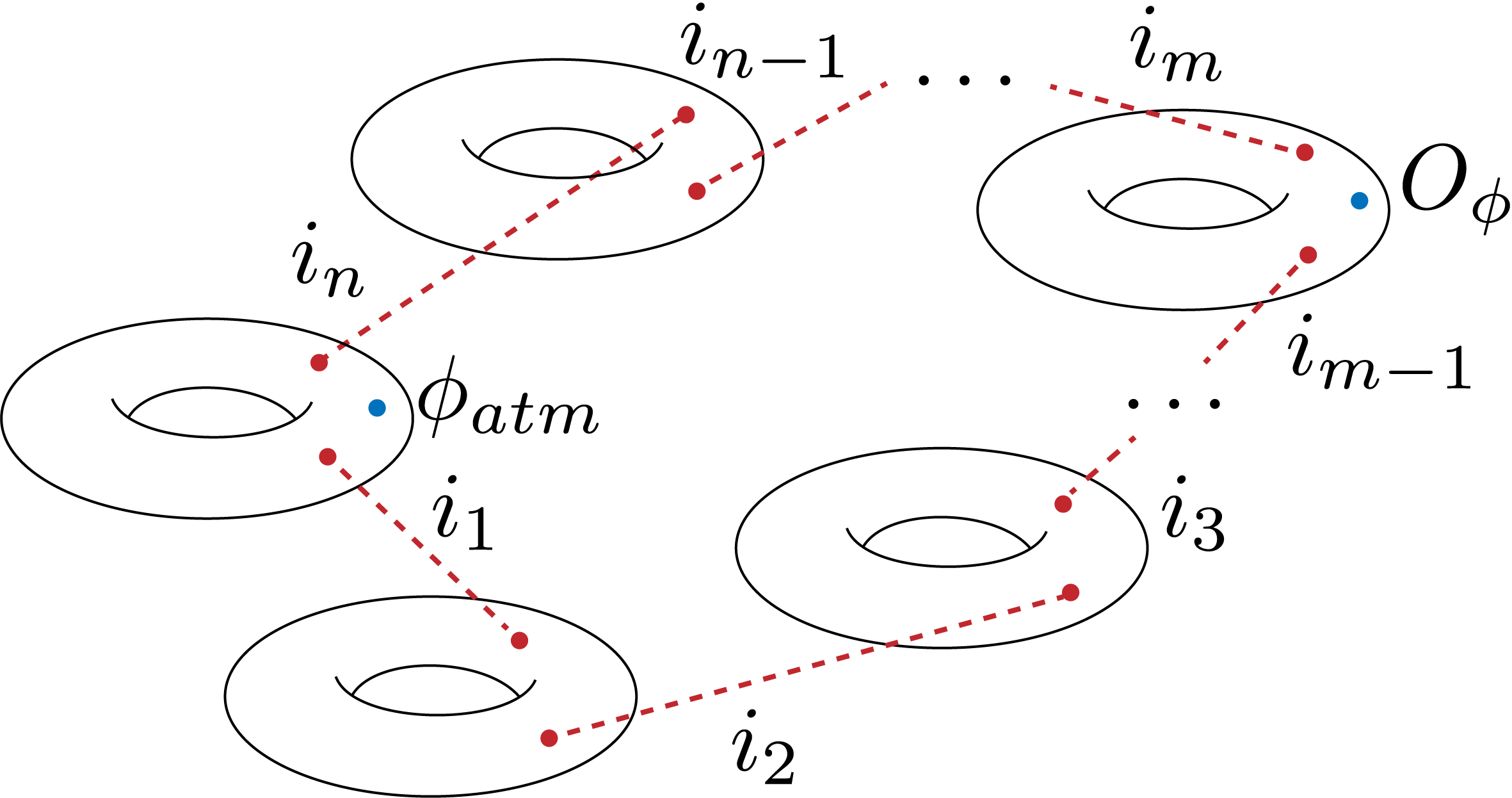}
    \caption{\footnotesize{Euclidean path integral computation of the modular replica correlation function (\ref{replicacalc}). The modular flowed correlation function (\ref{modflowedint}) that transports an operator $\phi$ to the interior of an eternal AdS black hole and is expected to exhibit the divergence (\ref{divergence}) when the observer hits the singularity, is obtained by an analytic continuation of the result of this path integral in the replica numbers $n,m$.}}
    \label{fig:twosidedBHreplica}
\end{figure}

More precisely, when $t \sim O(R_0)$, the double-scaled limit of (\ref{modflowedint}) diverges logarithmically at a \emph{finite rescaled modular time} $s_{div}=\frac{\tau_{div}}{\log S_{probe}}$, at which $r_\gamma(\tau_{div}) \sim R_p$, namely: 
\begin{equation}
    \exists \,\,\, s_{div}=\lim_{d.s.} \frac{\tau_{div}}{\log S_{probe}}<\infty \,\,\, \text{such that: }\, F(\tau_{div},t=O(R_0) ,\theta_0,\theta) \,\overset{d.s.}{\longrightarrow} \, \log \alpha =\infty \label{divergence}
\end{equation}
This is a divergence in a CFT correlator evaluated in the $N\to \infty$ limit (\ref{doublescaling2}) and it is a direct consequence of the presence of the black hole singularity in the semiclassical dual. It is independent of the precise location of the boundary operator which could, in fact, be spacelike separated from $\phi(r_\gamma(\tau_{div}), t_\gamma(\tau_{div}),\theta_0)$. Furthermore, the same logarithmic blow-up is expected for AdS black holes in any number of dimensions, as we argued above and in Appendix \ref{app:singularity}. It, therefore, appears to be a universal signature of the singularity in the CFT. Since the microscopic CFT calculation has no reason to be singular at a particular modular time when computed at any finite $N$, the CFT physics underlying this divergence is an important clue for the emergence and resolution of the black hole singularity. A detailed exploration of this property and its relation to previously identified singularity signatures \cite{Fidkowski:2003nf, Festuccia:2005pi} is extremely interesting and we hope to return to it in future work. 

\paragraph{Setting up the computation} It is illuminating to set up the CFT computation of the correlator (\ref{modflowedint}) which, as argued, is expected to signal the presence of the black hole singularity. The most natural way to approach the calculation is to employ a ``modular replica trick'', as was done in \cite{Gao:2021tzr}, and attempt to obtain (\ref{modflowedint}) from an analytic continuation of the correlator:
\begin{equation}
    \tilde{F}(n,m) = \text{Tr}_{sys}\left[ e^{-(n-m)K} \phi_0(r_0,0,\theta_0) e^{-mK} O(t,\theta) \right] \label{replicacalc}
\end{equation}
where $n,m \in \mathbb{N}$ and $\text{Tr}_{sys}$ refers to the trace over the Hilbert space ${\cal H}_L\otimes {\cal H}_{R}$ of the two holographic CFTs describing the background wormhole. The operators $e^{-kK}$ admit a simple Euclidean path integral representation since they are powers of the CFT density matrix of figure~\ref{fig:tfdobserver}. The replica correlator (\ref{replicacalc}) corresponds to the computation of the Euclidean path integral of figure~\ref{fig:twosidedBHreplica}, where each of the dotted lines represents a sum over all heavy operators in the energy window $[E_0,E_0+\Delta E]$ used in the construction of the microcanonical probe in Section \ref{sec:microcanonicalbh}. Interestingly, the computation reduces to a product of $n-2$ heavy-heavy torus two-point functions and two heavy-heavy-light torus three-point functions, where all the heavy operators are summed over a macroscopic window of conformal dimensions. Objects of this kind have been the focus of many recent works \cite{Kraus:2016nwo, Collier:2019weq, Belin:2020hea, Belin:2021ibv}, where asymptotic formulas for such statistical averages were derived. 

\begin{figure}
    \centering
    \includegraphics[scale=0.5]{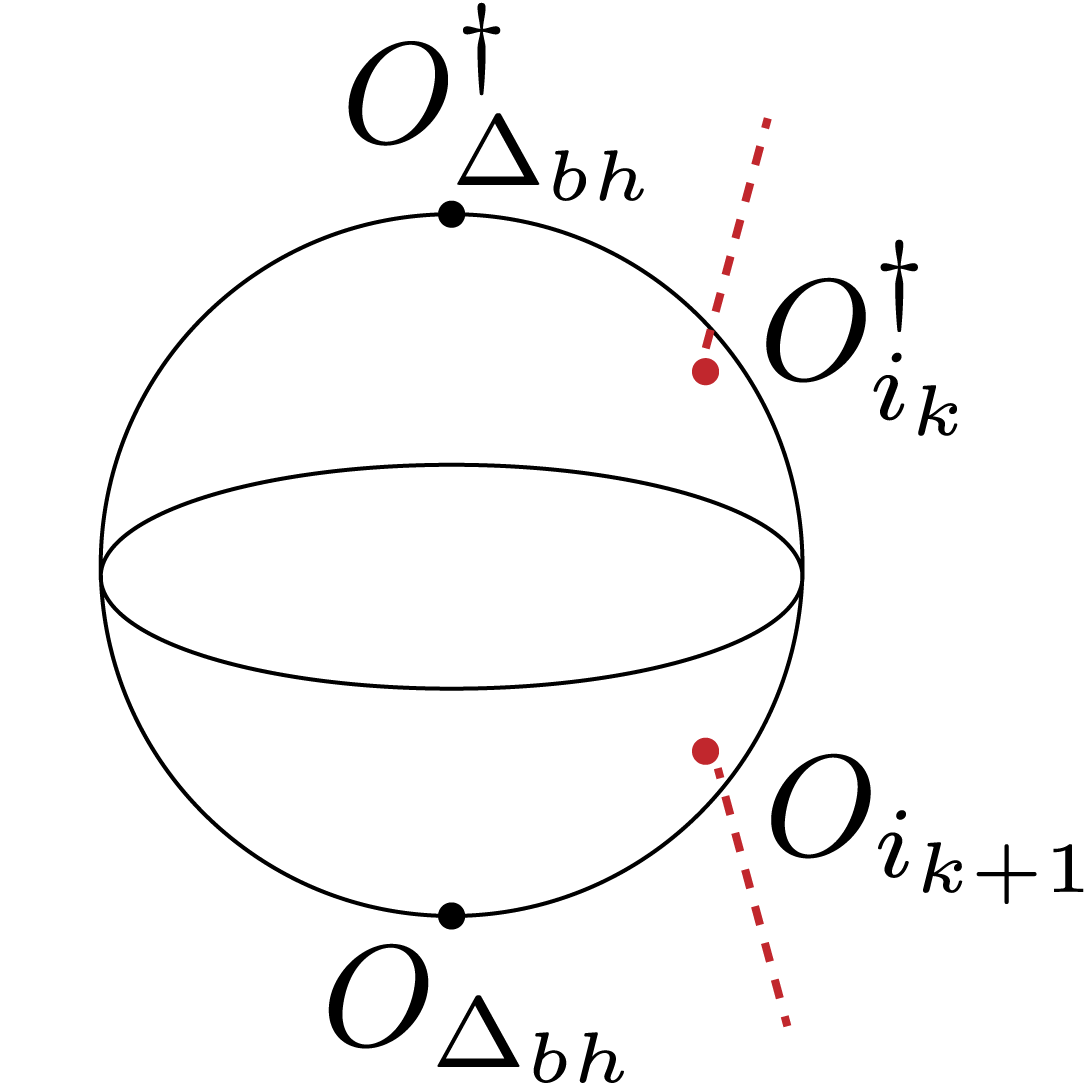}
    \caption{\footnotesize{The modular replica correlator (\ref{replicacalc}) whose analytic continuation gives the modular flow in a single-sided black hole microstate is identical to that of fig.~\ref{fig:twosidedBHreplica} with every torus 2-point function replaced by a sphere 4-point function, with dashed red lines denoting, again, a summation over operators in the microcanonical black hole energy window.}}
    \label{fig:singlesidedBHreplica}
\end{figure}

Analogously, we can set up the same computation for a probe in a single-sided black hole spacetime. The relevant modular operator $e^{-K}$ is obtained by the Euclidean path integral of figure~\ref{fig:euclideanprep} after tracing out the reference and the replica correlator (\ref{replicacalc}) admits a Euclidean path integral representation like in figure~\ref{fig:twosidedBHreplica} with all torus two-point functions replaced by all-heavy sphere four-point functions (figure~\ref{fig:singlesidedBHreplica}) and the two torus three-point functions by sphere five-point functions with one light insertion. Asymptotic expressions for such objects were also discussed in \cite{Collier:2019weq}. It is, therefore, conceivably within reach to explicitly perform the computation of (\ref{replicacalc}) directly in the CFT and understand how the black hole singularity manifests itself in the analytic continuation (\ref{modflowedint}), upon taking the double scaling limit (\ref{doublescaling2}).

\paragraph{Firewall typicality} Equally interestingly, a detailed CFT computation of the rescaled modular time $s_{div}$ at which $F$ diverges in a typical black hole microstate could be a concrete probe of the typicality of firewall states. A smooth geometric interior implies a correlator $F\sim O(1)$ for modular times $s>s_{hor}$ after the probe crosses the horizon which develops a singularity after a finite $s_{div}-s_{hor}\sim O(1)$ modular evolution beyond the horizon. Since modular time corresponds to the proper time along the infalling timelike geodesic, the typical value of $s_{div}$ is a measure of how deep inside the black hole interior the singularity lives. What would a firewall state look like from the point of view of the correlator (\ref{modflowedint})? If the behavior of the correlator near the singularity of the eternal black hole is a good guide for what happens when a firewall is encountered by our infalling observer, firewall states should yield an $s_{div}$ which is much closer to the moment of horizon crossing $s_{hor}$ than expected from the semiclassical computation. On the other hand, it is conceivable that the effect of firewalls on (\ref{modflowedint}) is distinct from that of the black hole singularity with the most likely alternative being that (\ref{modflowedint}) drops to nearly zero after $s\sim s_{hor}$. In either case, the expected behavior of this modular flowed two-point function would be disrupted in the particular microstate. We have, therefore, articulated a technical computation, i.e. the existence of a divergence in (\ref{modflowedint}) at a finite $s_{div}$ in the limit (\ref{doublescaling2}) and the typical value of $s_{div}$ in an ensemble of black hole microstates in an energy window, whose answer can illuminate the open problem of the typicality of firewalls.

\subsection{Local reconstruction in the black hole interior}\label{sec:interiorreconstruction}

Suppose we are now given an arbitrary single-sided black hole microstate, in which we insert our probe black hole observer near the boundary, via the path integral construction (fig.~\ref{fig:euclideanprep}). Can we reconstruct local operators in its interior using (\ref{interiorops})? As we have explained, this hinges on identifying our probe's local equilibrium state. If we try to apply the prescription of Section \ref{sec:nonequilibrium} for identifying $|\psi_{eq}\rangle$ in this case, however, we encounter a puzzle. We may start with a state with a macroscopic, empty interior for which modular flow is geometric and act on it with a unitary operator supported on the interior region that creates a series of high energy shocks, waiting to hit the infalling observer. The new state clearly violates local equilibrium but since such a unitary commutes with boundary operators within the code subspace due to bulk causality, the modular flowed correlator (\ref{modflowedint}) can easily be shown to be insensitive to it. Such unitaries must be dressed to features of the interior of the black hole, so as to avoid a large commutator with boundary operators. This is possible in states containing black holes of the type we are discussing that are generic spacetimes with no isometries and an explicit class of examples were recently discussed in \cite{Bahiru:2022oas}.

We are, therefore, unable to distinguish the local equilibrium state in the black hole interior using the criterion of the previous Section and the definition of the local operator basis in the interior becomes, once more, ambiguous. This problem is a version of the frozen vacuum puzzle of black hole interior reconstruction \cite{Bousso:2013ifa}. In this Section, we conjecture that \emph{operator complexity} provides the appropriate diagnostic that generalizes to this case.

We wish to distinguish the operator $\phi_{K_{eq}}(\tau)$ which, as we have argued, can be identified with a geometrically translated field along the bulk trajectory of the probe, from $\phi_{K_U}(\tau) = U\phi_{K_{eq}}(\tau) U^\dagger$ where $U$ is a bulk unitary excitation on top of the equilibrium state $|\psi_{eq}\rangle$ that could be entirely localized beyond the horizon of the background black hole. The physical phenomenon that can help us achieve this is the same as in Section \ref{sec:nonequilibrium}: An excitation $U$ that crosses paths with our probe gets absorbed by it, i.e. it scrambles, even if the latter happens behind a horizon. The fact that scrambling continues to take place near our probe even inside the bigger black hole is ensured by our double-scaling limit (\ref{doublescaling2}), in which the probe's scrambling time remains finite at infinite $N$ and can be chosen to be smaller than the $O(L_{AdS})$ time needed to hit the background singularity. Our challenge is, therefore, a technical rather than a conceptual one: The correlator (\ref{modflowedint}) which was sensitive to the scrambling effect in the black hole exterior ceases to be a good probe of this phenomenon behind the horizon and we need an alternative quantity to diagnose the relevant scrambling physics. 

We conjecture that this sought-after quantity is complexity; our claim is that the operators $\phi_{K_{eq}}(\tau)$ evolved with the \emph{equilibrium} modular Hamiltonian generate the \emph{smallest increase} of the state complexity among the $\phi_{K_U}(\tau)$ for all other modular Hamiltonians $K_U$ of states in the code subspace. The intuition for our proposal comes from observing that non-equilibrium modular flows generate out-of-time-order products of bulk operators in the vicinity of our probe black hole $U\phi(\tau)U^\dagger$ which generally create states of higher complexity than the local operator $\phi(\tau)$ due to the well-understood chaotic near horizon dynamics. The more speculative element of our conjecture is that this continues to be true when both $\phi(\tau)$ and $U$ are operators in the interior of the background black hole our probe gets absorbed by. 

To make our statement precise, let us define the measure of complexity we are interested in: $C[\phi]$ is the increase in the complexity of the state $|\psi\rangle\in {\cal H}_{code}$ when we excite it with $\phi$
\begin{equation}
    C_\psi[\phi] = C[\phi|\psi\rangle ] - C[|\psi\rangle] \label{complexity}
\end{equation}
where $C[|\psi\rangle ] $ is the circuit complexity of $|\psi\rangle$ relative to some reference state, e.g. the CFT vacuum.\footnote{Since all states in the code subspace built around $|\psi\rangle$ have the same geometric background, hence the same maximal bulk volume, the complexity of the excitation $\phi$ can be measured with reference to any state in ${\cal H}_{code}$.} Our conjecture can then be expressed as follows:

\paragraph{Conjectured definition of local equilibrium:} \emph{Let $|\psi\rangle$ be a state in ${\cal H}_{sys} \otimes {\cal H}_{ref}$ containing a big background black hole and a small probe black hole entangled with the reference, as in figures \ref{fig:euclideanprep} and \ref{fig:tfdobserver}, ${\cal H}_{code}^\psi$ a code subspace containing $|\psi\rangle$ and $\phi_{atm}$ the CFT dual of a local bulk operator near the probe at a given moment in time. The local equilibrium state in $ {\cal H}_{code}^\psi$ is a state $|\psi_{eq}\rangle = U |\psi\rangle$ for the unitary $U:{\cal H}^\psi_{code}\to {\cal H}_{code}^\psi$ that minimizes the complexity of $\phi_{K_U}(\tau_{scr})= e^{iK_U \tau_{scr}} \phi_{atm}e^{-iK_U \tau_{scr}}$, for $\tau_{scr}=\log S_{probe}$:
\begin{align}
 |\psi_{eq}\rangle = U_{eq}|\psi\rangle \quad \text{for} \quad U_{eq} = \text{argmin}_U\, C_\psi\left[\phi_{K_U}(\tau \to \tau_{scr} )\right] \nonumber\\
 \text{among all } U:{\cal H}^\psi_{code} \to {\cal H}^\psi_{code} \label{complexityequilibrium}
 \end{align}
if and only if the norm of the minimal complexity operator $\phi_{K_{eq}}(\tau_{scr})$ projected onto ${\cal H}_{code}^\psi$ satisfies
\begin{equation}
   \frac{||\phi_{K_{eq}}(\tau_{scr})||_{code} }{||\phi ||_{code}} = O(S_{probe}^0) \label{reconstructionreq}
\end{equation}}

The extra criterion (\ref{reconstructionreq}) can be thought of as a reconstructibility condition, since if it is not satisfied, the modular flowed operator cannot be represented in the code subspace in the double scaling limit (\ref{doublescaling2}). This is a crucial addition to the local equilibrium condition, because while the complexity minimization problem will always have a solution, not all code subspaces will have an admissible local equilibrium state! Code subspaces built about non-semiclassical states or in which the probe collides at high energies with background objects or enters high curvature regions are prime examples. The search for local equilibrium states must fail in these cases. Condition (\ref{reconstructionreq}) ensures the search for $|\psi\rangle_{eq}$ can fail.

\subsection*{Arguments for the conjecture}

Let us present some evidence for the local equilibrium definition above. The gist of our arguments is as follows. Consider first a CFT state describing just the probe black hole and a bulk excitation of fixed characteristic Schwarzschild energy $\delta E$. The latter is created by a local bulk operator $\phi$ smeared over a small causal diamond of time-width $\Delta t \sim \delta E^{-1}$, centered about some bulk point. Focusing on the near horizon geometry, for simplicity, the geometry is approximately Rindler and we can define the null coordinates $x^\pm = \rho \,e^{\pm \frac{2\pi t}{\beta}}$ where $\rho$ the geodesic distance from the bifurcation surface. As we move the fixed asymptotic energy excitation towards the probe black hole along an infalling trajectory, e.g. $x^+=x^+_0=const$, we will argue that the \emph{complexity} of the dual CFT excitation (measured as the difference between the complexities of the state with and without the excitation) is expected to monotonically increase. This is reflected in the bulk by (a) the increase in the expectation value of the operator's null energy (tracking the increase in the dual state's ``size'' due to introducing the excitation) and a related increase of the particle's backreaction on the maximal volume slice (corresponding to the increase of the dual state's complexity). This fact, will then be combined with the observation that operators $U\phi U^\dagger$ create particles closer to the probe's horizon than $\phi$ in the kinematic regime of (\ref{complexityequilibrium}) due to the infallers' shockwave effect, in order to support the claim that equilibirum modular flowed operators minimize the complexity measure (\ref{complexity}). We will then generalize this idea to situations when the probe is behind other causal horizons, by pointing out that the change in complexity appears to follow the same pattern when the probe is in the interior region of a bigger background black hole, which is the case the prescription of Section \ref{sec:nonequilibrium} runs into trouble.

\begin{figure}
    \centering
    \includegraphics[width= 7cm]{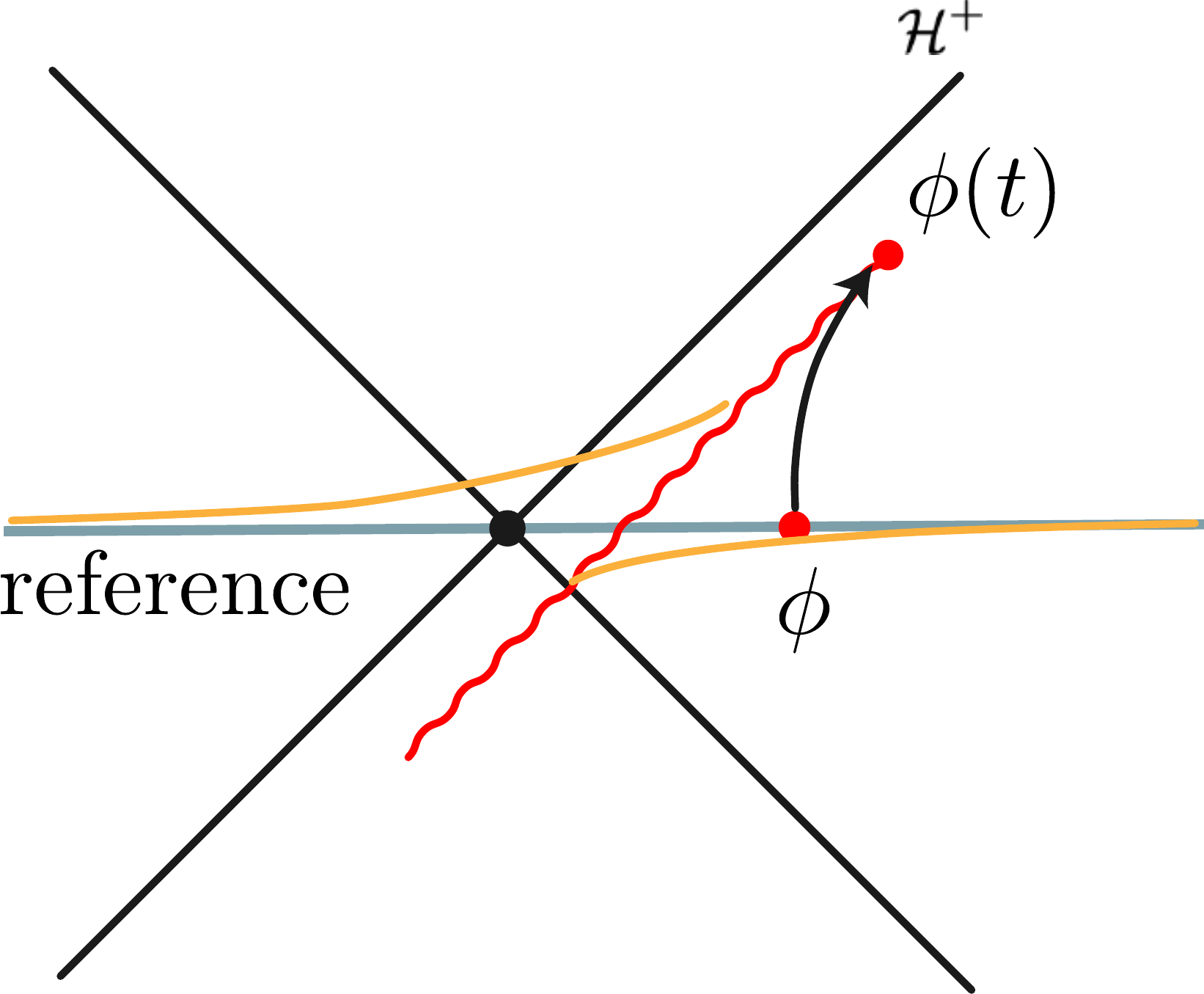} \includegraphics[width= 7cm]{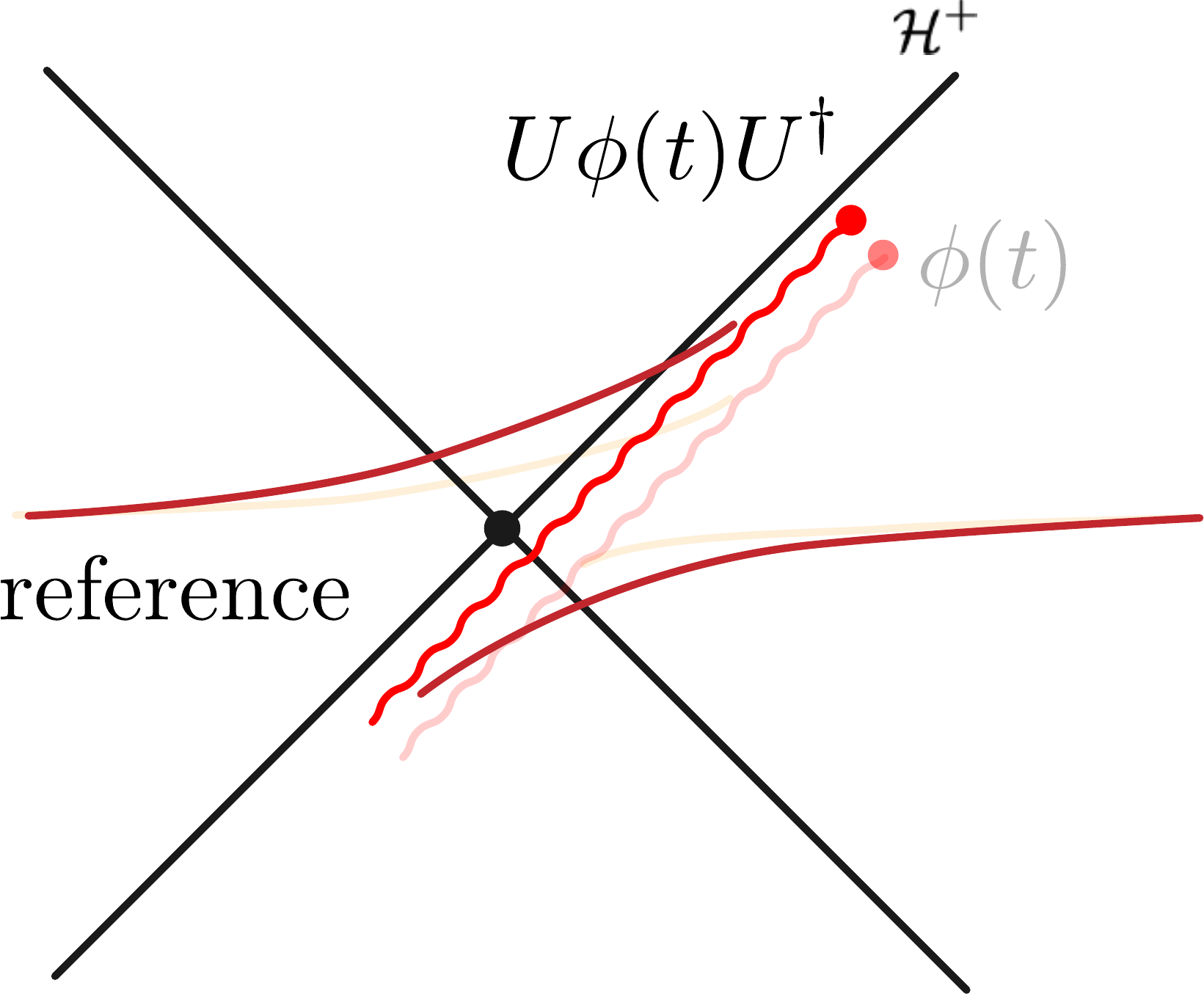}
    \caption{\footnotesize{\textbf{Left:} The geometry near the horizon of the probe black hole and the piece of the maximal volume slice  in that region (blue) that contains the initial atmosphere operator $\phi$. Acting with the Schwarzschild translated operator on the state backreacts on the maximal slice as $t\sim t_{scr}$, increasing its volume and the dual CFT state complexity. This increase is the complexity $C[\phi(t)]$ of the excitation created by equilibrium modular flow of $\phi$. \textbf{Right:} Non-equilibrium modular flows yield a Schwarzschild time translated field conjugated by unitaries describing local excitations on the initial slice. Due to the shockwave effect, $U\phi(t)U^\dagger$ lives closer to the probe's horizon, thus generating a higher energy shock that leads to a greater increase of the maximal volume than the equilibrium modular flowed operator.}}
    \label{fig:complexityargument}
\end{figure}

\paragraph{Evidence from operator size} Our first task is to establish that our complexity criterion is a good substitute of the criterion of Section \ref{sec:nonequilibrium} in the latter's regime of validity ---outside horizons. To make our argument precise, we can specialize to the AdS$_2$/SYK correspondence. Moreover, instead of focusing immediately on operator complexity, we may first look at the closely related notion of operator size. This is because, outside of black hole horizons, operator size and the complexity notion (\ref{complexity}) are closely related to each other. 

Reference \cite{Qi:2018bje} identified an SYK notion of operator ``size'', $S[O(t)]$, defined as the \emph{increase} in the average number of different fermion flavors needed to represent the state $O(t)\rho^{\frac{1}{2}}$ as compared to that of $\rho^{\frac{1}{2}}$, where $O(t)$ is the operator of interest and $\rho^{\frac{1}{2}}$ a background state ---in our case, the dual to the geometry of our probe black hole. This notion of operator size was found to be  holographically dual to the expectation value of the bulk radial momentum $P=P_+ -P_-$ of the excitation in the geometric background defined by $\rho^{\frac{1}{2}}$, namely $S[O]= |\langle O^\dagger \, P \, O\rangle|$, where $P_\pm$ the average null energy operators at the future and past horizons, generating translations $x^\pm  \to x^\pm +\epsilon^\pm$. This is a relation that is expected to hold in higher dimensional AdS/CFT setups as well \cite{Susskind:2018tei,Susskind:2020gnl}. 

We first use this correspondence to estimate the size of an excitation as we move it closer to the probe. Due to the exponential dependence of $x^\pm$ on Schwarzschild time, an excitation smeared over a fixed $\Delta t\sim \delta E^{-1}$ sized region becomes very localized in $x^-$ as we approach the horizon along a $x^+=x^+_0=const$ trajectory $\Delta x^- = \frac{2\pi \,x^-}{\beta} \Delta t$. By the uncertainty principle, the standard deviation of the conjugate momentum $\Delta P_-$ increases and since $P_\pm\geq 0$ by the Averaged Null Energy Condition, the dual operator size increases as: 
\begin{equation}
    S[\phi(\rho,t)]  \sim \Delta P_- \sim  \frac{\beta \,\delta E}{2\pi\,x^-}
\end{equation}
This quantity saturates at $x^- \sim L_{pl}$, when $S\sim N$ (or $N^2$ in higher dimensional gauge theories). This can be achieved either by bringing the operator at a Planckian geodesic distance from the horizon or by boosting Schwarzschild time beyond the scrambling time. It is worth mentioning here that, in contrast, operator complexity saturates at exponentially larger values, since it counts the number of elementary operations without allowing linear superpositions. Nonetheless, operator size is a good proxy for complexity in regimes of low complexity where causal reconstruction of the atmosphere suffices, like in the situation at hand.

To complete our argument, we want to now show that any $U\phi(t)U^\dagger$ is closer to the horizon than $\phi(t)$ and, thus, has a larger size (and complexity). This follows from recalling that the action of the operator $U\phi(x^+,x^-)U^\dagger$ on states $|\psi\rangle$ in the black hole code subspace is equivalent to \cite{tHooft:1987vrq,Kiem:1995iy,Polchinski:2015cea}:\footnote{when the center of mass energy of the $\phi(t)-U$ scattering is approaching the Planck energy but with an $O(\beta)$ impact parameter.}
\begin{align}
    U\phi(\rho,t) U^\dagger |\psi\rangle &= e^{iG}\left[\phi(\rho,t) U \right]U^\dagger |\psi\rangle +O(e^{-\frac{2\pi t}{\beta}},N^{-1}) \label{otop}
\end{align}
where $e^{iG}$ is a shorthand for the operation:
\begin{equation}
    e^{iG}\left[\phi_1(x_1^+,x_1^-,\omega_1) \phi_2(x_2^+,x_2^-,\omega_1)\right] = \exp\left[-16\pi iG_N  f(\omega_1,\omega_2) \partial_{x_1^-} \partial_{x_2^+}\right] \phi_1(x_1^+,x_1^-,\omega_1) \phi_2(x_2^+,x_2^-,\omega_1)
\end{equation}
where $(\nabla^2_{S^{d-1}}-1) f(\omega_1,\omega_2)= -2\pi \delta(\omega_1,\omega_2)$ the transverse graviton propagator on the horizon sphere, applied iteratively on larger products of local operators. $e^{iG}$ implements the shockwave-type gravitational interaction between the excitations $\phi_1$ and $\phi_2$ which for large (but not too large) time separations $|t_1-t_2|\gg \beta$ is the dominant effect \cite{tHooft:1987vrq}. This is because all other QFT effects dissipate due to quasinormal decay while the shockwave effect increases exponentially in $|t_1-t_2|$ up to the scrambling time $\beta \log S$. 

Using this property in (\ref{otop}) we find that non-equilibrium modular flowed operators create excitations closer to the horizon by an amount given by the Shapiro delay $\Delta x^-(\omega) \approx 16\pi G_N \int d\omega'\, f(\omega, \omega') \langle [P_+(\omega'),U] U^\dagger \rangle $. The increase in the operator size caused by this shift can be estimated fairly straightforwardly. Suppose we work in a Schwarzschild boost frame (i.e. pick an origin of Schwarzschild time in the near horizon region) in which the null energy of $U$ is $\langle [P_+,U]U^\dagger \rangle \sim O(\beta^{-1})$, hence $\Delta x^- \sim G_N \beta^{-d+1} $. The increase in operator size incurred by replacing $\phi(\rho,t)$ by $U\phi(\rho,t)U^\dagger$, for an operator at a geodesic distance $\rho \sim \beta$ from the horizon and at Schwarzschild time that approaches the scrambling time $t= t_{scr}-\sigma$, where $t_{scr}=\frac{\beta}{2\pi}\log S$, can then be estimated to be:
\begin{equation}
    \Delta S = S[U\phi(\rho,t)U^\dagger]- S[\phi(\rho,t)] \sim \frac{\delta E}{2\pi}\frac{G_N \beta^{-d+1}\,e^{\frac{2\pi t}{\beta}}}{1-G_N\beta^{-d+1}e^{\frac{2\pi t}{\beta}}} e^{\frac{2\pi t}{\beta}} \sim \left(\frac{\delta E}{2\pi}\frac{e^{-\frac{4\pi \sigma}{\beta}}}{1-e^{-\frac{2\pi \sigma}{\beta}}}\right) N^2 \label{sizechange}
\end{equation}
For $\beta\ll \sigma \ll t_{scr} $ which is the regime in which the relation (\ref{otop}) holds, the factor in the parenthesis in (\ref{sizechange}) is $\ll 1$ but it does not scale with $N$. We, therefore, conclude that the operator size of the general modular flowed operator $\phi_{K_U}(t)$ is $\Delta S = O(N^2)$ larger than the size of $\phi_{K_{eq}}(t)$ obtained by modular evolution of the local equilibrium state ---even when the perturbation $U$ away from local equilibrium consists of just a few thermal wavelength quanta. 

In conclusion, local equilibrium modular flow is the one that minimizes the size of the operator $\phi_{K_U}(t)$ as $t$ approaches $t_{scr}$, among the modular flows $K_U$ of all states in the same code subspace. Importantly, although different code subspace states differ only quantum mechanically while sharing the same classical background, the corresponding differences in operator size are macroscopic and scale with $N$. The minimal operator size is supporting the conjecture of minimal operator complexity, in this case.

\paragraph{Evidence from complexity-volume correspondence}
The increase in the state complexity caused by acting with the modular flowed operator can also be estimated more directly by looking at the response of the maximal volume of the bulk geometry to the presence of the particle created by $\phi_K(\tau)$ ---if a correspondence between state complexity and maximal bulk volumes is assumed to hold, at least parametrically. As we now explain, this also supports our conjecture that equilibrium modular flows produce the least complex excitations, among all modular flows in ${\cal H}_{code}$. The argument is summarized in figure~\ref{fig:complexityargument}.

We start by selecting the bulk maximal volume slice on which the initial atmosphere operator $\phi$ is local. This slice meets the boundary at some time $t_{init}$ and its volume $V_{init}$ estimates the complexity of the CFT state $|\psi\rangle_{init}$ at that time, $C_{init} \sim V_{init}$. Since $\phi$ is assumed to be a low energy field, the complexity of $\phi |\psi\rangle_{init}$ is also approximately $V_{init}$. In contrast, the complexity of the state $\phi(t)|\psi\rangle_{init}$ exhibits a discernible increase as the Schwarzschild time $t$ of the insertion approaches the scrambling time. The reason is the localization of the $\phi(t)$ wavepacket along the null direction, described earlier, and the corresponding increase in its null energy which renders the time translated particle a high energy shock in the frame of the initial maximal volume slice. The shock backreacts on the slice, inducing an increase in its volume, indicating the increase in complexity of the excited state. If the time translated local operator $\phi(t)$ is further conjugated by a unitary $U$ describing local excitations on the initial slice ---which is precisely the type of operator produced by non-equilibrium modular flows--- then by the shockwave effect $U\phi(t)U^\dagger$ is located \emph{closer} to the probe's horizon. This corresponds to a \emph{higher} energy shock than $\phi(t)$ from the point of view of the maximal slice and, by extension, yields a larger increase in the state complexity.\footnote{Notice here the presence of two distinct shockwave effects which are both caused by the redshift of the probe's near horizon region: The first shockwave is generated by the operator $\phi(t)$ and results in the deformation of the maximal volume slice at an earlier time. The second shockwave is generated by $U$, i.e. the initial quanta that disturb the atmosphere's local equilibrium, and is the shock that shifts $\phi(t)$ closer to the horizon, increasing its null energy. } Equilibrium modular flow, therefore, generates operators $\phi_{K_{eq}}(t)$ which induce the smallest complexity increase of the state $|\psi\rangle_{init}$, among all modular flowed operators $\phi_{K_U}(t)$ for $U: {\cal H}_{code} \to {\cal H}_{code}$.

\subsection*{Behind black hole horizons} Our actual setup of interest is slightly more complicated than the single black hole situation we have discussed so far. We have two relevant horizons, that of the background and of the probe black hole, respectively. The goal is to use the probe to detect the presence of unitary excitations in the interior of the background black hole; this feat is tantamount to resolving the frozen vacuum. It is straightforward to see, however, that the argument made above directly generalizes to this case. 

Unitary excitations in the interior of the background black hole that fall into our probe and disturb its local equilibrium, have exactly the same effect on the modular flowed operator $\phi_K(t)$ as before: For sufficiently large $t$ they simply shift the Schwarzschild time translated operator $\phi(t)$ closer to the probe's horizon. In order to diagnose this phenomenon using the prescription (\ref{complexityequilibrium}), we need to further argue that operators located in the \emph{interior} of the background black hole but \emph{outside} the probe's horizon become more complex when they are transported closer to the latter. To see this, we may repeat the complexity-volume argument made above.

Operators in the background black hole interior can have high complexity in the CFT \cite{Susskind:2020gnl, Haehl:2021emt}, but acting with them on the state does not necessarily result in a large increase of the state complexity because their backreaction on the maximal volume will typically be small. Due to the redshift factor in the atmosphere of our probe, however, otherwise low energy fields in the probe's atmosphere, e.g. $\phi$, can be converted into high energy shocks in the frame of the maximal volume slice, when translated in the probe's Schwarzschild time, $\phi(t)$, in a direct parallel to our earlier discussion. The rest of the argument, then, proceeds exactly as above: By shifting the shock $\phi(t)$ closer to the probe's horizon always increases its energy in the frame of the maximal volume slice, thus yielding a larger increase of its volume and, correspondingly, of the state's complexity. Since non-equilibrium modular flowed operators are always located closer to the probe's horizon than equilibrium ones, $K_{eq}$ minimizes the complexity of the modular flowed excitation. It would be very interesting to obtain direct CFT evidence for the behavior of complexity argued above. This, however, requires a working CFT definition of complexity that is sufficiently computationally tractable which lies beyond the scope of this work.

\section{Discussion}
\label{sec:discussion}
This paper develops a holographic reconstruction technique that aspires to unify operator reconstruction in the exterior and interior of black holes, resolve the ``frozen vacuum'' ambiguity of the latter \cite{Bousso:2013ifa} while, crucially, relying on first CFT principles, i.e. make no reference to the background geometry of the bulk spacetime other than such a semi-classical description exists. Conceptually, our proposal is a microscopic definition of the Hamiltonian generator of the time an internal semi-classical observer experiences in their rest frame. This is technically accomplished in two steps:
\begin{enumerate}
    \item A physical subsystem of the Universe is selected to be our observer. The observer's internal degrees of freedom are traced out, yielding a modular Hamiltonian for the rest of the Universe; in our framework, this is achieved by appending, entangling and subsequently tracing out the reference (fig.~\ref{fig:euclideanprep}). 
    \item This modular Hamiltonian depends on the precise state of the Universe and thus it differs for different QFT states about the same background geometry, i.e. in the code subspace ${\cal H}_{code}$. The modular Hamiltonian for the \emph{local equilibrium} state in ${\cal H}_{code}$ is identified with the proper time Hamiltonian of the observer. A key element of this paper is identifying a well-defined extremization principle for selecting this special local-equilibrium state in the CFT, discussed in Section \ref{sec:nonequilibrium} for exterior black hole reconstruction and in Section \ref{sec:interiorreconstruction} for a conjectured generalization that works in the interior.
\end{enumerate} 
In this Section, we discuss how our proposal informs and improves on other reconstruction approaches. We also summarize some important open questions raised in the text which appear to be within reach in the near future and conclude with speculation on the use of our framework in de Sitter quantum gravity.

\subsection*{Relation of our approach to other reconstruction methods}

\paragraph{HKLL, modular flow and Petz map} The standard method for defining local bulk operators in the holographic CFT is the HKLL method \cite{Hamilton:2006az}. These are CFT operators smeared over the boundary spacetime with a kernel that is determined, at leading order in $1/N$, by requiring that their two-point functions, computed in the CFT, reproduce the expected bulk two point-function; matching of the higher-point functions is then accomplished by perturbative $1/N$ corrections to these operators \cite{Kabat:2011rz, Kabat:2013wga} or the requirement of micro-causality \cite{Kabat:2015swa}. This extremely useful approach has two shortcomings: (a) It is a causal reconstruction, hence it cannot be used to define operators beyond the causal wedge of the boundary like the interior of black holes, and (b) it presupposes knowledge of the background geometry on which the bulk wave equation ought to be solved to obtain the bulk correlators required in the first step. In other words, the bulk semi-classical description of the system must \emph{already be known to us} in order to identify the, otherwise obscure, CFT operators that reproduce it. This problem appears innocuous about very symmetric states like the vacuum but is significant about general backgrounds; more importantly, this property of the HKLL map renders it somewhat uninformative, since no obvious CFT physics is leveraged to explain this emergent organization of the boundary degrees of freedom (see, however, \cite{Kabat:2016zzr}). This second limitation is, in a sense, more important since it survives in most other bulk reconstruction methods that manage to overcome the first, as we discuss shortly. 

Methods that extend the scope of reconstruction beyond the bulk regions causally accessible from the boundary and all the way to the edge of the entanglement wedge include the modular flow approach of \cite{Faulkner:2017vdd} and the Petz map construction of \cite{Cotler:2017erl, Chen:2019gbt}. Both methods are based on the following rationale: If the matrix elements of a bulk local operator $\phi$, inside the entanglement wedge of the boundary, with states in the code subspace ${\cal H}_{code}$ are known, then a CFT operator can be constructed that acts in the same way within ${\cal H}_{code}$. This allows reconstruction in the black hole interior \cite{Penington:2019kki} but, once again, one should pay attention to the assumed input: The bulk semi-classical description in the reconstruction region is explicitly utilized to define the matrix elements of $\phi$. Therefore, the CFT representation of $\phi$ provided by these reconstructions is, by design, destined to reproduce exactly the bulk QFT physics used to obtain the relevant correlators; if the latter fails, so does the former. One caveat in this argument is that in defining the CFT operator, the matrix elements are the same but the Hilbert spaces, i.e. the inner products between the relevant states, are different by virtue of the non-isometric property of the holographic map \cite{Akers:2022qdl, Kar:2022qkf}. Thus the two operators differ non-perturbatively. Puzzles of the black hole interior, however, exist already at the perturbative QFT level, for example whether a semi-classical behind the horizon region even exists for a particular black hole microstate, so the essence of our statement is unaffected by this subtlety.

In summary, all the approaches discussed so far do not constitute \emph{definitions} of bulk observables but instead they are \emph{maps} from the \emph{already defined QFT observables} in the bulk Hilbert space about a known semi-classical background (i.e. with correlation functions in ${\cal H}_{code}$ computed using the bulk QFT description) to operators acting on the non-perturbatively defined CFT Hilbert space. Our approach, therefore, comes with a clear advantage: It is a set of physically well-motivated CFT principles that \emph{define} the collective boundary excitations that make up the local fields in the holographic bulk, near the worldline of a bulk observer. No information about the specifics of the bulk background or the QFT equations of motion is inputted, except for the general assumption that such a description is ultimately available. Every step in our construction is fully articulated in the CFT.

An alternative method among the previous family of reconstruction approaches that mitigates the same problems is \cite{Kabat:2017mun, Kabat:2018smf, Callebaut:2022mqw}, which defines observables living at the intersection of RT surfaces as simultaneous zero modes of the corresponding modular Hamiltonians. This, in principle, goes beyond the causal wedge and does not use the background geometry as an input, since the existence or absence of simultaneous zero modes of different modular Hamiltonians indicates whether the dual RT surfaces intersect or not. This procedure has been technically implemented about the vacuum. Our method's advantage as compared to this case lies in its more straightforward technical implementation in more interesting setups, e.g. behind the horizon reconstruction \cite{Gao:2021tzr}, its ability to approach black hole singularities (Section \ref{sec:singularity}) and, conceptually, in providing not just a definition of bulk observables but also explaining the origin of the apparent dynamical laws in the reference frame of an internal observer.

\paragraph{Mirror operators and modular translated operators}
A second family of reconstruction techniques have attempted to \emph{define} bulk operators in the black hole interior in the CFT, using boundary principles. These include the Papadodimas-Raju (PR) mirror construction \cite{Papadodimas:2012aq, Papadodimas:2013jku, Papadodimas:2015xma}, the Verlinde-Verlinde proposal \cite{Verlinde:2012cy, Verlinde:2013qya} and related subsequent constructions of Nomura et al \cite{Nomura:2020ska, Murdia:2022giv} and the Leutheusser-Liu algebraic construction \cite{Leutheusser:2021frk, Leutheusser:2021qhd}. 

The first starts with the established HKLL reconstruction in the exterior region and proceeds to define the interior modes as the \emph{modular conjugation} of the exterior ones  ---roughly, the degrees of freedom that are thermally entangled with the exterior ones. Slightly more precisely, they are defined to act as the modular conjugation of the exterior operators on code subspace states but a separate requirement is made about their commutator with the boundary Hamiltonian, in order to be appropriately dressed gravitational observables. The ambiguous part of this prescription stems from the fact that the modular conjugation operator differs for different CFT states. For example, two states that differ by the action of a unitary operator in the black hole interior would lead to a different definition of the local interior operators (related by conjugation with the interior unitary), despite the fact that the states look identical from the asymptotic boundary due to bulk causality. One, therefore, needs to select a preferred state (or class of states) in each black hole code subspace about which interior reconstruction should be performed.

Papadodimas and Raju challenge the idea that interior unitaries that commute with all simple boundary observabes and, in particular, with the boundary Hamiltonian exist by virtue of the gravitational Gauss' law that requires appropriate ``dressing'' of all localized bulk operators to the asymptotic boundary. This allows them to choose the special reference states in every code subspace: They are the black hole states with sufficiently mild time-dependence, which they dub ``equilibrium''. However, good gravitational observables can be built by dressing local operators to \emph{any} feature of the background, even to the structure of the black hole interior state, as discussed in \cite{Iizuka:2021tut, Bahiru:2022oas}. It, therefore, seems quite likely to us that interior unitaries that commute with all simple boundary observables to all orders in perturbation theory can exist. This casts doubt on a foundational assumption of the PR construction, since their equilibrium condition is insensitive to interior unitaries with this alternative ``dressing'' and the proposal is rendered once again ambiguous.\footnote{Some further objections to the PR equilibrium notion were raised in \cite{Harlow:2014yoa} but reasonably satisfactory answers were provided in \cite{Papadodimas:2015jra}. A different persisting objection regarding typical states is discussed momentarily.} Another contentious aspect of the PR equilibrium notion is its prediction that typical black hole microstates are equilibrium and thus have smooth interiors \cite{deBoer:2018ibj,DeBoer:2019yoe}. This prediction goes contrary to reasonable alternative arguments made in \cite{Susskind:2015toa,Susskind:2020wwe} and, more recently, in \cite{Stanford:2022fdt} that support a different thesis. In short, while the PR proposal may ultimately be correct, it \emph{is} at present a choice and no satisfactory justification or a deeper underlying principle is currently available. This becomes a serious matter when contemplating its consequences: The states that will be dubbed ``PR equilibrium'' and will be used to define the interior modes will, by construction, have a smooth uneventful horizon in the infalling frame. The choice of PR equilibrium condition is, thus, equivalent to a choice of smooth horizon black hole microstates. We must, therefore, proceed with caution.

The works \cite{Verlinde:2012cy, Verlinde:2013qya} and \cite{Nomura:2020ska, Murdia:2022giv} provide a more microscopic justification for using modular conjugation to obtain the interior degrees of freedom, by combining the error correcting properties of the CFT subspace spanned by bulk QFT excitations on the black hole spacetime with the chaotic dynamics of the holographic CFT. The ``mirror operator'' formalism is an output of a more microscopic construction of interior operators, given in terms of the recovery map in an appropriately defined black hole code subspace ---with the mirror operators being an effective description of these operators when the recovery map is a sufficiently random operator.
With this choice of interior operators, the PR equilibrium condition is argued to be self-consistent. Nonetheless, it is once again unclear what prevents us from rotating the definition of the interior modes by a unitary $U$ acting only on the black hole interior and change by hand the interpretation of the states satisfying the equilibrium condition. Let us be clear about this: We are not saying that no argument for why this is not an allowed move exists; we are merely saying that we are not aware of an explicit physical argument against it and that understanding this aspect is an important question.

Our framework informs these discussions by providing a physically motivated definition of what the preferred equilibrium states in the relevant code subspace are. In contrast to the PR proposal, our notion of equilibrium is a local one, referring to the vicinity of a particular observer. The inclusion of the probe in our story was crucial for achieving this, since it was the scrambling properties of our probe black hole that we leveraged to select the ``local vacuum'' state in its neighborhood. It is also independent of whether a background black hole exists in the bulk or not; the local-equilibrium condition is needed for defining the proper time Hamiltonian of our observer in all cases, and it is not an extra element of the holographic dictionary when black holes are present. The relation between our equilibrium condition and the one appearing in these previous works is not clear but it is worth exploring more. At a more conceptual level, our approach also explains ``why'' a particular, rather complicated, CFT operator describes a local degree of freedom from the point of view of a bulk observer, by defining a continuous family of CFT operators (related by modular flow) that are local relative to the observer at different proper times.  

A more recent series of works \cite{Leutheusser:2021frk,Leutheusser:2021qhd} that is, perhaps, more conceptually related to our approach, used modular theory to define a unitary flow that evolves exterior operators to the interior of a two-sided black hole. The approach elaborated on ideas previously presented in \cite{DeBoer:2019kdj} and exploited the fact that the difference of modular Hamiltonians $K_{1,2}$ of included QFT subalgebras ${\cal A}_{1,2}$, e.g. the algebra ${\cal A}_1$ of operators in a causal wedge ${\cal W}_1$ in the black hole exterior and the one in a causal wedge ${\cal W}_2$ that is obtained by a null deformation of ${\cal W}_1$ and contained in it, defines the \emph{modular translation} operator $P=K_2-K_1$ whose flow propagates operators to the common future of ${\cal W}_1$ and its complement. Since the modular Hamiltonians depend explicitly on the choice of state, however, so does the modular translation. One then confronts the same obstacle as in the previous constructions: We can define infinitely many modular translation generators, one for every state in the code subspace. For example, we can consider the thermofield double state $|TFD\rangle$, with $K_{1,2}$ the modular Hamiltonians of two included causal wedges of the right bulk exterior and $P$ the modular translation, and also the state $U_L|TFD\rangle$, for which $K'_{1,2} = U_L K_{1,2} U^\dagger_L$ and, similarly, $P'=U_L P U_L$. These clearly define different flows. All of them produce the same abstract algebra of interior operators but what is unclear is which operators obtained this way define the basis of local fields in the interior. Is there for example a code subspace state for which the modular translation operator maps local exterior operators to local interior ones? Or if not, is there a separate principle for selecting the local basis of the interior algebra?

As discussed in detail in Section \ref{sec:infoparadox}, access to the local operator basis in the interior is key for diagnosing whether the infalling observer sees a smooth interior. That is because arbitrary unitary excitations of the interior can always be made invisible by rotating the interior algebra by the same unitary ---at the expense, of course, of the locality of the operators. Once again, a special choice of state needs to be made, in every black hole code subspace. For some cases, e.g. the eternal AdS black hole background, a natural special state exists, i.e the thermofield double state. It is unclear how to pick this state in more general situations, like long wormhole spacetimes or single-sided black holes. It is this ambiguity in the interior reconstruction that we are able to fix in a physically meaningful way using our approach.

\subsection*{Some concrete questions}
In the main text, we formulated a few interesting technical questions that can illuminate long-standing black hole puzzles using the ideas developed in this work. For the benefit of the reader, we summarize them here.
\begin{enumerate}
    \item Consider the modular flowed correlator 
    \begin{equation}
        \text{Tr}_{lr}\left[ \rho^{1-i\tau}O_r \rho^{i\tau} O_r\right]
    \end{equation}
    where $\rho$ is the density matrix defined by the Euclidean path integral of figure~\ref{fig:tfdobserver}, $O_r\in {\cal A}_{CFT_r}$ and the trace is over ${\cal H}_l\otimes {\cal H}_r$, which is supposed to propagate the operator $O_r$ along an infalling observer's geodesic in the eternal AdS black hole background. Does it exhibit the divergence (\ref{divergence}) in the double scaling limit (\ref{doublescaling2}), at the proper time $s_{div}=\frac{\tau_{div}}{\log S}$ when the observer hits the singularity? What CFT physics underlies this divergence and how does it get resolved in the finite $N$ theory?
    \item Does the same computation in a typical single-sided black hole microstate (namely, for $\rho$ given by the path integral of figure~\ref{fig:euclideanprep} after tracing out the reference) behave the same way? What is the expectation value and variance of the rescaled modular time $s_{div}$ at which the divergence occurs (i.e. proper time to the singularity) for black hole states in an energy window? This is a technical way of probing the question whether typical microstates have a macroscopic interior.
    \item Can the minimal complexity definition of local equilibrium states be established directly on the boundary in a simple example e.g. in the SYK model?
\end{enumerate}
Our framework is also potentially useful for diagnosing whether an entangled state between two CFTs corresponds to a semiclassical bulk wormhole, a topic recently explored in some detail in \cite{Engelhardt:2022qts}. The natural approach in this case is to send in a couple of infalling observers from the two exteriors and use their local equilibrium modular flows to propagate left and right observables behind the two horizons, respectively. If a shared semi-classical interior exists, the two sets of modular flowed operators should have identical action within the code subspace. This, of course, works when there exists a region that is causally accessible from both exteriors. For longer wormhole states, the causally inaccessible interior region cannot be probed this way, but the meaning of a semi-classical description of a region of space that cannot be probed by any observer is somewhat unclear.

\subsection*{De Sitter symmetries}
Our framework provides a set of rules for ascribing dynamical laws to a quantum system, i.e. a notion of internal time evolution, given (a) a Hilbert space, (b) a notion of ``simple'' operators that can be used to define operator complexity and (c) a decomposition of the system's degrees of freedom into observer and environment subsystems. Importantly, no externally defined Hamiltonian is needed. We may, therefore, expect it to be useful in quantum cosmology, e.g. in the microscopic description of de Sitter space, where no natural notion of time exists already in its semi-classical description, due to the absence of a ``cold'' asymptotic boundary. This perspective follows the spirit of earlier suggested approaches, e.g. \cite{Anninos:2011af}.

The recent work \cite{Chandrasekaran:2022cip} proposed a mathematical framework for defining gauge-invariant observables in a static patch of de Sitter space and explored the structure of their algebra. Interestingly, the spirit of their approach is closely aligned to ours: An extra system is appended to the de Sitter Universe's degrees of freedom which plays the role of the observer and the gravitational gauge constraints are imposed on the combined system. The observer's role is twofold: It selects a natural time evolution for the system, among the infinity of possible time evolutions of static patch observers  ---exactly as in our case--- and it provides an ``anchor'' where static patch observables can be gravitationally dressed to ---similarly, our atmosphere operators are defined by their location relative to the probe's horizon. The only difference of our observer is that it is ``internal'' to the system; it is a small black hole made out of the degrees of freedom of the Universe. Since black holes are universal gravitational objects, any theory of quantum de Sitter should include them as excitations.

Taking our construction seriously, then, expands on the discussion of \cite{Chandrasekaran:2022cip} in an interesting way. Given an initial observer and their static patch operator algebra, an infinite number of other semi-classical observers ---in relative motion to the original one--- can be introduced by inserting small black holes, as in Section~\ref{sec:framework}. The corresponding modular Hamiltonians (for the appropriate local equilibrium states) are identified with the Hamiltonians of these relatively boosted static patches. This is an important observation because different static patch Hamiltonians are mapped onto each other by the isometry group of de Sitter space. Our modular Hamiltonians must, therefore, obey the de Sitter algebra in the semi-classical limit. This is a very non-trivial property which very likely constrains the algebraic structure of the microscopic system. It may, therefore, offer a useful guide in the efforts to construct toy models of de Sitter quantum gravity.

\noindent \paragraph{Acknowledgements:} We have benefited greatly from conversations with a large number of people over the past couple of years. We thank Ahmed Almheiri, Panos Betzios, Raphael Bousso, Ven Chandrasekaran, Ping Gao, Arjun Kar, Adam Levine, Raghu Mahajan, Juan Maldacena, Charles Marteau, Yasunori Nomura, Kyriakos Papadodimas, Pratik Rath, Felipe Rosso, Moshe Rozali, Jamie Sully, Lenny Susskind, Mark Van Raamsdonk, Erik Verlinde and Ying Zhao for useful discussions and comments. LL is supported by the Simons Foundation via the It from Qubit Collaboration. The work of DLJ is supported in part by the DOE grants DE-SC0007870 and DE-SC0021013. JdB is supported by the European Research Council under the European Unions Seventh Framework Programme (FP7/2007-2013), ERC Grant agreement ADG 834878.

\appendix
\section{Near horizon modular flow in semiclassical states}
\label{app:modularflow}

Suppose that we start with a probe black hole in a state $|\psi\rangle$ of the type of figure~\ref{fig:euclideanprep} which contains particles destined to get absorbed by our probe. Consider also $\phi_{atm}(0,\vec{x})$ to be a local atmosphere field at some initial moment.  

Assuming the state $|\psi\rangle$ is semi-classical, the modular flow of the probe $K=VHV^\dagger$ can in general be expressed as $K=UH_{p.t.}U^\dagger$ where the unitary $U$ describes perturbative bulk excitations, assumed to be initially supported at spacelike separations from the atmosphere $A$, namely in the complementary causal diamond $C_{\bar{A}}$. The content of this rewriting is that part of the unitary $V$ is responsible for the new classical background our Black Hole will propagate in ---and thus is absorbed in the ``equilibrium'' component of $K$, i.e. the local Schwarzschild Hamiltonian $H_{p.t.}$, while the rest of it, $U$, encodes the additional QFT excitations. 

We should emphasize, here, that we have not yet explained \emph{how} to obtain this rewriting of $K$ starting from the original expression $K=VHV^\dagger$, we simply stipulated that such a rewriting exists. Our strategy is to compute the modular flow of $\phi_{atm}$ using expression $K=UH_{p.t.}U^\dagger$, identify a universal signature of the existence of infalling particles in $U$ and leverage it to propose a method for removing them, extracting in effect $H_{p.t.}$.

For concreteness we will make a Gaussian approximation to $U$ and write it as:
 \myeq{ U= \exp \left[ i\int_{{\bar{A}}} J(x) \phi(x)  + i\int_{{\bar{A}}} \int _{{\bar{A}}} F(x,y) \phi(x)\phi(y)\right] }
but this assumption will not be important. We will also neglect all QFT interactions with the exception of the field's coupling to gravity which will play a central role. The modular evolution then reads\footnote{dropping the subscript $atm$ to reduce clutter}
\myal{\phi_K(x, s) &= U e^{iH_{p.t}s} U^\dagger \phi(x,0) Ue^{-iH_{p.t}s} U^\dagger = U \phi_{H_{p.t}}(x,s) U^\dagger \nonumber\\
&= \phi_{H_{p.t}}(x,s) -i \frac{\delta U}{\delta \pi_{H_{p.t}}(x,s)} U^\dagger\nonumber\\
&= \phi_{H_{p.t}}(x,s) + \int_{y\in {\bar{A}}} J(y) \frac{\delta \phi(y,0)}{\delta \pi_{H_{p.t}}(x,s)} +    \int_{y\in {\bar{A}}}\int_{z\in {\bar{A}}} F(y,z) \frac{\delta }{\delta \pi_{H_{p.t}}(x,s)}\Big(\phi(y,0) \phi(z,0)\Big) \label{flow1}}
where the subscript denotes what operator we used to flow $\phi$. 

To compute (\ref{flow1}) we recall that, neglecting the coupling to gravity for a moment, we can express a free field in terms of the canonical variable pair $\phi$ and $\pi$ on a different Cauchy slice $\Sigma(s)$ as:
\myeq{\phi(x,0) = \int_{(y,s)\in D(x,0)\cap \Sigma(s)} \left( G_{ret}(x,0|y,s) \pi_{H_{p.t}}(y,s) + \partial_s G_{ret}(x,0|y,s) \phi_{H_{p.t}}(y,s)\right)}
where $D(x,0)$ denotes the lightcone of the point $(x,0)$. The field derivatives of this expression will give the free field contribution to $\phi_K$. The full result, should also include the contribution from the gravitational interaction. The precise form of this correction is complicated but thankfully not necessary for our discussion since most contributions to it are suppressed in $1/N$ and decaying in time. The correction of interest is the universal \emph{shockwave} part of the gravitational scattering. The latter is exponentially growing, leading to an $O(1)$ effect after $t\sim O(\beta \log N)$ and leaves a persistent imprint on the atmosphere operators in the form of a Shapiro delay. This can be straightforwardly incorporated in (\ref{flow1}) using the reasoning of 't Hooft \cite{tHooft:1987vrq} and Kiem, Verlinde, Verlinde \cite{Kiem:1995iy}, yielding overall:
\myal{\phi_K (x,s) &= \phi_{H_{p.t}}(x,s)  + \int \limits_{y \in {\bar{A}}\cap D(x,s)} J(y) G_{ret}(y,0|x,s) \nonumber\\
&+\int \limits_{y \in {\bar{A}}\cap D(x,s)} \int_{z\in {\bar{A}}}G_{ret}(y,0|x,s) F(y,z) \phi(z,0)+\int \limits_{y \in {\bar{A}}\cap D(x,s)} \int_{y\in {\bar{A}}}G_{ret}(z,0|x,s) F(y,z) \phi(y,0) \nonumber\\
&+  G_Ne^{2\pi s}  \int d\omega \, G^{grav}_{++--}(\omega| \omega(x,s)) \Big( U^\dagger [P_+(\omega), U]\Big) \, \frac{\partial}{\partial x^-} \phi_{H_{p.t}} (x, s) \label{phiK}}
where again $D(x,s)$ denotes the bulk lightcone of the point $(x,s)$. \\

It is useful to summarize the physical meaning of the different terms above:
\begin{itemize}
\item The first term is the Schwarzschild translation of $\phi$ near the horizon, generated by $H_{p.t}$. It is the geometric contribution to the modular flow which we are ultimately interested in extracting.
\item The second term is equal to $-\langle U^\dagger \phi_{H_{p.t}}(x,s) U\rangle$, which sets the expectation value of $\phi_K(x,s)$ in the state $|\psi\rangle$ to $0$ for all $s$, even in the cases where $\langle \phi_{H_{p.t}}(x,s)\rangle \neq 0$. This is a general fact about modular flow which follows from its defining property $K|\psi\rangle =0$.
\item The third and forth terms are non-local contributions to the modular flow of the local field $\phi(x)$. The appearance of such terms is very generic and they are a  consequence of bulk field entanglement in the chosen state $|\psi\rangle$. While they may seem hopelessly complicated due to their sensitive dependence on the precise state, it is important to observe that they all depend on the bulk causal propagator in the probe Black Hole background and, as a result, they decay in Schwarzschild time ---typically exponentially. Their effect is, thus, transient.
\item The last term encodes the gravitational interaction between $U$ and $\phi_{H_{p.t}}$ in the shockwave approximation, where $G^{grav}_{++--} (\omega_1| \omega_2) $ is the transverse graviton propagator at the horizon with $\omega(x,s)$ the solid angle of the bulk point $(x,s)$. This is negligible as long as $G_N e^{2\pi s} \ll 1$ assuming the horizon null energy of $U$ is $O(1)$, but becomes important at late Schwarzschild times. This is a persistent effect and it will be central in our analysis.
\end{itemize}

\paragraph{Persistent effects at large $s\sim \log N$} 

As is apparent from (\ref{phiK}), the presence of infalling matter that disturb the thermal equilibrium near our probe severs the direct link between modular flow and proper time evolution. Nevertheless, it was observed in \cite{tHooft:1987vrq} that most contributions to $\phi_K(s)$ are of a transient nature, due to quasinormal decay in the black hole atmosphere, reflected in the causal propagator property: 
\myeq{G_{ret}(y,0|x,s) \sim \exp\(-\frac{s}{\beta}\)}
In the $s\to \beta \log N$ limit, the terms that survive are the proper time flow $\phi_{H_{p.t.}}$ and the Shapiro delay which depends on the Averaged Null Energy of the background state. The large $s$ modular flow of an atmosphere field can, thus, always be expressed in terms of the ``equilibrium'' modular flow $H_{p.t.}$ corrected by a null shift with a \emph{positive} coefficient. Schematically:
\myeq{e^{iK_\psi s} \to  \exp \left[ iH_{p.t.}s +i  G_N\int P_+^{inf}(\omega) \,f(\omega, \omega') P_-(\omega')\, s \right] \label{result}}
The null shift on the right hand side reflects the fact that the event horizon of the black hole in the state $\psi$ is shifted relative to that of an equilibrium state $|\psi\rangle_{eq}\in {\cal H}^\psi_{code}$, due to the infalling excitations. The second term, turned on only by the presence of infalling energy and with a fixed sign coefficient, is responsible for the decay of the corrrelator (\ref{eqflow}) in Section \ref{sec:nonequilibrium} when the modular flow is not local equilibrium. Due to the dissipation of all other QFT effects at timescales comparable to the scrambling time, no other change of the code subspace state can increase the modular flowed correlator (\ref{eqflow}), other then removing the infalling energy that sources the second term in equation (\ref{result}).

\section{Wightman functions near black hole singularities}
\label{app:singularity}
\subsection{BTZ black hole}
We are interested in computing the Wightman 2-point function of a scalar field operator in black hole backgrounds. We are particularly interested in its behavior as one of the operators approaches the singularity. The first case we study is the 3-dimensional BTZ black hole with metric:
\begin{equation}
    ds^2= -(r^2-1)dt^2 +\frac{dr^2}{r^2-1} + r^2 d\theta^2 \label{btz}
\end{equation}
The horizon is at $r=L_{AdS}=1$, the asymptotic AdS boundary at $r\to \infty$ its temperature is set to $\beta=2\pi$ and the singularity at $r\to 0$. In this case, the singularity is of the conical type. 

\subsubsection*{Setting up}
We begin with a quick review of the details of quantizing a scalar field in the BTZ background. Canonical quantization about higher dimensional black holes proceeds analogously.

\paragraph{Outside the horizon} Suppose that $\phi$ is a massless scalar field on this background, whose equation of motion is:
\begin{equation}
    \Box \phi =0 \label{kleingordon}
\end{equation}
To canonically quantize the field, we introduce creation and annihilation operators for each AdS-Schwarzschild frequency $\omega$ and angular momentum $l$ 
\begin{equation}
   [a_{\omega,l}, a_{\omega', l'}^\dagger ] = \delta (\omega-\omega') \delta_{ll'} \label{ladderops}
\end{equation}
and define the local field operator $\phi(t,r,\theta)$ as:
\begin{equation}
    \phi(t,r,\theta) = \sum_{l=-\infty}^\infty \int_0^\infty \frac{d\omega}{\omega^{\frac{1}{2}}} \,a_{\omega, l}\,e^{-i\omega t} e^{il \theta} f_{\omega,l}(r) +\text{h.c.} \label{fieldop}
\end{equation}
The radial wavefunctions $f_{\omega,l}(r)$ are chosen to solve (\ref{kleingordon}) for each $\omega, l$, subject to the asymptotic boundary condition 
\begin{equation}
    f_{\omega,l}(r\to \infty ) \sim \frac{C(\omega, l)}{r^2} \label{asymptoticbc}
\end{equation}
which ensures its normalizability. Condition (\ref{asymptoticbc}) picks a unique solution to (\ref{kleingordon}) up to an overall normalization $C(\omega,l)$. This solution's behavior near the horizon is a superposition of ingoing and outgoing modes 
\begin{equation}
    f_{\omega,l}(r\to 1^+) \sim C(\omega,l) \left(A(\omega,l) (r-1)^{\frac{i\omega}{2}} +B(\omega,l) (r-1)^{-\frac{i\omega}{2}}\right) 
\end{equation}
with equal probability amplitudes $|A(\omega,l )|= |B(\omega,l)|$. The conventional normalization of $f_{\omega,l}(r)$ is, then, to select $C(\omega,l)= \left(A(\omega,l)B(\omega,l)\right)^{-\frac{1}{2}}$ so that
\begin{equation}
    f_{\omega,l}(r\to 1^+) \sim e^{i\delta(\omega,l)} (r-1)^{\frac{i\omega}{2}} +e^{-i\delta(\omega,l)} (r-1)^{-\frac{i\omega}{2}} \label{normalization}
\end{equation}
with $e^{i\delta(\omega,l)}= \left(\frac{A(\omega,l)}{B(\omega, l)}\right)^{\frac{1}{2}}$. This amounts to simply normalizing with respect to the usual Klein-Gordon inner product near the horizon. The relative phase $\delta(\omega,l)$ is physical and encodes information about the wave scattering in the exterior geometry of the black hole.

With this information, the Wightmann 2-point function can be readily computed, after selecting a state, via the formula
\begin{equation}
    \langle \phi(t,r,\theta) \phi(t',r',\theta')\rangle = \sum_{l,l'} \int\frac{d\omega \,d\omega'}{\sqrt{\omega \omega'}} \, e^{-i(\omega t-\omega't')}e^{i(l\theta-l'\theta')} f_{\omega,l}(r) f_{\omega',l'}^*(r') \langle a_{\omega,l} a^\dagger_{\omega',l'}\rangle +\text{h.c.} \label{propagator}
\end{equation}
In our case, the state of interest will be the Hartle-Hawking state with occupation numbers:
\begin{equation}
    \langle a^\dagger_{\omega,l}a_{\omega',l'}\rangle_{HH}= \frac{e^{-\beta\omega}}{1-e^{-\beta\omega}} \delta(\omega-\omega') \delta_{ll'} \label{HHstate}
\end{equation}
and $\beta=2\pi$ for the BTZ geometry (\ref{btz}). The boundary-to-boundary 2-point function in this state becomes simply:
\begin{equation}
  \langle O(t,\theta) O(t',\theta')\rangle_{HH}  = \sum_{l=-\infty}^\infty e^{il(\theta-\theta')}\int_{-\infty}^\infty \frac{d\omega}{\omega} \frac{e^{2\pi \omega}}{e^{2\pi \omega}-1} e^{-i\omega(t-t')} C^2(\omega,l) \label{btobexpression}
\end{equation}
All these quantities will be computed explicitly later in this Section.

\paragraph{Behind the horizon} In the black hole interior region, the general solution to the equations of motion is the same one obtained above but the quantization of the field proceeds in a different way. That is because $r$ becomes the time-like direction while $t$ becomes spacelike. A definite positive frequency wavefunction is, therefore, the solution that behaves as $(r-1)^{\frac{i\omega}{2}}$ near the horizon (and $(r-1)^{-\frac{i\omega}{2}}$ for negative frequencies). We, thus, impose the boundary condition
\begin{equation}
   g_{\omega,l}(r\to 1^-) \sim \Tilde{c}(\omega,l) (1-r)^{-\frac{i\omega}{2}} \label{intbc}
\end{equation}
which fixes the solution up to the overall normalization $\Tilde{c}(\omega,l)$. This is, in turn, fixed by setting it equal to the previously obtained phase $e^{-i\delta(\omega,l)}$ of the exterior left-movers in (\ref{normalization}), for continuity 
\begin{equation}
    \tilde{c}(\omega,l) = e^{-i\delta(\omega,l)} \label{intnorm}
\end{equation}
The positive frequency wavefunctions are then $g^*_{\omega,l}(r)$ whose phase $e^{i\delta(\omega,l)}$ matches that of the exterior right-movers in (\ref{normalization}).

However, there is an additional complication: Since $t$ is a spacelike direction, the $e^{\pm i \omega t}$ wavefunctions describe interior left and right movers respectively. But, since there is no other boundary condition to constrain their relative weight, each of them requires its own ladder operators. This means that 2 pairs of creation-annihilation operators, $(b^{(l)}_{\omega, l},b^{(l)}_{\omega, l}\,^\dagger)$ and $(b^{(r)}_{\omega, l},b^{(r)}_{\omega, l}\,^\dagger)$, are needed to describe the black hole interior and the field operator reads:
\begin{align}
    \phi(t,r,\theta)&=\sum_{l=-\infty}^\infty e^{il \theta}\int_0^\infty \frac{d\omega}{\omega^{\frac{1}{2}}} \,\left(b^{(l)}_{\omega, l}\,e^{-i\omega t}  g_{\omega,l}(r) +b^{(r)}_{\omega,l}\,e^{i\omega t}  g_{\omega,l}(r) + \text{h.c.} \right) \label{fieldopint}
\end{align}
The action of the interior ladder operators is determined by demanding continuity of the field across the future horizon, in the Hartle-Hawking state. Using the boundary condition (\ref{intbc}), (\ref{intnorm}), the field operator (\ref{fieldopint}) near ${\cal H}^+$  becomes:
\begin{equation}
    \phi(t,r\to 1^-,\theta) \sim \sum_{l=-\infty}^\infty e^{il \theta}\int_0^\infty \frac{d\omega}{\omega^{\frac{1}{2}}} \,\left(b^{(l)}_{\omega, l}\,e^{-i\delta(\omega,l)}e^{-i\omega (t+\frac{1}{2}\log(1-r))} + b^{(r)}_{\omega,l}\, e^{-i\delta(\omega,l)}e^{i\omega (t-\frac{1}{2}\log(1-r))}  + \text{h.c.} \right) \label{nearhorizon-}
\end{equation}
An ingoing (left-moving) light-ray near the horizon corresponds to the trajectory $t+\frac{1}{2}\log|r-1|=v=const$. Comparing (\ref{nearhorizon-}) to the near horizon behavior of the exterior modes (\ref{fieldop}) with (\ref{normalization}), continuity of the field action on $|HH\rangle$ across the horizon along a $v=const$ ingoing light-ray necessitates two properties: (a) the operator identification
\begin{equation}
     b_{\omega,l}^{(l)} = a_{\omega,l}\quad \quad\quad      b_{\omega,l}^{(l)}\,^\dagger = a_{\omega,l}^\dagger \label{identification1}
\end{equation}
which, physically, is the requirement that all in-going (left-moving) flux of exterior particles arriving at ${\cal H}^+$ become left-moving excitations in the black hole interior, and (b) the relation:
\begin{align}
    b_{\omega,l}^{(r)}|HH\rangle &= e^{-\pi \omega} a_{\omega,l}^\dagger|HH\rangle\quad \quad b_{\omega,l}^{(r)}\,^\dagger|HH\rangle = e^{\pi \omega} a_{\omega,l}|HH\rangle \label{identification2}
\end{align}
which mathematically follows from the discontinuity in the principal value of the outgoing wavefunction evaluated at constant $v$, $e^{-i\omega \log(r-1)}$, across the branch point $r=1$. A more physical way of understanding this is as the requirement of smoothness across the horizon in $|HH\rangle$, enforced by demanding that the field correlator between two points on the opposite side of the horizon has the same coincidence singularity as a flat space 2-point function. Unlike (\ref{identification1}), this is not an operator equation but rather a property of the Hartle-Hawking state ---the demand that interior and exterior out-going modes are thermally entangled. The above relations imply the occupation numbers:
\begin{align}
    \langle b_{\omega,l}^{(r)}\,^\dagger b_{\omega',l'}^{(r)} \rangle_{HH}&= \frac{e^{-2\pi\omega}}{1-e^{-2\pi\omega}} \delta_{ll'}\delta(\omega-\omega')  \quad \quad \quad
    \langle b_{\omega,l}^{(r)}a_{\omega',l'} \rangle_{HH}&= \frac{e^{-\pi\omega}}{1-e^{-2\pi\omega}}\delta_{ll'}\delta(\omega-\omega') \label{intoccupationnumbers}
\end{align}

\subsubsection*{BTZ computations}
\paragraph{Normalizable modes}
In the BTZ case, all the above information can be computed exactly. The radial mode equation is
\begin{equation}
   r^2 \left(r^2-1\right) f''(r)+ r \left(3 r^2-1\right) f'(r)+f(r) \left(\frac{r^2 \omega ^2}{r^2-1}-l^2\right)=0
\end{equation} 
admitting the general solution
\begin{align}
   f_{\omega,l}(r)&= c_1\left(r^2-1\right)^{-\frac{i \omega }{2}} \, r^{-i l}  \, _2F_1\left(-\frac{i}{2}  (l+\omega ),-\frac{i}{2}  (2 i+l+\omega );1-i l;r^2\right) \nonumber\\
   &+c_2\left(r^2-1\right)^{-\frac{i \omega }{2}} \, r^{i l} \, _2F_1\left(\frac{i}{2}  (l-\omega ),\frac{i}{2}  (-2 i+l-\omega );i l+1;r^2\right) \label{gensol}
\end{align}
Near the asymptotic boundary, this behaves as
\begin{equation}
    f_{\omega,l}(r\to \infty)\to  \frac{c_2 e^{\frac{\pi  (l-\omega)}{2}} \Gamma (i l+1)}{\Gamma \left(\frac{i}{2}  (-2 i+l-\omega )\right) \Gamma \left(\frac{i}{2}  (-2 i+l+\omega )\right)}+\frac{c_1 e^{-\frac{\pi  (l+\omega)}{2} }\Gamma (1-i l)}{\Gamma \left(-\frac{i}{2}  (2 i+l-\omega )\right) \Gamma \left(-\frac{i}{2}  (2 i+l+\omega )\right)} +O(r^{-2})
\end{equation}
and cancellation of the non-normalizable finite term fixes $c_2$ in terms of $c_1$:
\begin{equation}
    c_2=-c_1\frac{ e^{-\pi l} \Gamma (1-i l) \Gamma \left(\frac{i}{2}  (-2 i+l-\omega )\right) \Gamma \left(\frac{i}{2}  (-2 i+l+\omega )\right)}{\Gamma (i l+1) \Gamma \left(-\frac{i}{2}  (2 i+l-\omega )\right) \Gamma \left(-\frac{i}{2}  (2 i+l+\omega )\right)} \label{c2ofc1}
\end{equation}
The radial wavefunction, then, correctly decays asymptotically as (\ref{asymptoticbc}) with coefficient:
\begin{equation}
    C(\omega,l)= c_1\frac{2  \sinh (\pi  l) \Gamma (1-i l) \Gamma \left(\frac{i}{2}  (-2 i+l+\omega )\right)}{\left(e^{\pi  l}-e^{\pi  \omega }\right) \Gamma \left(-\frac{i}{2}  (l-\omega )\right)}
\end{equation} 
In order to fix the overall normalization $c_1$ of the wavefunction, we  look at its near horizon behavior which reads:
\begin{align}
    f_{\omega,l}(r\to 1^+) &\sim  c_1 \,(r-1)^{-\frac{i \omega}{2}}\frac{2^{-\frac{i \omega }{2}} \left(e^{ l \pi }-e^{-l \pi }\right) \pi   \text{csch}(\pi  \omega ) \Gamma (1-i l)}{\left(e^{l \pi }-e^{\pi  \omega }\right) \omega  \Gamma \left(-\frac{i}{2}  (l-\omega )\right) \Gamma \left(-\frac{i}{2}  (l-\omega +2 i)\right) \Gamma (-i \omega )}\nonumber\\
    &+c_1\,(r-1)^{\frac{i \omega}{2} } \frac{2^{\frac{i \omega }{2}} e^{-\pi  (l+\omega )} \pi  \text{csch}(\pi  \omega ) \Gamma (1-i l) \left(\text{csch}\left(\frac{1}{2} \pi  (l+\omega )\right) \sinh \left(\frac{1}{2} \pi  (l-\omega )\right)+e^{l \pi }\right)}{\omega  \Gamma (i \omega ) \Gamma \left(-\frac{i}{2}  (l+\omega )\right) \Gamma \left(-\frac{i}{2}  (l+\omega +2 i)\right)}
\end{align}
The normalization condition (\ref{normalization}), then, fixes $c_1(\omega,l)$ to
\begin{align}
   c_1&=\frac{e^{\pi  l} \omega  \sqrt{\left(e^{\pi  l}-e^{\pi  \omega }\right) \left(e^{\pi  (l+\omega )}-1\right)} (\coth (\pi  l)-1) \sinh (\pi  \omega )}{2 \pi  \Gamma (1-i l)}\times\nonumber\\
   &\times\Big|\Gamma (-i \omega )\Big| \sqrt{\Gamma \left(-\frac{i}{2}  (l-\omega )\right)\Gamma \left(-\frac{i}{2}  (l+\omega )\right)\Gamma \left(-\frac{i}{2}  (2 i+l-\omega )\right)\Gamma \left(-\frac{i}{2}  (2 i+l+\omega )\right)} \label{c1}
\end{align}
from which the phase $\delta(\omega,l)$ can be computed.

Finally, having determined all parameters of the general solution, the asymptotic behavior of the radial wavefunction is given by (\ref{asymptoticbc}) with coefficient $C(\omega,l)$ equal to:
\begin{equation}
   C(\omega,l)=\sqrt{\frac{\pi  \omega (l-\omega ) (l+\omega )  )}{2\left(e^{\pi  (l-\omega)}-1\right) \left(1-e^{-\pi  (l+\omega )}\right)} \frac{\left(e^{2\pi  \omega}-1\right)}{e^{2\pi\omega}}} \label{Comega}
\end{equation}
It is manifest that $C(\omega,l)$ is even in both $l$ and $\omega$. This is an important property which can be used to confirm that the boundary-to-boundary 2-point correlator obtained from (\ref{btobexpression}) satisfies the KMS relation:
\begin{equation}
\langle O(t-2\pi i,\theta) O(t',\theta')\rangle_{HH}= \langle O(t',\theta')O(t,\theta) \rangle_{HH}
\end{equation}
which in frequency space becomes:
\begin{align}
    G(-\omega,l)&=e^{-2\pi\omega}G(\omega,l)\\
   \text{for } G(\omega,l)&= \int_0^{2\pi}d\theta\,\int_{-\infty}^\infty dt e^{i\omega t} e^{-il\theta}\langle O(t,\theta) O(0,0)\rangle_{HH}=\frac{\pi(l-\omega)(l+\omega)}{2\left(e^{\pi  (l-\omega)}-1\right) \left(1-e^{-\pi  (l+\omega )}\right)} \nonumber
\end{align}
The poles of this expression give the correct quasinormal frequencies and its functional form is consistent with \cite{Birmingham:2001dt} in the infinite volume limit.

\paragraph{Interior modes}
We now want to construct the interior modes $g_{\omega,l}(r)$ and study their behavior near the singularity. The boundary condition (\ref{intbc}) for the negative frequency modes selects the solution (\ref{gensol}) with coefficients obeying:
\begin{align}
    c^{g}_2&=-\frac{c_1^g \Gamma (1-i l) \Gamma \left(\frac{i}{2} (l-\omega )\right) \Gamma \left(\frac{i}{2}  (-2 i+l-\omega )\right)}{\Gamma (i l+1) \Gamma \left(-\frac{i}{2}  (l+\omega )\right) \Gamma \left(-\frac{i}{2}  (2 i+l+\omega )\right)} \label{c2g}
\end{align}
while further normalizing it according to (\ref{intnorm}) fixes the remaining coefficient to:
\begin{align}
    c_1^g&=-\frac{\omega  \sqrt{e^{\pi  l }-e^{-\pi\omega}}  \big|\Gamma (-i \omega )\big|  \sinh ^2\left(\frac{1}{2} \pi  (l-\omega )\right)\text{csch}(\pi  l)}{\pi  \sqrt{\left(e^{\pi  l}-e^{\pi  \omega }\right)}  \Gamma (1-i l)}\times\nonumber\\
    &\times \sqrt{\Gamma \left(-\frac{1}{2} i (l-\omega )\right) \Gamma \left(-\frac{1}{2} i (l+\omega )\right) \Gamma \left(-\frac{1}{2} i (2 i+l-\omega )\right) \Gamma \left(-\frac{1}{2} i (2 i+l+\omega )\right)} \label{c1g}
\end{align}

\paragraph{Behavior near the singularity} For all non-vanishing angular momenta $l\neq 0$ the radial wavefunctions are irregular but bounded near the singularity:
\begin{equation}
    g_{\omega,l\neq 0}(r\to 0) \sim c_2^g(\omega,l) e^{\pi\omega} \,r^{il}+c_1^g(\omega,l) e^{\pi\omega} \,r^{-il} 
\end{equation}
The only exception is the zero angular momentum mode. Its precise expression can be obtained by taking the $l\to 0$ limit of the general solution (\ref{gensol}) with coefficients (\ref{c2g}) and (\ref{c1g}). As $r\to 0$, the negative frequency modes diverge logarithmically:
\begin{equation}
    g_{\omega, l=0} (r\to 0) \sim \gamma(\omega) \log r \label{zeromode}
\end{equation}
with coefficient:
\begin{equation}
    \gamma(\omega) = \sqrt{\frac{\omega}{\pi}}\frac{e^{\frac{\pi\omega}{2}}-e^{-\frac{\pi\omega}{2}}}{\sqrt{\sinh \pi \omega}} \label{gamma}
\end{equation}
The positive frequency wavefunctions $g^*_{\omega,l=0}$ have the same limit.

The implication of this divergence for the Wightman propagator is clear. The divergent amplitude of the S-wave wavefunction makes it the dominant contribution to the propagator (\ref{propagator}) when one of the field insertions approaches the singularity. The 2-point function is, therefore, approaching a rotation invariant and logarithmicallty divergent function, for any location of the second insertion
\begin{equation}
    \langle\phi(r\to 0, t,\theta) \phi(r',t,\theta')\rangle \to  F(r',t'-t)\,\log r
\end{equation}

More precisely, the bulk-to-boundary correlator in the limit where the bulk point approaches the singularity and $\Delta t=t'-t$ reads:
\begin{align} \langle \phi(t,r\to 0,\theta) O(t',\theta')\rangle_{HH} & \to \int_{0}^\infty \frac{d\omega}{\omega} \Bigg[\left(\langle a_{\omega,0} b_{\omega,0}^{(r)}\rangle_{HH} e^{-i\omega \Delta t} +\langle a_{\omega,0}^\dagger a_{\omega,0}\rangle_{HH} e^{i\omega \Delta t} + \text{h.c.}  \right) \times \nonumber\\
&\times \gamma(\omega) C(\omega,0)  \Bigg]\log r
   \label{2ptexpression2}
\end{align}
Plugging in the occupation number expectation values (\ref{HHstate}), (\ref{intoccupationnumbers}) as well as the results (\ref{Comega}), (\ref{gamma}) which yield $\gamma(\omega)C(\omega,0) = \omega^2 $ we get:
\begin{align}
    \langle \phi(t,r\to 0,\theta) O(t',\theta')\rangle_{HH} & \to \int_{0}^\infty d\omega\,\omega \Bigg[\left(\frac{e^{-\pi \omega}}{1-e^{-2\pi\omega}} e^{-i\omega \Delta t} +\frac{e^{-2\pi \omega}}{1-e^{-2\pi\omega}} e^{i\omega \Delta t} \right)  +\nonumber\\
  &+ \left( \frac{ 1}{1-e^{-2\pi\omega}} e^{-i\omega \Delta t} +\frac{e^{-\pi \omega}}{1-e^{-2\pi\omega}} e^{i\omega \Delta t}  \right)  \Bigg]\log r \nonumber\\
  &= \left[\int_{-\infty}^\infty d\omega \frac{\omega\,e^{-\pi\omega}}{1-e^{-2\pi\omega}}e^{-i\omega (t'-t)}+\int_{-\infty}^\infty d\omega \frac{\omega\,}{1-e^{-2\pi\omega}}e^{-i\omega (t'-t)} \right] \log r \nonumber\\
  &=-\frac{\log r}{\sinh^{2}\left(t'-t\right)} \label{2ptexpression3}
\end{align}
The is indeed the expression derived in the main text by computing the propagator via the method of images. What the computation in this Appendix clarifies is that the logarithmic divergence in the correlator is a direct consiquence of the divergence of the individual zero angular momentum modes in the interior.

As was discussed in \cite{Hamilton:2006fh, Hamilton:2007wj}, this divergence should be interpreted as a signature of the black hole singularity in the CFT. The CFT dual of a local bulk field operator becomes non-normalizable as we approach the black hole singularity. The apparent non-normalizability of these operators is likely tied to the $N\to \infty$ limit required to make contact with the semiclassical holographic bulk description ---hinting at a possible framework for the resolution of the singularity at finite $N$.

\subsection{5D AdS-Schwarzschild black hole}
The analysis for higher dimensional black holes is conceptually identical to our BTZ discussion, albeit technically less tractable. The metric in AdS-Schwarzschild coordinates is:
\begin{equation}
    ds^2=-\frac{\left(r^2-1\right) \left(r^2+2\right) dt^2}{r^2}+\frac{r^2 dr^2}{\left(r^2-1\right) \left(r^2+2\right)}+ r^2\left( d\chi^2 \sin ^2\theta  \sin ^2\varphi + d\varphi^2 \sin ^2\theta + d\theta ^2\right)
\end{equation}
where the horizon is at $r=L_{AdS}=1$.

The rules of the quantization are the same, the formal expression for the boundary-to-boundary propagator is (\ref{btobexpression}) ---with the angular wavefunctions replaced by the spherical harmonics $Y_I$ on $S^3$, where $I=(l,m,m')$--- and the bulk-to-boundary 2-point function when the bulk point approaches the singularity becomes 
\begin{align} \langle \phi(t,r\to 0,\Omega) O(t',\Omega')\rangle_{HH} & \to \sum_{I} Y_I(\Omega')Y^*_I(\Omega)\int_{0}^\infty \frac{d\omega}{\omega} \left(\langle a_{\omega,0} b_{\omega,0}^{(r)}\rangle_{HH} e^{-i\omega \Delta t} +\langle a_{\omega,0}^\dagger a_{\omega,0}\rangle_{HH} e^{i\omega \Delta t} \right) \times \nonumber\\
&\times g_{\omega,l}(r\to 0) C(\omega,l) + \text{h.c.}  
   \label{2ptexpression5d}
\end{align}
The occupation numbers are again given by (\ref{HHstate}) and (\ref{intoccupationnumbers}). Since $C(\omega)$ is related to the Fourier transform of the boundary 2-point function, the only data we need to obtain is the behavior of the scalar field modes in the black hole interior $g_{\omega,l}$ as they approach $r=0$. This is what we turn our attention to now.

Since the angular momentum is a conserved charge of the CFT, it is of little interest for our purposes. We can simply work in the fixed angular momentum sector by considering instead the correlator of the local bulk field $\phi(t,r\to 0,\Omega)$ with the $l=0$ mode of the boundary operator $O(t) = \int d\Omega \,O(t,\Omega)$. The wave equation for the $l=0$ radial wavefunction reads:
\begin{equation}
 \left(r^4+r^2-2\right) r f_{\omega,l=0}''(r)+ \left(5 r^4+3 r^2-2\right) f_{\omega,l=0}'(r)+\frac{r^5 \omega^2 }{r^4+r^2-2}f_{\omega,l=0}(r)  =0 \label{KG5d}
\end{equation}
The $\omega=0$ solution can be obtained analytically
\begin{equation}
    f_{\omega=0,l=0}(r) = c_2+c_1 \left(\frac{1}{6} \log \left(r^2-1\right)+\frac{1}{12} \log \left(r^2+2\right)-\frac{\log (r)}{2}\right)
\end{equation}
and demanding normalizability
\begin{equation}
    f(r\to \infty) \to \frac{C}{r^4} \label{bc5d}
\end{equation}
sets $c_2=0$. This special mode, therefore, exhibits logarithmic dicergences at the black hole singularity $r=0$ and at the horizon $r=1$.

The horizon divergence is an uninteresting feature of the $\omega=0$ mode that disappears for all other solutions. The general near horizon behavior can be obtained by solving (\ref{KG5d}) locally about $r=1$ using the Frobenius expansion:
\begin{equation}
    f_{\omega,l=0}= (r-1)^{-\frac{i\omega}{6}} \sum_{n=0}^\infty c_n(\omega) (r-1)^n + (r-1)^{\frac{i\omega}{6}} \sum_{n=0}^\infty d_n(\omega) (r-1)^n
\end{equation}
The series coefficients $c_n(\omega),d_n(\omega)$ for $n\geq 1$ are all recursively determined in terms of $c_0(\omega)$ and $d_0(\omega)$ by demanding that the equations of motion are satisfied order by order in the $(r-1)$ expansion. The two remaining coefficients are, then, fixed by the asymptotic boundary condition (\ref{bc5d}) and the near horizon normalization, as for the BTZ. The resulting coefficients will once again reduce to a relative phase $c_0(\omega)=e^{-i\delta(\omega)}$, $d_0(\omega)=e^{i\delta(\omega)}$ which, however, we will not determine explicitly in this case.

The divergence near the black hole singularity, on the other hand, persists for all $\omega$ (and all $l$ ---in contrast to the BTZ example). To determine the negative frequency interior modes $g_{\omega, l=0}(r)$ we need to find the solution of (\ref{KG5d}) that approaches
\begin{equation}
    g_{\omega,l=0}(r\to 1^-) \sim e^{-i\delta(\omega)}(1-r)^{-\frac{i\omega}{6}} \label{horbc5d0}
\end{equation}
near the horizon. Employing the Frobenius method for locally solving the radial wave equation again, this time about $r=0$, we find that the general solution for $g(r)$ has the form:
\begin{equation}
    g_{\omega,l=0}(r) = (1-r)^{-\frac{i\omega}{6}}\left(\log r \sum_{n=0}^\infty a_n(\omega) r^n +  \sum_{n=0}^\infty b_n(\omega) r^n\right) \label{r0frobenius}
\end{equation}
where we factored out a $(1-r)^{-\frac{i\omega}{6}}$ term for convenience ---since it is analytic about $r=0$ for $r<1$, it can alternatively be absorbed in the series. It is straightforward to show that all coefficients in this expansion are recursively determined in terms of $a_0(\omega)$ and $b_0(\omega)$ by satisfying (\ref{KG5d}) order by order in $r$. If the series is convergent for $r<1$, the remaining 2 coefficients, $a_0(\omega)$ and $b_0(\omega)$, are then determined by demanding (\ref{horbc5d0}):\footnote{The structure of the recurrence relations guarantee that the $b_n$s depend on both $a_0$ and $b_0$, whereas the $a_n$s only depend on $a_0$.}
\begin{equation}
     \sum_{n=0}^\infty b_n(\omega)=e^{-i\delta(\omega)}  \label{horbc5d}
\end{equation}

Is it possible that enforcing (\ref{horbc5d}) sets $a_0=0$, killing the logarithmically divergent solution? The answer is no. The solution satisfying the boundary condition (\ref{horbc5d0}) and its complex conjugate linearly span the set of all solutions. If $g_{\omega,l=0}$ does not contain the logarithmic solution near $r=0$, then neither does $g_{\omega,l=0}^*$, nor any linear combination of the two. But, as we just showed, the general solution is of the form (\ref{r0frobenius}) near $r=0$ and we have arrived at a contradiction. Hence, $g_{\omega,l=0}(r)$ must be logarithmically divergent at $r=0$.

Putting everything together, the correlator between a local operator near the black hole singularity and a boundary operator with zero angular momentum reads:
\begin{align} \langle \phi(t,r\to 0,\Omega) O(t')\rangle_{HH} &\to \left[\int_{-\infty}^\infty \frac{d\omega}{\omega} \frac{e^{-\pi\omega}+1}{1-e^{-2\pi\omega}}e^{-i\omega (t'-t)} a_0(\omega) C(\omega,0) \right] \log r
   \label{2ptexpression5dfinal}
\end{align}
where we used the fact that both $a_0(\omega)$ and $C(\omega)$ are even. The precise time dependence of the 2-point function can, in principle, be determined as described above but it would require numerical analysis of the differential equations that goes beyond the scope of our work. The key observation is the appearance of the logarithmic divergence in the correlator in the 5D AdS-Schwarzschild black hole, which indicates that operators close to the singularity become non-normalizable CFT operators. The same feature persists in other dimensions as well.

\addcontentsline{toc}{section}{References}
\bibliography{References}
\bibliographystyle{JHEP}

\end{document}